\documentclass[twocolumn,twocolappendix]{aastex631}

\usepackage{CJK}
\usepackage{graphicx}
\usepackage{xcolor}
\usepackage{amsmath}
\usepackage{hyperref}
\usepackage{natbib}

\newcommand{\LCDM}{$\Lambda$-CDM}
\newcommand{\MSun}{\mathrm{M}_\odot}
\newcommand{\SM}{M_\star}
\newcommand{\SFR}{\dot{M}_\star}

\defcitealias{LNB15}{LNB15}

\begin{document} 

\title{SF-R You Sure? The Conflicting Role of Star Formation Rates in Constraining the Evolution of Milky Way Analogues in Cosmological Simulations}

\author[0000-0003-0147-9562]{Alicia M. Savelli}
\affiliation{Canadian Institute for Theoretical Astrophysics, University of Toronto, 60 St. George Street, Toronto, ON M5S 3H8, Canada}
\affiliation{David A. Dunlap Department of Astronomy \& Astrophysics, University of Toronto, 50 St. George Street, Toronto, ON M5S 3H4, Canada}
\affiliation{Dunlap Institute for Astronomy \& Astrophysics, University of Toronto, 50 St. George Street, Toronto, ON M5S 3H4, Canada}
\affiliation{Department of Physics, Brock University, 1812 Sir Isaac Brock Way, St. Catharines, ON L2S 3A1, Canada}

\author[0000-0003-2573-9832]{Joshua S. Speagle (\begin{CJK*}{UTF8}{gbsn}沈佳士\ignorespacesafterend\end{CJK*})}
\affiliation{Department of Statistical Sciences, University of Toronto, 9th Floor, Ontario Power Building, 700 University Ave, Toronto, ON M5G 1Z5, Canada}
\affiliation{David A. Dunlap Department of Astronomy \& Astrophysics, University of Toronto, 50 St. George Street, Toronto, ON M5S 3H4, Canada}
\affiliation{Dunlap Institute for Astronomy \& Astrophysics, University of Toronto, 50 St. George Street, Toronto, ON M5S 3H4, Canada}
\affiliation{Data Sciences Institute, University of Toronto, 17th Floor, Ontario Power Building, 700 University Ave, Toronto, ON M5G 1Z5, Canada}

\author[0000-0001-8108-0935]{J. Ted Mackereth}
\affiliation{Canadian Institute for Theoretical Astrophysics, University of Toronto, 60 St. George Street, Toronto, ON M5S 3H8, Canada}
\affiliation{Dunlap Institute for Astronomy \& Astrophysics, University of Toronto, 50 St. George Street, Toronto, ON M5S 3H4, Canada}
\affiliation{David A. Dunlap Department of Astronomy \& Astrophysics, University of Toronto, 50 St. George Street, Toronto, ON M5S 3H4, Canada}

\author[0000-0002-8659-3729]{Norman Murray}
\affiliation{Canadian Institute for Theoretical Astrophysics, University of Toronto, 60 St. George Street, Toronto, ON M5S 3H8, Canada}
\affiliation{David A. Dunlap Department of Astronomy \& Astrophysics, University of Toronto, 50 St. George Street, Toronto, ON M5S 3H4, Canada}
\affiliation{Department of Physics, 60 St. George Street, Toronto, ON M5S 1A7, Canada}

\author[0000-0001-9298-3523]{Kartheik G. Iyer}
\altaffiliation{Hubble Fellow}
\affiliation{Columbia Astrophysics Laboratory, Columbia University, 550 West 120th Street, New York, NY 10027, USA}
\affiliation{Dunlap Institute for Astronomy \& Astrophysics, University of Toronto, 50 St. George Street, Toronto, ON M5S 3H4, Canada}

\correspondingauthor{Alicia M. Savelli}
\email{alicia.savelli@utoronto.ca}

\begin{abstract}
Milky Way analogues (MWAs) have long been studied by astronomers to place our Galaxy within an extragalactic context. With the power of cosmological simulations, we are now able to not only characterize MWAs today, but also watch as they evolve through cosmic time. We use the EAGLE and IllustrisTNG simulations to study a group of MWAs defined by their stellar mass (SM) and star formation rate (SFR). We trace these galaxies back along their evolution to investigate the star forming and mass assembly tracks taken by a galaxy to become a MWA today in light of these chosen parameters. We also take mock-observations of ``MWAs” at $z>0$ and trace them forwards in time to determine if galaxies that looked similar to the Milky Way earlier in their evolution still look like the Milky Way today, thus quantifying a selection efficiency which could inform future observational studies of MWAs.  We find that most galaxies with Milky Way-SM follow a similar evolution regardless of present-day SFR, although MWAs in IllustrisTNG generally have not quenched, leading to star formation histories that produce ``too-blue" galaxies today. Additionally, we find contamination by MWA-``imposters" in our mock-observations, with low selection efficiency at high redshift due to the tight constraint requiring convergence to the Milky Way’s present-day SFR. Our work suggests present-day SM may suffice as a stand-alone selection parameter and helps to clarify how MWAs \textit{should} be selected, and thus will be an important reference for future studies of both simulated and observed MWAs.

\end{abstract}


\section{Introduction \label{sec:intro}} 
\subsection{Milky Way analogues}

The classification of galaxies into different subgroups has been a long standing practice in astronomy since Hubble's original tuning fork \citep{Hubble1926}.  While Hubble categorized galaxies by their morphological structure into elliptical types and spiral types, nearly 100 years later astronomers continue to subdivide galaxies based on their properties in an effort to understand their evolution \citep[e.g.,][]{Conselice2014}.  One group of galaxies of particular interest to astronomers is Milky Way analogues (MWAs) -- galaxies with similar properties to our own Milky Way (MW).

Our Galaxy supplies us with a wealth of information which we are able to retrieve through surveys like Gaia~\citep{Gaia} and SDSS~\citep{SDSS} to help us understand processes such as star formation and stellar evolution, distributions of gas and dust, and chemical enrichment of the interstellar medium (ISM); however, MWAs offer additional information and perspectives not available to us in our own Galaxy.  

One key advantage of studying MWAs is the ability to place our Galaxy within an extragalactic context and examine how we compare to the broader galaxy population.  This idea motivates the question of whether the MW is a ``typical" galaxy of its type, sparking a growing movement to quantify the MW's uniqueness and efficacy to act as a baseline for our understanding of galaxy properties and evolution \citep[e.g.,][]{Flynn+2006,Hammer+2007,Yin+2009}. 

MWAs also provide valuable insight into our own Galaxy, allowing us to ``fill in the blanks".  Some of the MW's properties must be measured using alternate methods than those used for external galaxies -- and some properties cannot be directly measured at all -- given the nature of our placement within the dusty Galactic plane, causing light to be absorbed and scattered away from our instruments \citep[e.g.,][]{BrandtDraine2012,Marshall+2008}.  MWAs can then be used to infer some of these missing properties, following the Copernican assumption that the MW should not be special among galaxies sharing similar ``calibration properties".  Then, it can be assumed that a group of galaxies calibrated to match in such a way on given parameters should also match on other related properties \citep{Bottinelli+1985,deVacCorwin1986}.  Such a technique has been successfully used in the past to constrain the MW's colour \citep[e.g.,][]{LNB15,Fielder+2021}.   

Finally, we see the MW as it exists in a single moment in time, today.  While some of our Galaxy's history can be recovered through galactic archaeology, such as its merger history with galactic dynamics and chemical enrichment \citep[e.g.,][]{DeasonBelokurov2024}, this still does not paint an overall picture of the MW's evolution in every respect.  MWAs observed at higher redshift may present some insight into how the MW could have behaved in the past; however, these still offer only single snapshots in time, potentially acting as a ``baby book" filled with MW siblings, but providing no information on one single galaxy's continuous evolution \citep[e.g.,][]{Patel+2013,vanDokkum+2013,Papovich+2015,Tan+2024}.  Yet another caveat concerning observed, high-redshift MWAs is the uncertainty in the galaxies' present-day properties and behaviours; if a MWA's evolution diverged from that of the MW at some point between when that galaxy was observed and today, it may not be accurately informative of our Galaxy's history \citep[e.g.,][]{Kruijssen+2019,HortaSchiavon2024}.       

\subsection{Cosmological simulations}

Cosmological simulations may resolve some of the aforementioned obstacles in understanding the MW.  These are large-scale simulations that follow the growth of the cosmos from the initial quantum fluctuations, through the formation of large-scale structure and galaxies, and evolved to compare to observations of the Universe today \citep{SommervilleDave2015,Vogelsberger+2020}.  With these Universes-in-a-box, measurements of galaxy properties are simplified to read-outs of specified quantities, offering both consistency and precision not attainable with observations.  Further, cosmological simulations provide the opportunity of time-reversal, making it possible to follow a single galaxy's quasi-continuous evolution through cosmic time \citep{Springel+2005}.  

Although cosmological simulations are very promising contributions to the field of galaxy evolution, we note that they come with a few disadvantages that should not be overlooked.  First, due to the large range in scales spanned in cosmological evolution -- from the large-scale structure of the cosmic web, down to individual galaxies, down further to the individual stars and gas constituents making up those galaxies, and finally down to the elements, atoms, and particles making up those constituents -- it is not currently computationally possible to simulate all of the relevant processes at once.  Simulations thus have finite resolution limits and must turn to subgrid physics to model process on scales below those limits \citep[e.g.,][]{EAGLE2}.  Additionally, the properties we measure computationally will not be perfectly comparable to observations, although it is common to take ``mock observations" of a simulation attempting to mimic true observational techniques \citep[e.g.,][]{Tang+2021}.    

A number of different ``flavours" of cosmological simulations exist, each employing their own physics and numerics on a variety of scales.  Two main traditional contenders for numerical approaches are \textit{hydrodynamical models}, which simultaneously solve for gravity using $N$-body methods and evolve the coupled gas as a fluid -- e.g., EAGLE \citep{EAGLE1,EAGLE2}, Illustris \citep{Illustris1,Illustris2} \& IllustrisTNG \citep{TNG1,TNG2}, Mufasa \citep{Mufasa} \& Simba \citep{Simba} -- and \textit{semi-analytical models}, which evolve the dark matter and gas components separately using simplified flow equations describing bulk motions in the simulation box -- e.g., Santa Cruz SAM \citep{SCSAM1,SCSAM2,SCSAM3,SCSAM4} (see e.g., \citet{SommervilleDave2015,Vogelsberger+2020} for reviews).  A newer contender is \textit{semi-empirical models}, which are tuned specifically to match a range of observations -- e.g., UniverseMachine \citep{UM1,UM2}.  \textit{Zoom-in} simulations also exist, which evolve the Universe in a cosmological box with low resolution, then ``zoom in" on select galaxies and rerun the simulation at much higher resolution to capture important physical processes more accurately than possible with subgrid models
-- e.g., FIRE-2 \citep{FIRE}.  Finally, there are \textit{bespoke} zoom-in simulations which aim to reproduce the MW exactly for a given set of properties -- e.g., \citet{Guedes+2011,Wang+2015,Grand+2017,Font+2020,Agertz+2015,Garrison-Kimmel+2014,Sawala+2016,Libeskind+2020,Semenov+2014}.  With many excellent simulations at our disposal, ultimately for this work, we choose to make use of EAGLE and IllustrisTNG, as they are both large-scale hydrodynamical models with comparable but slightly different subgrid physics.

\subsection{Purpose and Outline\label{sec:Qs}}

There is no absolute or precise definition of MWAs; the selection criteria are subjective and dependent on the specific research questions under investigation. Since two groups of MWAs with distinct definitions may be comprised of different galaxies, one's choice in selection parameters may result in different evolutionary pathways between the groups and thus influence the conclusions drawn. 

This is exactly the aim of this work -- use the results of the evolutionary analysis of a group of galaxies to refine the selection criteria for MWAs of a specific type.  We establish this selection refinement for a group of MWAs in cosmological simulations, and subsequently assess its implications on observational studies.

Accordingly, the primary question motivating this work is then: how does the initial selection of MWAs bias our results?  Further, how \textit{should} MWAs be selected?  To lead our analysis, we introduce the following guiding questions:
\begin{enumerate}
    \item \textbf{What defines a Milky Way analogue?} The definition of a MWA includes both the choice in selection parameters as well as the specifics of the selection technique;
    
    \item \textbf{How do the selection criteria used to pick out Milky Way analogues alter the sample of evolutionary tracks that are followed?} To answer this question, in addition to our sample of MWAs, we consider control groups for each selection parameter. We then compare the evolutionary histories of the control groups to that of the MWAs to discover the role each parameter plays in ``making" a MWA; and
    
    \item \textbf{How can the evolutionary histories of simulated Milky Way analogues better inform future observations?} The results of our evolutionary analysis will provide insight for how we expect the MW to have looked or behaved in its past, which can then be used to guide the selection of observed MWAs.
\end{enumerate}

This paper will address each of the above inquiries in turn, and thus takes the following form.  We begin with a brief description of the cosmological simulations from which we draw our galaxy sample, as well as their relevant subgrid routines and physical models in Section~\ref{sec:data}.  In Section~\ref{sec:defining_MWAs}, we address the first of the guiding questions and specify the criteria (Section~\ref{sec:selection_parameters}) and techniques (Section~\ref{sec:selection_method}) used to select MWAs in the simulations, and highlight some relevant properties at $z=0$ (Sections~\ref{sec:top10}, \ref{sec:colour}).  In Section~\ref{sec:highz}, we examine the second guiding question and trace MWAs back to higher redshift to identify their progenitors (Section~\ref{sec:traceback}) and study and explain their evolutionary histories (Section~\ref{sec:hists}).  Section~\ref{sec:obs} then discusses the third question in outlining mock observations of MWAs in the simulations (Sections~\ref{sec:OSA},~\ref{lowz}) in order to quantify the efficiency of MWA selection at higher redshift (Section~\ref{sec:fosa}).  Finally, in Section~\ref{sec:disc}, we respond to our initial guiding questions having contextualized our results (Section~\ref{sec:As}) and discuss further implications in Sections~\ref{sec:SAs} and~\ref{sec:SFRbad}, wrapping up with conclusions and a summary in Section~\ref{sec:end}.  More key properties of MWAs not otherwise referenced in the main text can be found in Appendix~\ref{app:z=0}.

\section{Data \label{sec:data}} 
We consider galaxies from two large-scale, cosmological, (magneto-)hydrodynamical simulations: EAGLE and IllustrisTNG.  This section provides a brief overview of the simulations, their input parameters, relevant subgrid physics, and calibration to observables.  For a more comprehensive description of the prescriptions used in the simulations, we direct the reader to the methods papers referenced herein.  

\subsection{The EAGLE simulations \label{sec:eagle}} 

The EAGLE (Evolution and Assembly of GaLaxies and their Environments) project \citep{EAGLE1,EAGLE2} consists of a suite of cosmological, hydrodynamical simulations of the formation and evolution of galaxies and their gas, stars, black holes, and dark matter.  The evolution is modelled within the \LCDM\ universe as measured by the 2013 Planck cosmology \citep{Planck}.  The simulations were performed using a modified version of the $N$-Body gravity solver Tree-PM and smoothed particle hydrodynamics (SPH) code GADGET-3, most recently described in \cite{GADGET}.  The main modifications to the code are the pressure-entropy formulation of SPH~\citep{Hopkins2013}, the time stepping \citep{ DurierDallaVecchia2012}, and the subgrid physics -- which are briefly described in the following subsection.  

We pull our galaxy sample from the \texttt{Ref-L100N1504} simulation.  This is the largest-volume simulation, run within a periodic cube of side $L=100$ cMpc with an initially equal number $N=1504^3$ of collisionless dark matter particles of mass $m_\mathrm{dm}=9.70\times10^6\ \MSun$ and SPH particles of mass $m_\mathrm{g}=1.81\times10^6\ \MSun$, corresponding to an intermediate resolution.  The naming convention ``Ref-" denotes that the simulation is the ``reference" model, rather than one of the models run with recalibrated parameter values for the subgrid feedback mechanisms to better match observations of the present-day Galaxy Stellar Mass Function (GSMF) \citep{Baldry+2012, LiWhite2009}.  In addition to the GSMF, EAGLE was also tuned to broadly reproduce observations of galaxy sizes \citep{Shen+2003, Baldry+2012}, the stellar mass-halo mass relation \citep{UM1,Moster+2013}, and the stellar mass-black hole mass relation \citep{McConnellMa2013}.

\subsubsection{Relevant EAGLE subgrid physics}
Subgrid prescriptions are implemented to model star formation; heat, energy, and momentum injection into, as well as metal enrichment of, the interstellar medium (ISM) gas due to stellar winds, radiation, and supernovae; and the formation, accretion, and feedback of active galactic nuclei (AGN). Contrary to common practice for most earlier cosmological, hydrodynamical simulations, EAGLE employs only one type of feedback each for stars and for AGN -- stochastic thermal feedback, which the authors view as a simpler and more natural prescription.  Here we briefly summarize the physical model prescriptions relevant to this work -- the formation and feedback of stars and black holes. 

Star formation in the simulations is stochastic, where gas particles above a certain metallicity-dependent density threshold \citep{Schaye2004} -- meant to mimic cold, dense gas -- may be converted into star particles at a pressure-dependent rate following \citet{SchayeDallaVecchia2008} and using a Chabrier initial mass function \citep{Chabrier2003}.  Mass and nucleosynthesized metals are returned to the ISM following the stellar mass loss scheme in \citet{Wiersma2009b}, and then released mass and metals are transferred from stellar particles to their SPH neighbours at each time step.  As in \citet{Wiersma2009a}, smoothed abundances are used to compute element-by-element rates of the radiative cooling and photoheating of gas for H, He, C, N, O, Ne, Mg, Si, S, Ca, \& Fe. The stellar feedback prescription captures the effects of stellar winds, radiation pressure on dust grains, and supernovae collectively  rather than independently, as they cannot be properly distinguished at the resolution of the simulation.  The feedback is thermal, with a fixed temperature jump, and stochastic, with a probability that an SPH neighbour is heated related to the average energy injection of supernovae.  The amount of feedback is dependent on local gas metallicity and density, and the efficiency is tuned to reproduce the $z\sim0$ GSMF \citep{Baldry+2012,LiWhite2009}. 

Supermassive black holes (BHs) are seeded into the centres of halos with total mass greater than $10^{10}\ \MSun/h$, which are identified with a regularly run friends-of-friends algorithm.  The BHs grow by mergers with other BHs and by gas accretion with a rate that remains below the Eddington limit and is influenced by the BH mass and velocity relative to the surrounding gas, the local density and temperature, and the angular momentum of the accreting gas with respect to the BH. Similarly to stellar feedback, AGN feedback is also implemented thermally with a fixed temperature -- corresponding more closely to the ``quasar" feedback mode only and not the ``radio" mode -- and stochastically, with a probability related to the available reservoir of feedback energy, the fixed temperature jump, and the density of the the surrounding gas.  The AGN feedback was calibrated to reproduce the $z=0$ stellar mass-black hole mass scaling relation \citep{McConnellMa2013}. 

\subsection{The IllustrisTNG project\label{sec:TNG}}
IllustrisTNG~\citep[hereafter TNG]{TNG1,TNG2} is ``The Next Generation" of the original Illustris project, with improvements upon Illustris in the form of advanced numerical techniques and new physics prescriptions (particularly for seed magnetic fields and feedback from AGN), yielding reduced tensions between the simulation results and observational constraints.  The result is a set of cosmological, magnetohydrodynamical simulations of the formation and evolution of galaxies, following the same components as EAGLE of gas, stars, BHs, and dark matter in a \LCDM\ cosmology consistent with Planck~\citep{Planck2015} measurements.  The moving-mesh code Arepo~\citep{Arepo} is used to evolve the magnetohydrodynamic equations coupled with self-gravity, also computed using a Tree-PM solver.  This inclusion of magnetic fields is one of the main differences between TNG and the EAGLE simulations.

Our galaxy sample is collected from TNG100\footnote{We choose TNG100 as it is the closest TNG simulation in size and resolution to \texttt{Ref-L100N1504} in EAGLE.}~\citep{TNG100-0,TNG100-1,TNG100-2,TNG100-3,TNG100-4,TNG100-5}.  The suffix ``100'' denotes the size of the periodic box of this run, a cube of side $L=106.5$ cMpc with initially $N=1820^3$ each of collisionless dark matter particles of mass $m_\mathrm{dm}=7.5\times10^6\ \MSun$ and baryonic particles of mass $m_\mathrm{baryon}=1.4\times10^6\ \MSun$.  TNG is calibrated on observations of present-day galaxy properties and statistics including the GSMF \citep{Baldry+2008,Baldry+2012,Bernardi+2013,DSouza+2015}, the stellar mass-halo mass relation \citep{UM1,Moster+2013}, the stellar mass-stellar size relation \citep{Baldry+2012,Shen+2003}, the black hole-galaxy mass relation \citep{Kormendy+2013,McConnellMa2013}, and the total gas mass content within the virial radius of massive groups \citep{Giodini+2009,Lovisari+2015}.  In addition to these $z\sim0$ constraints, TNG is tuned to reproduce the overall shape of the cosmic star formation rate density at $z\lesssim10$ \citep{UM1,Oesch+2015}.

\subsubsection{Relevant IllustrisTNG subgrid physics}
TNG includes subgrid routines to model many physical processes in galaxy evolution, including stellar formation, evolution, and feedback; chemical enrichment and primordial and metal-line cooling; the formation, accretion, merging, and multi-mode feedback of supermassive black holes (BHs); and the seeding and amplification of cosmic magnetic fields.  Here, we provide a concise summary of the prescriptions for star formation, stellar feedback, and AGN feedback -- the physics most relevant to this work. 

A stochastic model is implemented for star formation, where a star forms when a gas cell reaches a given density threshold following the Kennicutt-Schmidt relation \citep{Schmidt1959,Kennicutt1989} and assuming a Chabrier initial mass function \citep{Chabrier2003}.  Stellar populations return mass and metals to the ISM through supernovae of type Ia and type II as well as through AGB stars, and the scheme follows the production and evolution of H, He, C, N O, Ne, Mg, Si, Fe, \& ``other metals" (the sum of all metals not individually tracked).  Once metal enriched, the gas cools radiatively in an ionizing UV background including self-shielding corrections.  Stellar feedback is launched directly from star forming gas to drive isotropic galactic outflows.  Unlike EAGLE, the feedback is kinetic and stochastic, with gas cells either fully or partially converting into wind particles which decouple from the gas until they either leave the ISM or have reached a maximum travel time, at which point they recouple and deposit mass, momentum, metals, and thermal energy to their new host cell. 

BHs form in halos of mass $5\times10^{10}\ \MSun h^{-1}$, accrete gas from neighbouring cells, and are merged if found to be within the accretion or feedback region of another BH.  Unlike EAGLE, TNG employs two methods of AGN feedback: a low-accretion, kinetic, ``radio" mode and a high-accretion, thermal, ``quasar" mode.  While the thermal mode remains the same as in the original Illustris project, in which thermal energy is injected into the immediately surrounding gas, TNG implements a new model for the kinetic feedback, which implements a kinetic BH-driven wind model shown to be responsible for quenching intermediate- and high-mass halo galaxies in the simulation~\citep{TNG1}.  

\section{Defining Milky Way analogues \label{sec:defining_MWAs}}

The first question formulating this study is \textit{what defines a ``Milky Way Analogue"?} There is no single, clear-cut answer to this question; rather, the definition often depends on the specifics of what is being studied.  Thus, the selection criteria for analogues will vary depending on the research question at hand.  

The definition of a MWA is comprised of two components: first, the chosen properties to resemble the MW -- the selection parameters -- and second, some metric to measure the similarity of analogues to the MW itself.  We will address both of these criteria thoroughly and in succession, beginning with identifying and motivating our choice in selection parameters for our specific science case, then outlining the method used to calculate our ``Milky Way-ness" metric.  Once our group of analogues is identified, we perform a deeper examination into our most MW-like galaxies.  Particularly, we investigate some of the key properties in our analogue group at $z=0$, focusing mainly on how efficiently the simulations are reproducing the MW's colour, with a more detailed look into other properties in Appendix~\ref{app:z=0}, including chemical enrichment and bulge-to-total ratio.

\subsection{Selection parameters \label{sec:selection_parameters}}
\begin{table*}[t]
    \begin{tabular}[t]{ccccccc}
        \hline
         &
        \multicolumn{3}{c}
        {\begin{tabular}{c}
            \textbf{Sample Size (EAGLE)}\\Total galaxies: 40312
        \end{tabular}} &
        \multicolumn{3}{c}
        {\begin{tabular}{c}
            \textbf{Sample Size (TNG)}\\Total galaxies: 53939
        \end{tabular}}\\ 
        \hline \hline
        Weighting Method & MWAs & SM cont. & SFR cont. & MWAs &  SM cont. & SFR cont.
        \\
        \hline \hline
        
            \begin{tabular}{c}
                \textbf{Method 1:}\\\textbf{Binary weights ($\gamma_1$)}
            \end{tabular} 
            & 302 & 1294 & 877 & 264 & 2525 & 2300 \\\hline
            \begin{tabular}{c}
                \textbf{Method 2:}\\\textbf{Distance weights ($\gamma_2$)}
            \end{tabular} 
            & 75 & 647 & 490 & 74 & 1290 & 1303 \\\hline
            \begin{tabular}{c}
                \textbf{Method 3:}\\\textbf{Population weights ($\gamma_3$)}
            \end{tabular} 
            & \textbf{60} & 585 & 475 & \textbf{62} & 1176 & 1278 \\\hline
    \end{tabular}
    \caption{(Effective) sample sizes of MWAs and control groups for the three selection methods in EAGLE and TNG. Effective sample sizes for the MWAs using the primary weighting method used in this paper -- population weights -- have been bolded for clarity.}
    \label{tab:methods}
\end{table*}
As this work focuses primarily on the star formation histories (SFHs) of MWAs, we have chosen our selection parameters to be stellar mass (SM) and star formation rate (SFR).  We make our selection at $z=0$ and trace our MWA group back through time, using the SFR of their main progenitors to map out SFHs.  Essentially, we aim to study the evolutionary paths required to produce a group of galaxies with present-day SM and SFR fixed to be similar to that of the MW.

We note that this definition of MWA may not be consistent with some of the MW's other present-day properties.  However, since these properties are not relevant to our science case of investigating SFHs, it is not necessary for our analogues to match the MW on these properties and therefore any inconsistencies of this nature do not impact nor lessen our results.  Thus, any mention of a MW ``analogue" throughout the remainder of this paper specifically references our own definition of MWA -- a galaxy similar to the MW in terms of its present-day SM and SFR.  Any conclusions we draw should not be extrapolated to MWAs of any other type, defined in their own way to address a different research question.

We study the impact of both SM and SFR on analogue selection and evolutionary histories by considering three distinct analogue groups: one group of \textit{Milky Way Analogues} selected on both SM ($\SM$) and SFR ($\SFR$), the \textit{SM-controlled} group selected on only SM, and the \textit{SFR-controlled} group selected on only SFR.  The control groups allow for a more careful treatment of the individual contributions of each parameter to the main MWA sample selection; for example, comparing the SM-controlled sample to the MWAs allows us to observe the changes to the sample's properties when excluding or including SFR, in order to illustrate the role that SFR plays in MWA evolution and MWA selection today.  Thus, the three analogue groups allow for an in-depth analysis of the functions of SM and SFR in evolving MWAs, both individually and in combination. 

The values used for the MW's present-day SM and SFR in the selection come from Licquia, Newman, and Brinchmann~\citep[hereafter LNB15]{LNB15}, which employed a hierarchical Bayesian analysis method on past measurements of SM and SFR of the MW, re-normalized to be directly comparable to external galaxies, obtaining results of:
\begin{align}
    M_{\star\mathrm{,MW}}&=6.08\pm1.14\times10^{10}\ \mathrm{M_\odot},\label{eq:LNB-SM}\\
    \dot{M}_{\star\mathrm{,MW}} &= 1.65 \pm 0.19\ \MSun\ \mathrm{yr}^{-1}.
    \label{eq:LNB-SFR}
\end{align}

\subsection{Selection methods \label{sec:selection_method}}

We select our analogue samples from the EAGLE and TNG cosmological hydrodynamic simulations at redshift $z=0$, corresponding to \texttt{Snapnum = 28} in EAGLE and \texttt{Snapnum = 99} in TNG100.  To improve computational efficiency, we perform a cut to include only galaxies with $\SM\geq 10^8\ \MSun$, as the very low-SM galaxies are numerous in both simulations and are not relevant to our study.  The result is a total sample of 40,312 galaxies in EAGLE and 53,939 galaxies in TNG.  

From this greater sample, we identify MWAs and both control groups by calculating for each galaxy a metric of its similarity to the MW amongst the background of the greater galaxy sample.  This metric corresponds to a weighting system which we call the \textit{Milky Way-ness} (MW-ness) factor, $\gamma$.  In order to pinpoint the optimal selection technique, we consider a total of three methods as outlined below, with each subsequent method an improvement on the previous.

The first method we consider to select MWAs is a simple box cut, or binary weighting system, using specified intervals on SM and SFR and which we call $\gamma_1$.  The second method calculates a MW-ness metric, which we denote $\gamma_2$, using $\chi^2$ weights to measure the distribution of galaxies relative to the MW itself.  The third and final method is an adjustment to the second method, by applying the $\chi^2$ weights and dividing by a probability density function (pdf) that accounts for a data skew towards low-SM, low-SFR galaxies.  This results in an improved measure of MW-ness, $\gamma_3$, measuring the distribution of galaxies from the MW compared to the background number density.  In addition to MWAs, we also identify SM- and SFR-controlled samples using each method.  All three selection methods are explained in detail in the following sections.  The primary method utilized in this paper is the third weighting method, the $\chi^2$ weights adjusted against the background galaxy population, and thus from now on we denote the weights by $\gamma\equiv\gamma_3$ unless otherwise specified.  

\begin{figure*}[htbp!]
    \centering
    \includegraphics[width=\textwidth]{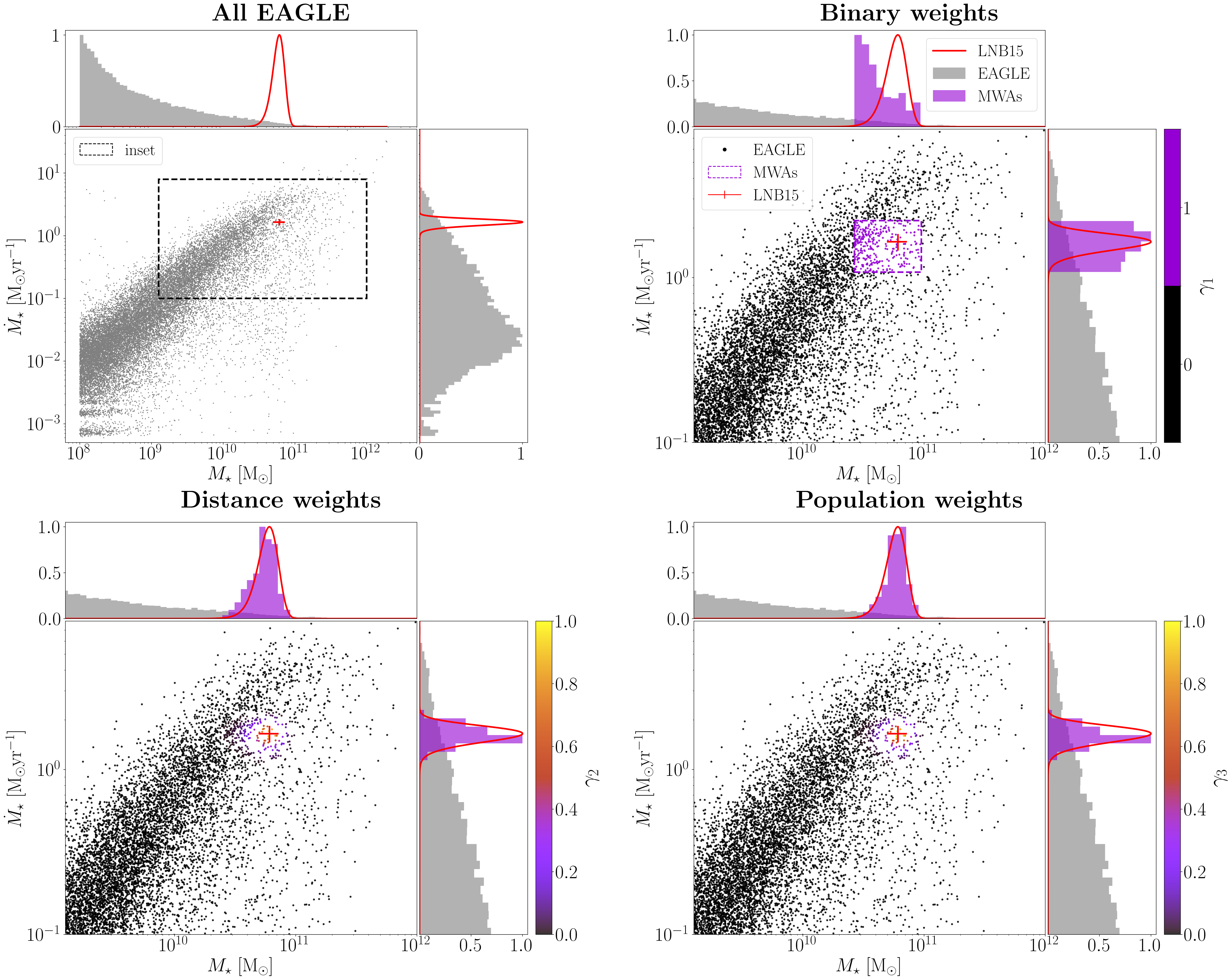}
    \caption{
    SFR vs SM corner plots illustrating the three MWA selection methods in EAGLE: the full background sample (\textbf{top left}), and zoomed-in distributions showing the binary weights ($\gamma_1$; \textbf{top right}), distance weights ($\gamma_2$; \textbf{bottom left}), and population weights ($\gamma_3$; \textbf{bottom right}).  The dashed box in the background sample panel indicates the location of the insets showing the three weighting systems.  The scatter plots are coloured by MW-ness, indicated by the violet box cut for the binary weights $\gamma_1$ and the colour bars for $\gamma_2$ and $\gamma_3$.  Histograms show the distribution of SM and SFR over all galaxies in grey and MWAs in violet.  To compare to the MW itself, the \citetalias{LNB15} values for the MW's SM and SFR are indicated by the red crossbars in the scatter plot, and red Gaussian curves in the histograms.  The optimal population selection method will have a distribution of MWAs that fits the red Gaussian curves exactly.  Thus, this paper uses method 3: population weights to select the populations of analogue groups, and so $\gamma\equiv\gamma_3$ hereafter unless otherwise specified.
    }
    \label{fig:selection}  
\end{figure*}

A visual comparison of the three methods is illustrated in
Figure~\ref{fig:selection} using galaxies from the EAGLE simulation, 
with corner plots in SFR vs. SM space.  This includes scatter plots coloured by $\gamma_1,\ \gamma_2,\ \gamma_3$ overlaid with a red crossbar depicting \citetalias{LNB15}'s measurements for the MW, as well as histograms showing the distribution of SM (top) and SFR (right) over the greater galaxy sample in grey and over the MWAs in violet, compared to a red Gaussian distribution calculated with \citetalias{LNB15}'s measurements.  Ideally, the distribution of MWAs should fit this red Gaussian, as a favourable selection should pick out more galaxies with values for SM and SFR closer to the MW's.  Additionally, the resulting galaxy sample sizes from each selection method are summarized in Table~\ref{tab:methods}.

\subsubsection{Method 1: Binary weights \label{sec:method1}}
The first method of MWA selection we consider is a box cut on SM and SFR of $\rho_p\pm3\Delta_p$, where $\rho_p$ and $\Delta_p$  are respectively the value and uncertainty reported by \citetalias{LNB15} for each parameter $p\in\{\textrm{SM,\ SFR}\}$.  This cut corresponds to a binary weighting system, $\gamma_1$, with galaxies inside the box assigned a weighting of 1 and galaxies outside the box assigned a weighting of 0:
\begin{equation}
    \begin{cases}
        \gamma_1 = 1 & \textrm{if }\SM\in\rho_\mathrm{SM}\pm3\Delta_\mathrm{SM}\\
        &\cap\ \SFR\in\rho_\mathrm{SFR}\pm3\Delta_\mathrm{SFR},\\
        \gamma_1=0 & \textrm{otherwise.}
    \end{cases}
\end{equation}

This box cut method results in a total of 302 MWA galaxies in EAGLE and 264 MWA galaxies in TNG.  Similar 1D cuts using only SM result in a SM-controlled sample size of 1294 in EAGLE and 2525 in TNG.  Likewise, the SFR-controlled sample sizes are 877 in EAGLE and 2300 in TNG.  

This is the simplest method for determining MWAs; however, the box cut does not allow for a metric of MW-ness and so all MWAs are considered equal, with no one galaxy considered more similar to the MW than any other.  From Figure~\ref{fig:selection}, it is also clear that this method poorly reproduces a Gaussian distribution of analogues, with a strong skew towards galaxies of lower SM and a close-to-even spread of galaxies in SFR.   

\subsubsection{Method 2: Distance weights \label{sec:method2}}
To introduce a MW-ness metric for measuring each individual galaxy's similarity to the MW, we apply a $\chi^2$ distribution to the entire galaxy sample:
\begin{equation}
    \chi_i^2=\left(\frac{M_{\star,i}-\rho_\textrm{SM}}{\Delta_\textrm{SM}}\right)^2+\left(\frac{\dot{M}_{\star,i}-\rho_\textrm{SFR}}{\Delta_\textrm{SFR}}\right)^2,
    \label{eq:chi2}
\end{equation}
where $M_{\star,i}(z)$ and $\dot{M}_{\star,i}(z)$ are respectively the SM and SFR of all galaxies in the sample, $\rho_p$, $\Delta_p$ are defined as above for binary weights, and the subscript $i\in[0,n]$ denotes the $i^\textrm{th}$ galaxy in the sample of $n$ total galaxies.  The MWAs are then identified by the weights, $W_{i}$, from the $\chi^2$ distribution in Equation~\ref{eq:chi2}:
\begin{equation}
    \gamma_2=\left\{W_{i}\right\}=\left\{\exp\left(-\frac{\chi_i^2}{2}\right)\right\}.
    \label{eq:W0}
\end{equation}
The resulting effective sample sizes, $N$, are calculated from
\begin{equation}
    N=\frac{\left(\sum_{i=0}^n W_i\right)^2}{\sum_{i=0}^n\left(W_i^2\right)},
    \label{eq:ess}
\end{equation}
to be 75 MWAs in EAGLE and 74 MWAs in TNG.  The weights for the control samples are calculated similarly, with only the relevant pair of $\{\rho_p,\Delta_p\}$, resulting in SM-controlled effective sample sizes of 647 for EAGLE and 1290 for TNG, and SFR-controlled effective sample sizes of 490 in EAGLE and  1303 in TNG. 

We call this method the \textit{distance weights}, as they are related to how ``far away" each individual MWA is from the MW itself in SM-SFR space.  These weights now provide us with a metric of MW-ness, which we call $\gamma_2$, on a continuous scale of 0 to 1, with 1 being the most MW-like and 0 being the least.  This scale is clearly seen in Figure~\ref{fig:selection}, as the galaxies more similar to the MW are coloured with a higher $\gamma_2$.  This method also produces a better fit to the Gaussian distributions in SM and SFR than the binary weights, making the $\chi^2$ method of MWA selection preferable; however, there is still a noticeable skew towards low-SM/low-SFR galaxies in the MWA group. 

\subsubsection{Method 3: Population weights \label{sec:method3}}
To optimize our MW-ness metric so that our MWA group better fits a Gaussian distribution, we must find a correction to our $\gamma_2$ weights that accounts for the skew towards low-SM or low-SFR galaxies.  To accomplish this, we calculate a pdf from a Gaussian kernel density estimation (KDE) on two dimensions (SM and SFR), using \texttt{scipy.stats.gaussian\_kde(X).pdf(X)} on the SM and SFR data \texttt{X}.  We then calculate our new $\gamma_3$ weights by dividing the $\gamma_2$ weights calculated in Equation~\ref{eq:W0} by this pdf:
\begin{equation}
    \gamma_3=\frac{\gamma_2}{\textrm{pdf}}=\left\{\exp\left(-\frac{\chi_i^2}{2}\right)/\textrm{pdf}\right\},
    \label{eq:W}
\end{equation}
resulting in an effective sample size of 60 MWAs in EAGLE and 62 MWAs in TNG.  A SM-controlled sample (identified with $\gamma_\textrm{3,SM}$) is found similarly by calculating one-dimensional $\chi^2$ weights  and Gaussian KDE using only SM, with an effective sample size of 585 galaxies in EAGLE and 1176 galaxies in TNG.  The SFR-controlled sample (identified with $\gamma_\textrm{3,SFR}$) is found in the same way, with sample sizes of 475 galaxies in EAGLE and 1278 galaxies in TNG.

We now have a method of determining MWAs that includes a metric of MW-ness and that offsets the large skew towards low-SM or low-SFR galaxies.  The MWA distribution in Figure~\ref{fig:selection} now better fits the red Gaussian, and the colouring indicates that galaxies most similar to the MW have $\gamma_3$ values concentrated around the MW.

To maintain a normalized weighting system with $\gamma_3=1$ corresponding to the MW itself, we include the \citetalias{LNB15} MW values for SM and SFR in our calculation, and scale $\gamma_3$ accordingly.  This results in a few EAGLE and TNG galaxies with $\gamma_\textrm{3,SM}>1$ or $\gamma_\textrm{3,SFR}>1$, so that it is no longer true that higher weights are more MW-like.  This weighting system compromise between the MW-ness of an individual galaxy and the number density of background galaxies with similar SM or SFR.  This technique reduces bias based on properties of an overwhelming background sample, and thus is useful for population averages, and so we name this method the \textit{population weights}.  The adjusted $\chi^2$ is then the method used to determine MWA populations in this paper, and so from now on we denote $\gamma\equiv\gamma_3$ unless specified otherwise.  However, we briefly note that when considering how MW-like an \textit{individual galaxy} is, rather than a population, the original $\chi^2$ distance weights $(\gamma_2)$ are preferred.  The distance weights are more effective in this case since they tell exactly how similar any one individual galaxy is to the MW itself in terms of its SM and SFR without accounting for the background population, maintaining that higher $\gamma_2$ corresponds to more MW-like. 

\begin{figure*}[htbp!]
    \centering
    \includegraphics[width=0.85\textwidth]{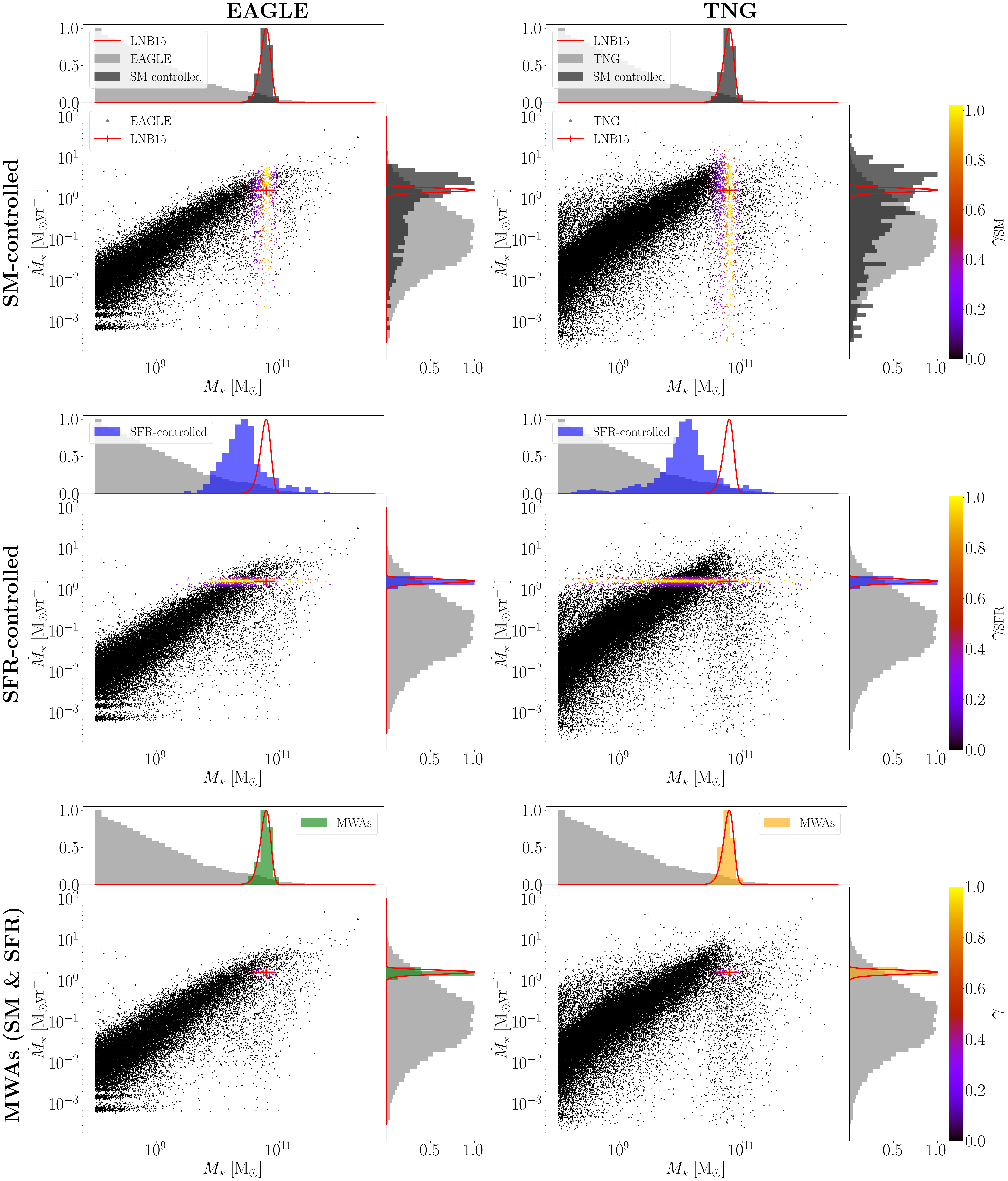}
    \caption{SFR vs SM corner plots with the same layout as Figure~\ref{fig:selection} for EAGLE (\textbf{left column}) and TNG (\textbf{right column}), showing the SM-controlled sample (\textbf{top row}; black histogram, points coloured by $\gamma_\textrm{SM}$ weights), SFR-controlled sample (\textbf{middle row}; blue histogram, points coloured by $\gamma_\textrm{SFR}$ weights), and MWAs (\textbf{bottom row}; green histogram for EAGLE, orange histogram for TNG, points coloured by $\gamma$ weights).  The red crossbars indicating the observed SM and SFR for the MW have been enlarged by a factor of 2 for better visibility.  The discretized striping displayed at low $\SFR$ in EAGLE is a numerical artifact resulting from resolution limits in the simulation.
    }
    \label{fig:analogues}
\end{figure*}

The associated selections based on the population weights are shown in Figure~\ref{fig:analogues}
for the SM-controlled sample (top), SFR-controlled sample (middle), and MWAs (bottom) from EAGLE (left) and TNG (right).
\begin{table}
    \centering
    \footnotesize
    \begin{tabular}[t]{@{}ccccc}
        \hline
        \multicolumn{5}{c}{\textbf{EAGLE}} \\ 
        \hline \hline
        MWA & ID & $\gamma_2$ &  $\textrm{SM}$  $(10^{10}\textrm{M}_\odot)$ & $\textrm{SFR}$  $(\textrm{M}_\odot \ \textrm{yr}^{-1})$ \\
        \hline \hline
        
        1 & 17418547 &    0.95706 &           6.01187 &         1.70513 \\\hline
     2 & 12755176 &    0.95521 &           5.93398 &         1.59788 \\\hline
     3 & 17407790 &    0.92156 &           5.96665 &         1.57556 \\\hline
     4 & 17757339 &    0.83705 &           6.37937 &         1.75175 \\\hline
     5 & 18345325 &    0.80959 &           5.90895 &         1.52984 \\\hline
     6 & 16594174 &    0.79594 &           6.79353 &         1.60167 \\\hline
     7 & 17404785 &    0.77068 &           5.29447 &         1.69082 \\\hline
     8 & 16689343 &    0.72383 &           5.47974 &         1.76544 \\\hline
     9 &  8077031 &    0.70894 &           6.97856 &         1.60093 \\\hline
    10 & 18115428 &    0.70710 &           5.36788 &         1.75458 \\\hline\\
        
        \hline
        \multicolumn{5}{c}{\textbf{TNG}}\\
        \hline \hline
        MWA & ID & $\gamma_2$ &  $\textrm{SM}$  $(10^{10}\textrm{M}_\odot)$ & $\textrm{SFR}$  $(\textrm{M}_\odot \ \textrm{yr}^{-1})$ \\
        \hline \hline
       1 & 499463 &    0.97837 &           5.85001 &         1.63954 \\\hline
     2 & 499113 &    0.89140 &           6.13212 &         1.74069 \\\hline
     3 & 448840 &    0.86887 &           5.55839 &         1.70090 \\\hline
     4 & 429041 &    0.84422 &           5.41839 &         1.64178 \\\hline
     5 & 427607 &    0.81674 &           6.69126 &         1.71509 \\\hline
     6 & 134852 &    0.78845 &           6.55614 &         1.54577 \\\hline
     7 & 389345 &    0.73348 &           5.63844 &         1.51976 \\\hline
     8 & 454539 &    0.73164 &           5.47979 &         1.76204 \\\hline
     9 & 453393 &    0.67647 &           5.55057 &         1.79295 \\\hline
    10 & 295079 &    0.67138 &           5.17341 &         1.72704 \\\hline
    \end{tabular}
    \caption{IDs and properties of the top 10 MWAs in EAGLE and TNG listed in descending order.}
    \label{tab:top10}
\end{table}
\subsection{Top 10 MWAs \label{sec:top10}}
Introducing a metric of MW-ness makes it straightforward to pick out the top 10 most MW-like galaxies in both simulations.  Recall that measuring the MW-ness of individual galaxies is best done using the distance weights $(\gamma_2)$.  In EAGLE, the top 10 MWAs have a $\gamma_2$ factor between 0.71 and 0.96; in TNG, the $\gamma_2$ factor lies between 0.67 and 0.98.  The nearly equivalent effective sample sizes of MWAs in both EAGLE and TNG along with this similar range in $\gamma_2$ for the most MW-like galaxies indicates a similar distribution of MW-ness across both simulations, which can also be seen in the bottom pannel of Figure~\ref{fig:analogues}.  The top 10 MWA galaxy IDs, $\gamma_2$s, SMs, and SFRs are summarized in Table~\ref{tab:top10} for both simulations.

\begin{figure*}
    \centering
    {\includegraphics[width=\linewidth]{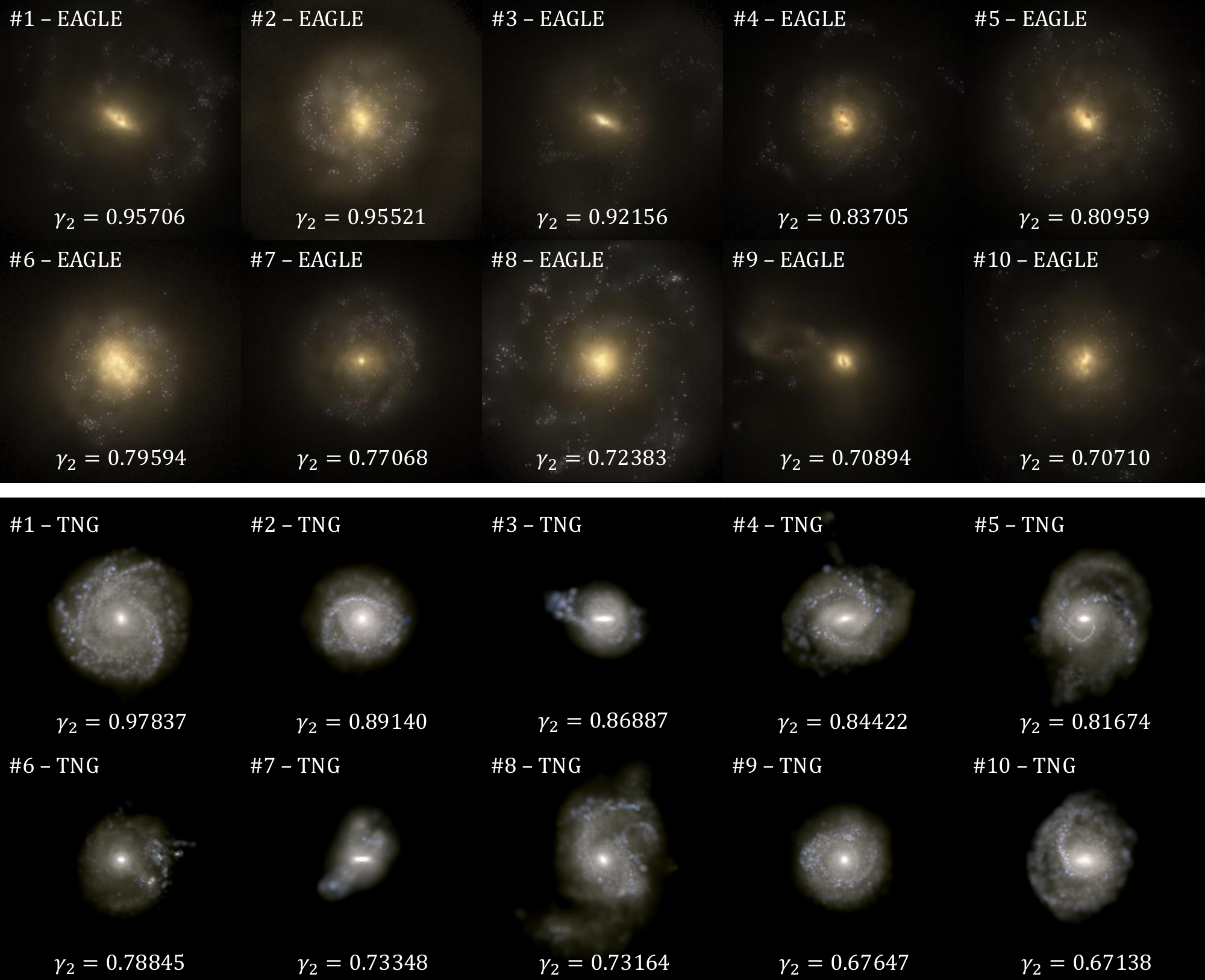}}
    \caption{False-colour face-on images of the top 10 MWAs in EAGLE (\textbf{top}) and TNG (\textbf{bottom}).  All galaxies show a disk structure, as expected.  While the MWAs in TNG are bluer than those in EAGLE (see Section~\ref{sec:colour}), the difference in colour in these images is partly a result of the different filters applied by each simulation (see text for details).}
    \label{fig:top10}
\end{figure*}

False-colour face-on stellar light composites of the top 10 MWAs from both EAGLE and TNG are shown in Figure~\ref{fig:top10}.
The images from both simulations show majority spiral galaxies as expected, suggesting that the combination of SM and SFR may act somewhat as a proxy for general morphology.  The EAGLE MWAs appear ``redder" than those from TNG, but this is partially an artifact of the filters applied by the simulations to produce the mock observations.  EAGLE uses SDSS effective filters u (3543 \AA), g (4770 \AA), and r (7625 \AA), while TNG uses NIRcam filters f070W, f115W, and f200W for their mock images.

\subsection{Some MWAs are redder than others\label{sec:colour}}

As a simple check of our selection method and to see how the simulation output compares to observation, we analyze some of the present-day properties of our MWA group.  We begin here with an investigation into the colour of our sample of MWAs in both simulations, but note that more properties including chemical enrichment and bulge-to-total ratio are analyzed in Appendix~\ref{app:z=0}.  The comparison $g-r$ colour used comes again from \citetalias{LNB15}, where they have calculated a value of $g-r=0.682^{+0.066}_{-0.056}$ from a box cut of galaxies with similar SM and SFR to the MW, since these quantities are strongly correlated with a galaxy's colour and so their colour is expected to match the MW's as well. 

\begin{figure*}[hbtp!]
    \centering
    \plottwo{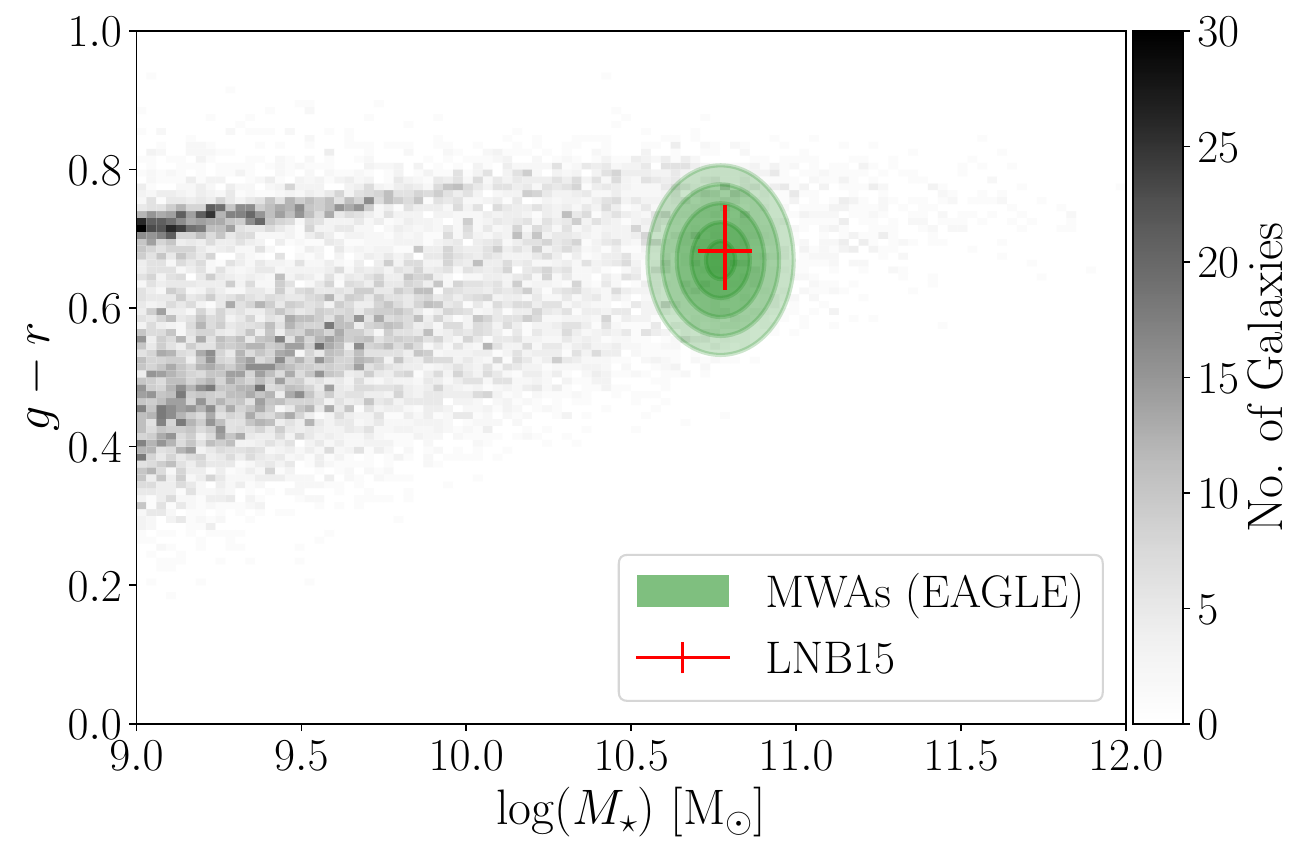}{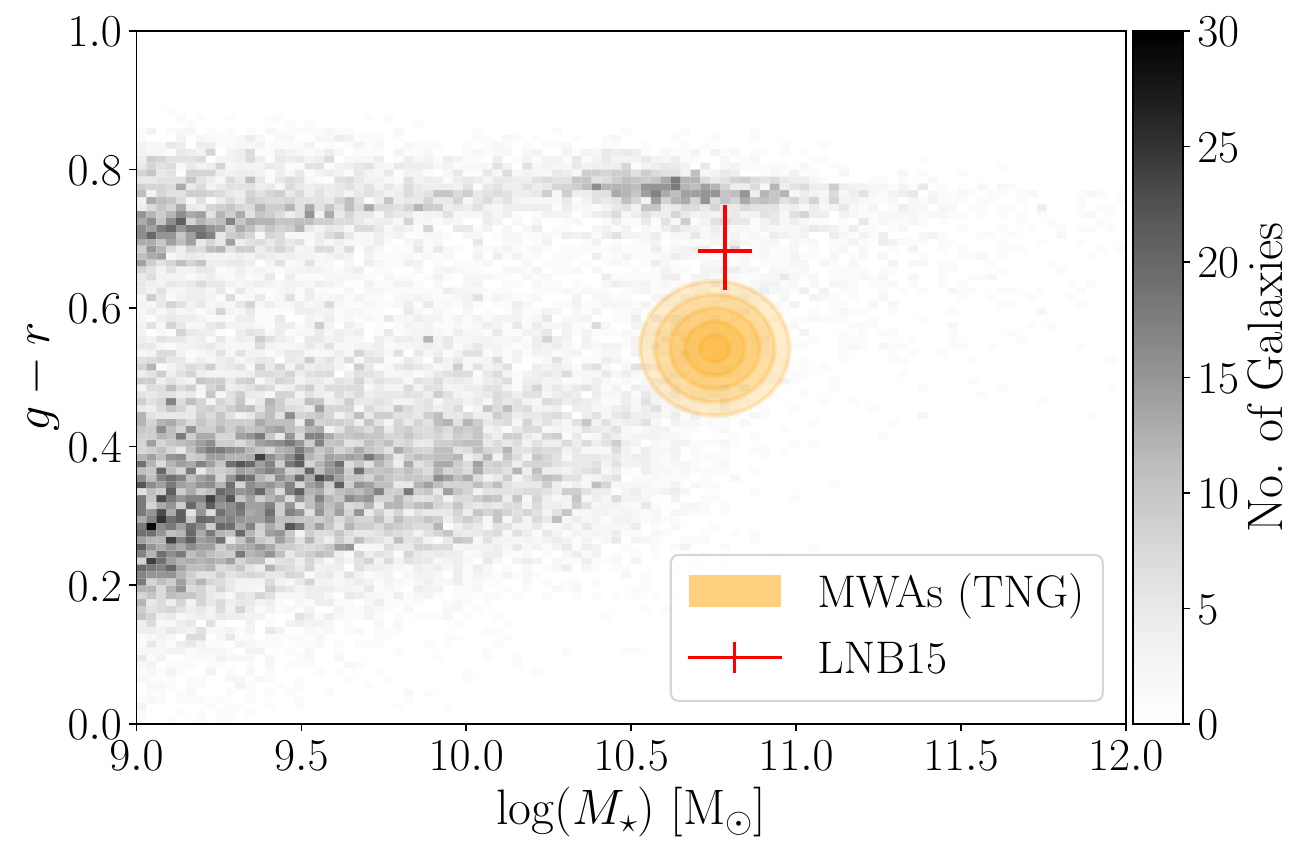}
    \caption{$g-r$ colour vs SM in EAGLE (\textbf{left}) and TNG (\textbf{right}).  The plot shows the density of galaxies in this space overlaid with green and orange shaded contours indicating the distribution of MWAs in EAGLE and TNG, respectively, with red crossbars indicating the position of the MW as found from the values in \citetalias{LNB15} ($g-r=0.682^{+0.066}_{-0.056}$).  Both simulations display clear bimodality in colour space, with a stronger separation in modes in TNG than in EAGLE.  The MWAs in both simulations also appear more red than blue, in agreement with the literature presenting the MW as a red spiral \citepalias[][\citealt{Fielder+2021}]{LNB15}.  EAGLE predicts the MW's colour nearly perfectly ($g-r=0.669\pm0.055$), whereas the MWAs in TNG are bluer than expected ($g-r=0.542\pm0.039$).}
    \label{fig:colour}
\end{figure*}

Figure~\ref{fig:colour}
shows the distribution of simulated galaxies in $g-r$ colour vs SM space, with contours indicating the position of MWAs in this space in EAGLE (green) and TNG (orange), where the simulated MWA $g-r$ values are calculated from $\gamma$-weighted averages and standard deviations to be $g-r=0.669\pm0.055$ in EAGLE and $g-r=0.542\pm0.039$ in TNG.  Also overlaid are the \citetalias{LNB15} values of $g-r$ colour and SM for the MW indicated by the red crossbars.  For the contours and crossbars, an average between the asymmetric uncertainties on the MW's value was taken.  The graphic for TNG includes dust, while the graphic for EAGLE was created by compiling galaxies from the tables \texttt{RefL0100N1504\_DustyMagnitudes} (with dust) and \texttt{RefL0100N1504\_Magnitudes} (without dust), as the simulation which includes dust is missing a significant number of galaxies, and so these gaps were filled in with the $g-r$ and SM values from the simulation run without dust.  We note that this substitution of $g-r$ data from \texttt{RefL0100N1504\_Magnitudes} where galaxies are missing in  \texttt{RefL0100N1504\_DustyMagnitudes} has no effect on our results, as majority of the missing galaxies have very low $\gamma$ weights.

Both EAGLE and TNG show clear bi-modality in colour magnitude, although the bi-modality is stronger in TNG.  In both simulations, the MWAs are shown to be more red than blue, in good agreement with~\citetalias{LNB15} and \citet{Fielder+2021}, indicating that MWAs are red(-der) spirals in both simulations; however, EAGLE seems to predict almost perfectly the MW's colour, whereas in TNG the analogues appear to be considerably bluer than the MW.  The very small discrepancy between the colour of MWAs in EAGLE and the true value for the MW is likely due to a lack of dust in the simulation.  The larger discrepancy in TNG MWAs and observations is not so easily explained, but is found in Section~\ref{sec:hists} to likely be a result of lack of quenching. In order to fully understand the cause we turn our investigation to the SFHs of our three analogue groups.

\section{Milky Way analogues at higher redshift\label{sec:highz}}

In this section, we address our second guiding question: \textit{how do the selection criteria used to pick out Milky Way analogues alter the sample of evolutionary tracks that are followed?} Specifically, we want to examine how MWAs evolve over time, and what role each selection parameter plays in that evolution -- do galaxies that have a MW-SFR today evolve the same way as galaxies that have a MW-SM today?  First, we will outline how MWAs are tracked back in time through their evolution. Then, we will analyze their SFHs and stellar mass assembly histories (SMHs) against the two control samples.  This analysis will show the track a progenitor galaxy takes to evolve into a MWA today, and how that track differs when fixed to present-day SM, SFR, or both.

\subsection{Tracing Milky Way analogues back in time\label{sec:traceback}}
We define \textit{Milky Way analogue progenitors} at redshift $z$ (MWAs$(z)$) as the main progenitors of our MWAs from redshift 0 (MWAs$(0)$) traced along their main progenitor branch (mpb) back to redshift $z$.  To identify main progenitors at higher redshift, both EAGLE and TNG use the \texttt{LHaloTree} algorithm \citep{Springel+2005}.  We note that TNG offers the \texttt{Sublink} algorithm for identifying main progenitors as well, but we use \texttt{LHaloTree} here for consistency with our EAGLE analysis.

\begin{figure*}[hb]
    \centering
    \includegraphics[width=\linewidth]{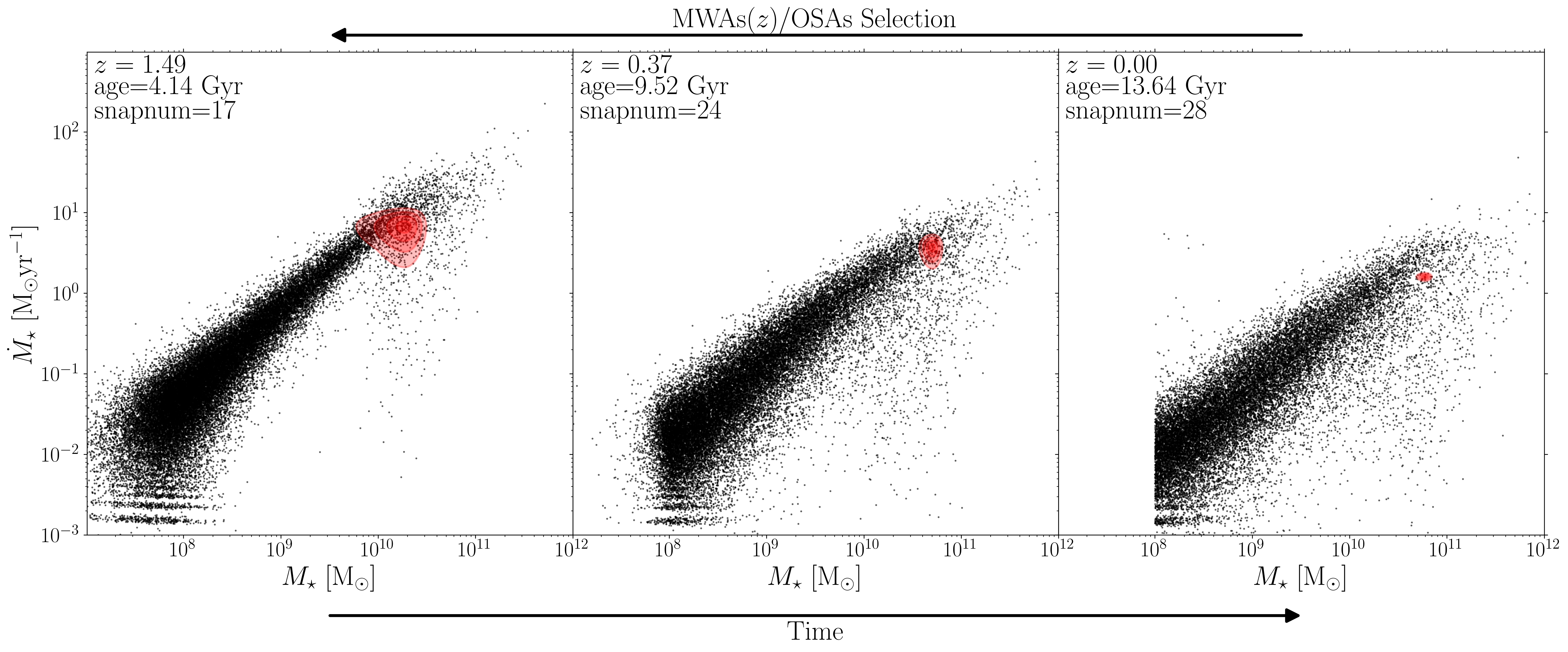} 
    \caption{The distribution of Milky Way analogue progenitors (MWAs$(z)$) at increasing redshift from right to left panels in EAGLE; also, an illustration of the input mean $\mu_p$ and standard deviation $\sigma_p$ for the observationally selected analogues (OSAs; see Section~\ref{sec:OSA}).  The distribution of MWAs$(z)$ at various snapshots is illustrated by the placement and size -- respectively indicating the average and standard deviation in SM and SFR of the MWAs$(z)$ -- of the red contours, which overlay a plot of SFR vs SM for all EAGLE galaxies at that snapshot.  The redshift, age of the universe, and \texttt{SnapNum} in EAGLE are all indicated in the top left of each panel.  The increasing span of the distribution at higher redshift illustrates the concept of progenitor contamination, as the larger standard deviation of the MWAs$(z)$ makes for a larger uncertainty input to calculate the OSA weights.  This can be thought of as the ``pseudo selection box" widening to pick out more OSAs at earlier times in the simulation, many of which do not end up as MWAs today. }
    \label{fig:ellipses}
\end{figure*}

The distribution of MWAs($z$) at increasing redshift is illustrated in Figure~\ref{fig:ellipses}
by the red blob overlaying a scatter of SFR vs SM of all galaxies in EAGLE at various snapshots.  The red blob is a series of elliptical contours centred on the average SM and SFR of the MWAs$(z)$ at that snapshot, with the width and height of the contour proportional to the standard deviation in SM and SFR of the MWAs$(z)$, respectively.  Opacity scales with standard deviation as well, with the most translucent contours corresponding to a standard deviation of $\sigma=2$.  The elliptical contours appear morphed as a result of plotting in logarithmic space.  A similar distribution of MWAs$(z)$ can be shown in TNG, though is not included here.

Notably, the size of the blob, and thus the breadth of the distribution, increases with increasing redshift (from right to left in the panels in Figure~\ref{fig:ellipses}).  This does not mean that all galaxies contained by the blob are MWA progenitors at that snapshot; rather, the blob shows the increasing spread of progenitors at earlier simulation times.  We discuss the consequences of this expanding distribution in detail in Section~\ref{sec:obs}.

\subsection{Evolutionary histories of Milky Way analogues\label{sec:hists}}
\begin{figure*}[hbtp!]
    \centering
    \plottwo{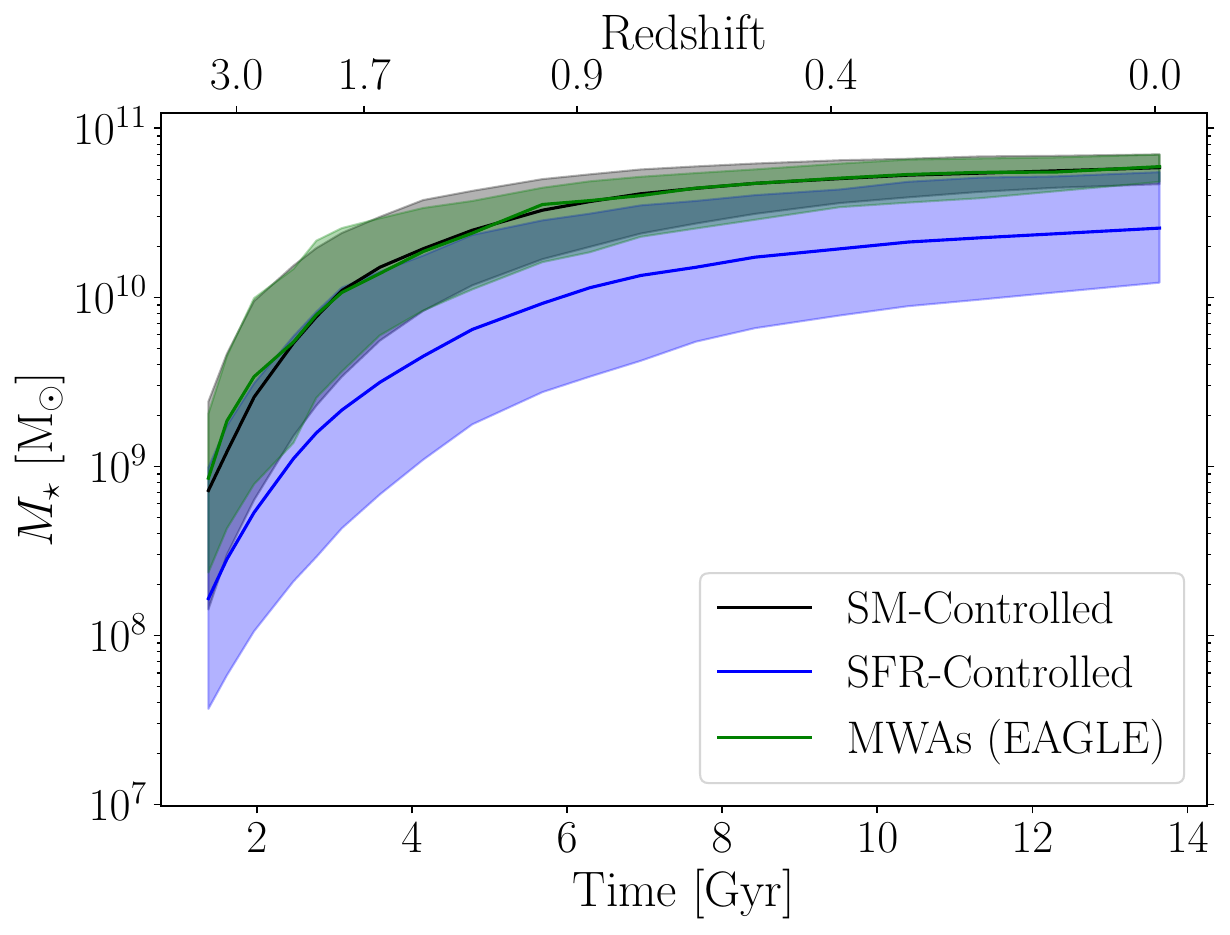}{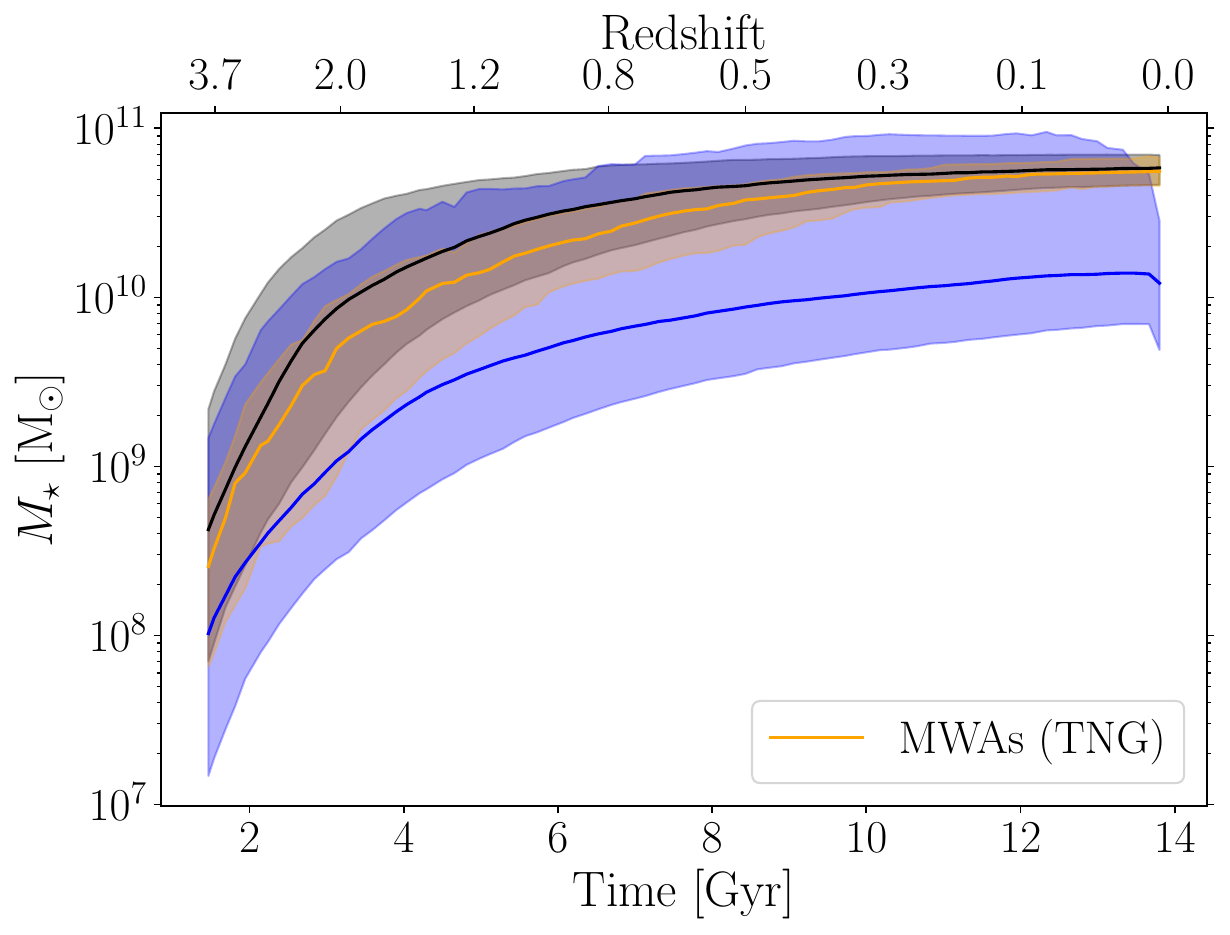}
    \plottwo{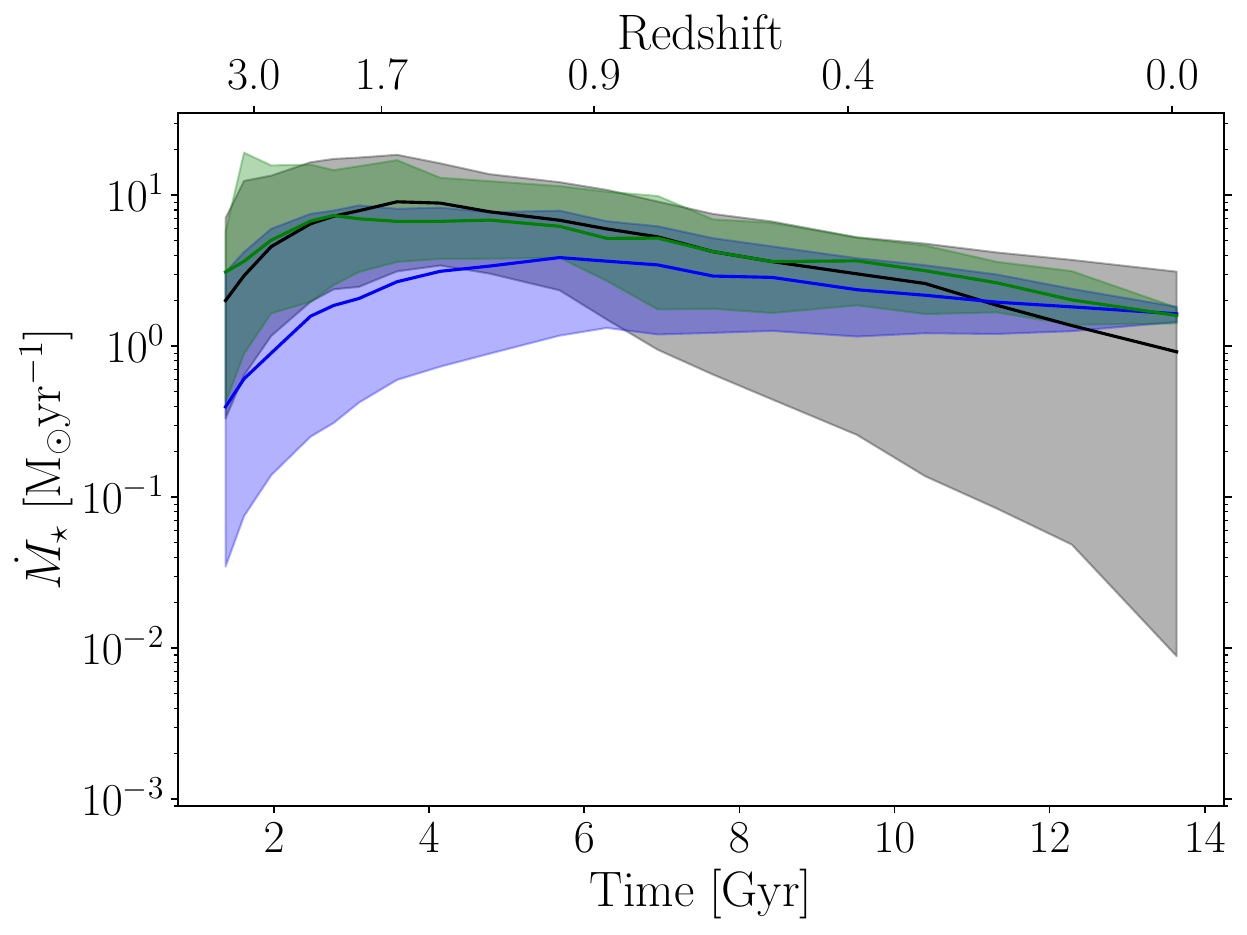}{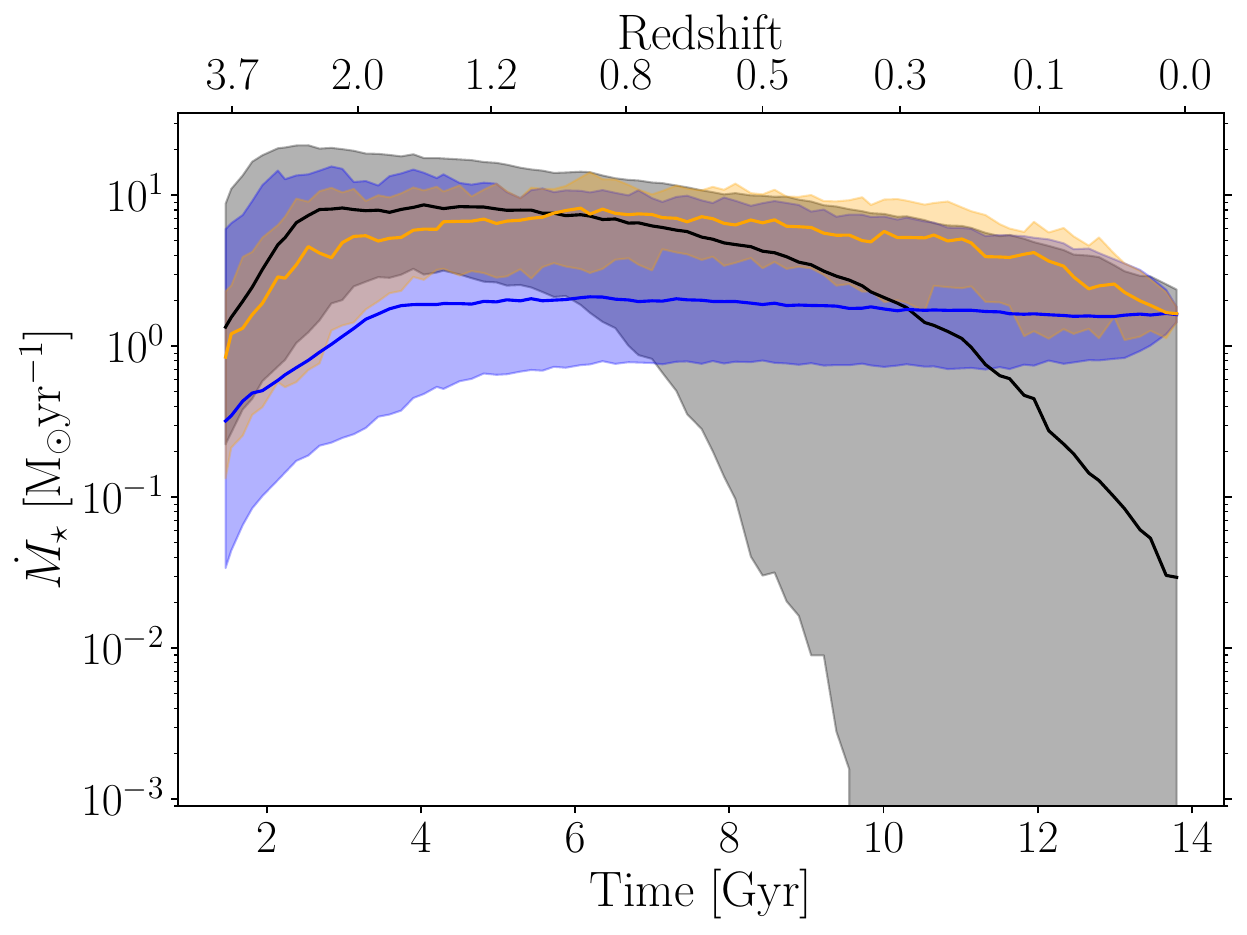}
    \caption{The stellar mass assembly histories (SMHs; \textbf{top row}) and star forming histories (SFHs; \textbf{bottom row}) in EAGLE (\textbf{left column}) and TNG (\textbf{right column}) for the MWAs($z$) -- the main progenitors of the present-day MWAs(0).  In all panels, the solid line indicates the median and the shaded region indicates the $1\sigma$ spread of the SM-controlled sample in black, the SFR-controlled sample in blue, and the MWAs in green (EAGLE) and orange (TNG).  In both simulations, the MWA SMHs follow very closely to those of the SM-controlled sample, indicating that present-day SM holds more information about MWA assembly histories than present-day SFR -- this is especially true in EAGLE, where the MWAs trace the SM-controlled group almost perfectly.  The SFHs also show a similar trend between simulations, in that the MWAs follow more closely to the SM-controlled group at early times and the SFR-controlled group at late times.  We also see that the SFHs of the MWAs in TNG remain elevated, while the SM-controlled sample rapidly quenches; whereas in EAGLE, both the MWAs and SM-controlled group exhibit slow quenching, with the MWAs remaining only slightly elevated at later times.  In EAGLE, the distribution of MWA SFRs going back in time is seen to broaden, indicating the divergence of the group's SFR at earlier times due to high SFR-variability in EAGLE.  In this figure and all figures illustrating our galaxy samples' evolution, we have made a cut just before Time $\sim2$ Gyr for visual clarity.}
    \label{fig:hists}
\end{figure*}

Now that we have tracked our MWAs back through their evolution, we can begin to dissect their histories.  The SFHs and SMHs for both EAGLE and TNG are shown in Figure~\ref{fig:hists}.
These plots show the average and the spread in the SFR and SM as functions of time of the SM-controlled group (black), the SFR-controlled group (blue), and the MWAs$(z)$ (green, EAGLE; orange, TNG). The solid line indicates the weighted median of the group at each time step, and the shaded region confines the $16^\textrm{th}$ to $84^\textrm{th}$ weighted quantiles, where the weights applied are the population weights calculated from the adjusted $\chi^2$ distribution in Equation~\eqref{eq:W}.  Essentially, we have taken our analogue group that was selected on the MW's present-day SM and SFR (the MWAs(0)) and the resulting MW-ness distribution at $z=0$, traced all galaxies back along their mpbs, and re-applied the $\gamma$ weights at each time step to identify the MWAs$(z)$ at that snapshot.  A similar approach was repeated with the control groups.  

A few insights into MWA evolution can be extracted from these plots.  Comparing the results of the two simulations to each other, we point out a few similarities and differences.  The similarities across both simulations provide general insight into the evolution of MWAs, while the discrepancies highlight differences in the underlying subgrid physics. 

\subsubsection{Similarities between EAGLE and TNG evolutionary histories}
Beginning with the SMHs (top row of Figure~\ref{fig:hists}), we see that the first similarity across both simulations is how the MWAs$(z)$ follow the path of the SM-controlled sample more closely than the SFR-controlled sample.  In EAGLE especially, the mass assembly of the MWAs$(z)$ and the SM-controlled group is nearly identical, while the tracks are still close -- although distinct -- in TNG.  However, in both simulations, the MWAs$(z)$ and the SFR-controlled group share minimal overlap, suggesting that present-day SM is a better indicator of a MWA's mass assembly than SFR.  In other words, galaxies with similar present-day SM to the MW evolve along the same track for their SMHs, while galaxies that evolve to the MW's present-day SFR follow different tracks and end up with comparatively too-low SM.

Another commonality between the simulations appears in the SFHs (bottom row of Figure~\ref{fig:hists}), in which the MWAs$(z)$ follow more closely to the SM-controlled group in early times and switch to follow more closely to the SFR-controlled group at late times.  This crossover is demonstrated in Figure~\ref{fig:hists} by the intersection between the medians of the two control samples in the SFH plots, with MWAs$(z)$ evolution tracing the SM-controlled sample before this point, but following the SFR-controlled sample after it.  This transition gives the MWAs their characteristic SFH profile.  At early times when tracing the SM-controlled group, the MWAs$(z)$ are elevated above the SFR-controlled sample; this early burst in SFR is required to evolve to a higher SM at present day, as seen in the SMHs where the MWAs(0) end their evolution with far more SM than the SFR-controlled sample (the green and orange lines representing MWA evolution end above the blue line representing the SFR-controlled sample).  At late times, when following more closely to SFR, the MWAs$(z)$ are elevated above the quenching SM-controlled group.  

We note that while both simulations see this crossover, in TNG the point at which the medians of the two control samples intersect -- and thus when the MWAs$(z)$ begin following more closely to the SFR-controlled group -- happens sooner and leaves a larger separation between control samples than in EAGLE.  This earlier and wider distinction between the control groups is due to the rapid quenching in the SM-controlled group in TNG, which paired with the clearer distinction between the MWAs$(z)$ and the SM-controlled group in the SMHs compared to EAGLE, demonstrates that SFR is a more important selection parameter in TNG than in EAGLE.  

Conversely, in EAGLE this point of divergence occurs at $\sim9$ Gyr, after which the MWAs$(z)$ remain slightly elevated while the SM-controlled sample begins to slowly quench.  This late-time point of SFR divergence between the MWAs$(z)$ and the SM-controlled group at $\sim9$ Gyr corresponds to a point in the SMH when the MWAs$(z)$ have already assembled $\sim80\%$ of their stellar mass; thus, the elevated SFR in the MWAs$(z)$ group at later times has negligible effect on the mass assembly, since these galaxies have already formed the vast majority of their stars by this point.  This all suggest that in EAGLE, MWAs essentially follow the same evolutionary track whether or not present-day SFR is included in the selection criteria; that is, galaxies with present-day SM similar to the MW all evolve along the same track, regardless of their present-day SFR.

\subsubsection{Differences between EAGLE and TNG evolutionary histories}
Despite this general trend in the MWA SFH profiles across both simulations, there are two key differences in MWA evolution setting EAGLE and TNG apart.  First, as mentioned above, the elevation of MWAs$(z)$ SFR above the SM-controlled sample at late times is larger in TNG than in EAGLE.  In EAGLE, both the MWAs$(z)$ and SM-controlled group display a downwards trend in SFR over time, indicating a slow quenching.  In TNG, however, the MWAs$(z)$ maintain a higher, nearly constant SFR for a longer period of time, while the SM-controlled group rapidly quenches.  It seems then that in TNG, the MWAs are the MW-SM galaxies ``lucky" enough to avoid strong quenching.  Returning to Figure~\ref{fig:colour}, we now have an explanation for the discrepancy between the MW's colour from \citetalias{LNB15} and the colour of TNG's MWAs; since MWAs in TNG exhibit little quenching, they remain star forming and thus ``bluer" as compared to EAGLE's MWAs, the latter of which do experience some quenching and thus appear more ``red and dead".

The second key difference is in the spread of the distribution of MWA SFHs.  In EAGLE, the spread in MWA(0) SFR is very narrow at present day (this is by design, since the analogues have been selected to be contained within this small cut in SFR), but broadens going back in time.  This broadening reflects the growing size of the MWAs$(z)$ distribution as also depicted by the red blob in Figure~\ref{fig:ellipses}.  However, we do not see this same behaviour in TNG; the spread in TNG MWAs' SFHs remains relatively constant throughout time.  

To interpret the broadening of the distribution of EAGLE MWAs' SFHs and the comparatively consistent spread for those in TNG in Figure~\ref{fig:hists}, we consider the \textit{population dispersion} of the MWAs$(z)$ in each simulation.  This broadening effect going backwards in time corresponds to the SFHs of the individual MWAs spreading out from one another at earlier times, and is what we call the population dispersion.  

\begin{figure}
    \centering
    \includegraphics[width=0.94\linewidth]{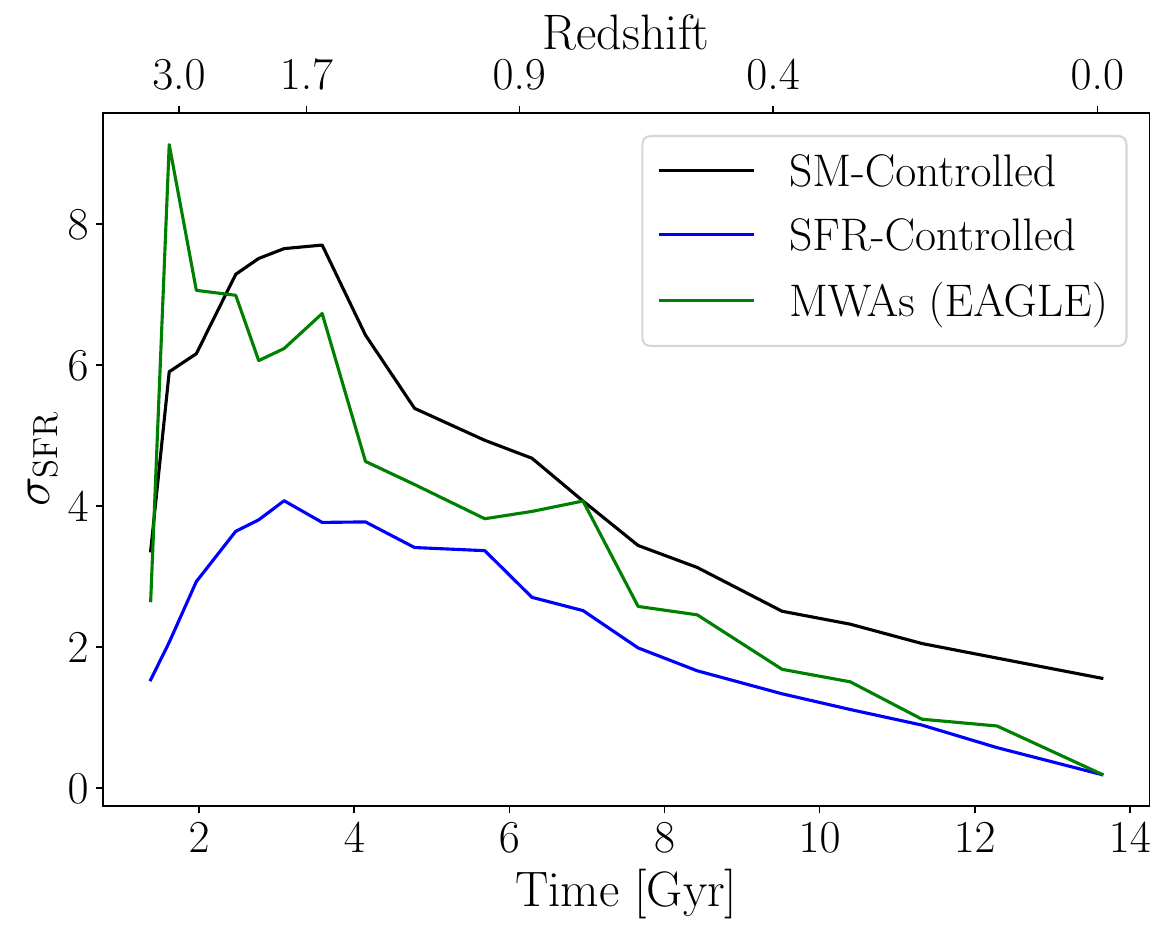}
    \includegraphics[width=0.94\linewidth]{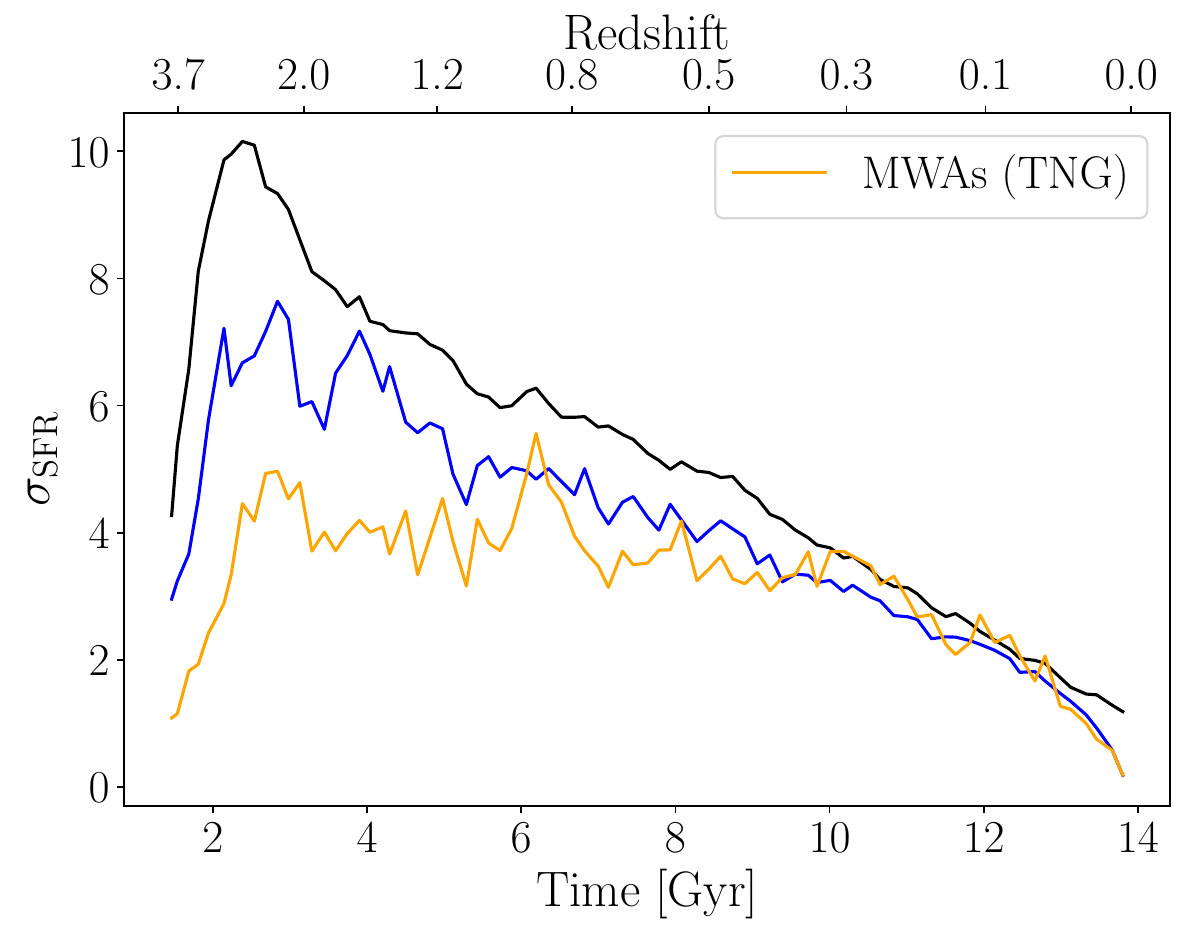}
    \caption{The population dispersion, given by the standard deviation in the SFRs of each analogue group in EAGLE (\textbf{top}) and TNG (\textbf{bottom}).  The line colours are the same as in Figure~\ref{fig:hists}.  In EAGLE, the population dispersion of the MWAs$(z)$ and the SM-controlled sample follow a similar, monotonic increase going back each time step from present day.  This increase in the population dispersion is due to the high SFR-variability known to exist between time steps in EAGLE, with a given galaxy's SFR being quite bursty across cosmic time.  In comparison, the MWAs$(z)$ in TNG show a shallow increase in population dispersion going back a few Gyr from today, before levelling off, indicating less bursty behaviour than in EAGLE.  This explains the star forming histories in Figure~\ref{fig:hists}, which show  broadening in EAGLE's MWA star forming histories going back in time due to this increase in population dispersion, and a comparatively constant spread for TNG's MWAs.}
    \label{fig:dsfr}
\end{figure}

We quantify the population dispersion by plotting the standard deviation in SFR at each time step, $\sigma_\textrm{SFR}$, in Figure~\ref{fig:dsfr}.  
The black lines show $\sigma_\textrm{SFR}$ for the SM-controlled group, the blue lines for the SFR-controlled group, and the green and orange lines for the MWAs$(z)$ in EAGLE and in TNG, respectively.  In EAGLE, $\sigma_\textrm{SFR}$ for the MWAs$(z)$ is seen to be monotonically increasing going backwards in time (the green line has a nearly constant negative slope), whereas in TNG $\sigma_\textrm{SFR}$ for the MWAs$(z)$ increases going back in time for only a few Gyr before plateauing.  Thus, population dispersion in EAGLE is quite high; however, in TNG, the MWAs' SFHs evolve more cohesively at early times, indicated by a lower dispersion in SFR.

A possible cause for such a high population dispersion in EAGLE is SFR variability between snapshots.  \citet{Iyer2020} has shown in their Figure 11 that EAGLE is indeed ``burstier" than IllustrisTNG and exhibits high SFR-variability on short timescales.  This bursty behaviour is what is causing the ``pseudo selection box" to expand; as the SFR is jumping around between time steps, the distribution is forced to broaden going back in time to capture the increased spread in SFR.  We discuss more consequences of the high population dispersion for EAGLE in the following section.

In summary, both simulations show comparable trends in MWA evolution.  MWA mass assembly in both EAGLE and TNG track their respective SM-controlled group, with the requirement for MWAs to evolve to the MW's present-day SFR seemingly negligible.  Both simulations also produce MWA SFHs that track the SM-controlled group at early times and crossover to follow more closely to the SFR-controlled group at late times, with the MWAs(0) ending with an elevated SFR in both EAGLE and TNG as compared to the SM-controlled group.  On the other hand, the dissimilarities between MWA evolution in the two simulations requires further follow-up analysis to explain why TNG's MWA SFHs remain elevated while the SM-controlled group rapidly quenches.  This question involves a deep dive into the subgrid models in each simulation and thus will be the focus of a future publication.

\section{Progenitor contamination at higher redshift\label{sec:obs}}

In this section, we address our final guiding question: \textit{How can our results better inform observational studies of Milky Way analogues?}  Previously, we noted that the spread of the distribution of MWAs$(z)$ in Figure~\ref{fig:ellipses} increases with increasing redshift.  If the red blob shown in Figure~\ref{fig:ellipses} is instead considered as a ``pseudo selection box" used to identify ``MWAs" at higher redshift, we can define a method of selecting ``MWAs" in the simulations analogous to selecting ``MWAs" observationally.  Here, we write ``MWAs" in quotations as the selection process for MWAs in observational studies will result in a very different group of galaxies compared to the method we have defined for MWA selection in the simulations, as we will soon show.  Here, we will analyze the selection efficiency of these ``observationally selected analogues", and the resulting consequences for observational studies of MWAs. 

\subsection{Observationally selected analogues\label{sec:OSA}}
We define \textit{observationally selected analogues} (OSAs) as galaxies similar to the MWAs$(z)$ at higher redshift in terms of the MWAs'$(z)$ SM$(z)$ and SFR$(z)$ at that redshift.  In other words, these are the analogues of the MWA progenitors, mimicking how observers might select MW-like galaxies at higher redshift based on the SM and SFR we would expect for the MW at that redshift.

We now formally introduce a system for finding OSAs based on the $\gamma$ factor for all galaxies at $z=0$.  We trace our MWAs from $z=0$ (which we have previously defined as ``MWAs(0)") back in time to a an earlier snapshot, corresponding to a higher redshift $z$ (which we have previously defined as ``MWAs$(z)$").  We compute the weighted average and standard deviation in both the SM$(z)$ and SFR$(z)$ on the MWAs$(z)$ group.  These values are used in a new $\chi^2$ distribution, similar to that defined in Equation~(\ref{eq:chi2}):
\begin{multline}
    \chi_{i}(z)^2=\left(\frac{M_{\star,i}(z)-\mu_\textrm{SM}(z)}{\sigma_\textrm{SM}(z)}\right)^2\\+\left(\frac{\dot{M}_{\star,i}(z)-\mu_\textrm{SFR}(z)}{\sigma_\textrm{SFR}(z)}\right)^2,
    \label{eq:chi2-n}
\end{multline}
where $M_{\star,i}(z)$ and $\dot{M}_{\star,i}(z)$ are respectively the SM and SFR of each galaxy in the sample at redshift $z$, the subscript $i\in[0,n]$ denotes the $i^\textrm{th}$ galaxy in the sample of $n$ total galaxies, $\mu_\textrm{SM}(z)$ and $\mu_\textrm{SFR}(z)$ are the average SM and SFR of the MWAs$(z)$ group, respectively, and $\sigma_\textrm{SM}(z)$ and $\sigma_\textrm{SFR}(z)$ are the standard deviation in the SM and SFR of the MWAs$(z)$ group, respectively.  The weights at $z$, which we call $\gamma_2(z)$, are then calculated as before:
\begin{equation}
    \gamma_2(z)=\left\{W_i(z)\right\}=\left\{\exp\left(-\frac{\chi_i(z)^2}{2}\right)\right\}.
    \label{eq:W-z}
\end{equation}
These weights, $\gamma_2(z)$, now define our OSAs at redshift $z$.  Note our choice here to use the original distance weights ($\gamma_2$) and not the population weights ($\gamma_3$) since we are dealing with the MW-ness of individual galaxies as they evolve through time (the criterion we defined for application of the original distance method), rather than the MW-ness of a population (the criterion we defined for application of the population method).  This process is repeated at each snapshot in the simulation.

We return to Figure~\ref{fig:ellipses}, and note that the red blob overlaying the plot of SFR vs SM now acts as a ``pseudo selection box" picking out OSAs at earlier snapshots based on the properties of the MWAs$(z)$ at that snapshot.  This is exactly how ``MWAs" are identified in observational studies.

\subsection{Tracing observationally selected analogues forwards in time\label{lowz}}

It is worth noting again that the ``pseudo selection box" in Figure~\ref{fig:ellipses} increases in size with each previous snapshot\footnote{We stress the adjective ``pseudo" as the OSAs in this study are indeed selected using the distance weights discussed in Section~\ref{sec:method2} rather than the traditional box cut method, and we say ``selection box" here only as an illustrative tool.}.  This indicates that with increasing redshift, more and more galaxies are being selected as OSAs that are not part of the MWAs(0) group -- in other words, many OSAs do not become MWAs today.  We introduce the term \textit{contamination} to describe this concept of non-MWAs(0) galaxies (MWA-``imposters") being considered OSAs at higher redshift. 

This OSA contamination poses two main consequences for astronomers interested in MWAs.  In the context of the study of galaxy evolution, the expansion of the ``pseudo selection box" to include non-MWAs(0) galaxies implies that the MW is not really special in its evolutionary history -- there may be no clear distinction between a galaxy that evolves similarly to the MW and a galaxy that turns out like the MW at present day.  In the context of observations, this result suggests that only a small number of what we currently consider to be ``MWAs" at higher redshift are actually similar to the MW today.  Thus, observational studies of ``MWAs" may not provide much insight into the evolutionary history of our own galaxy, but might be alright for studying the general evolution of galaxies of a particular SM today (see Section~\ref{sec:SFRbad}).

\subsection{Quantifying contamination with the selection efficiency of observationally selected analogues\label{sec:fosa}}
\begin{figure*}[hb!]
    \centering
    {\includegraphics[width=\linewidth]{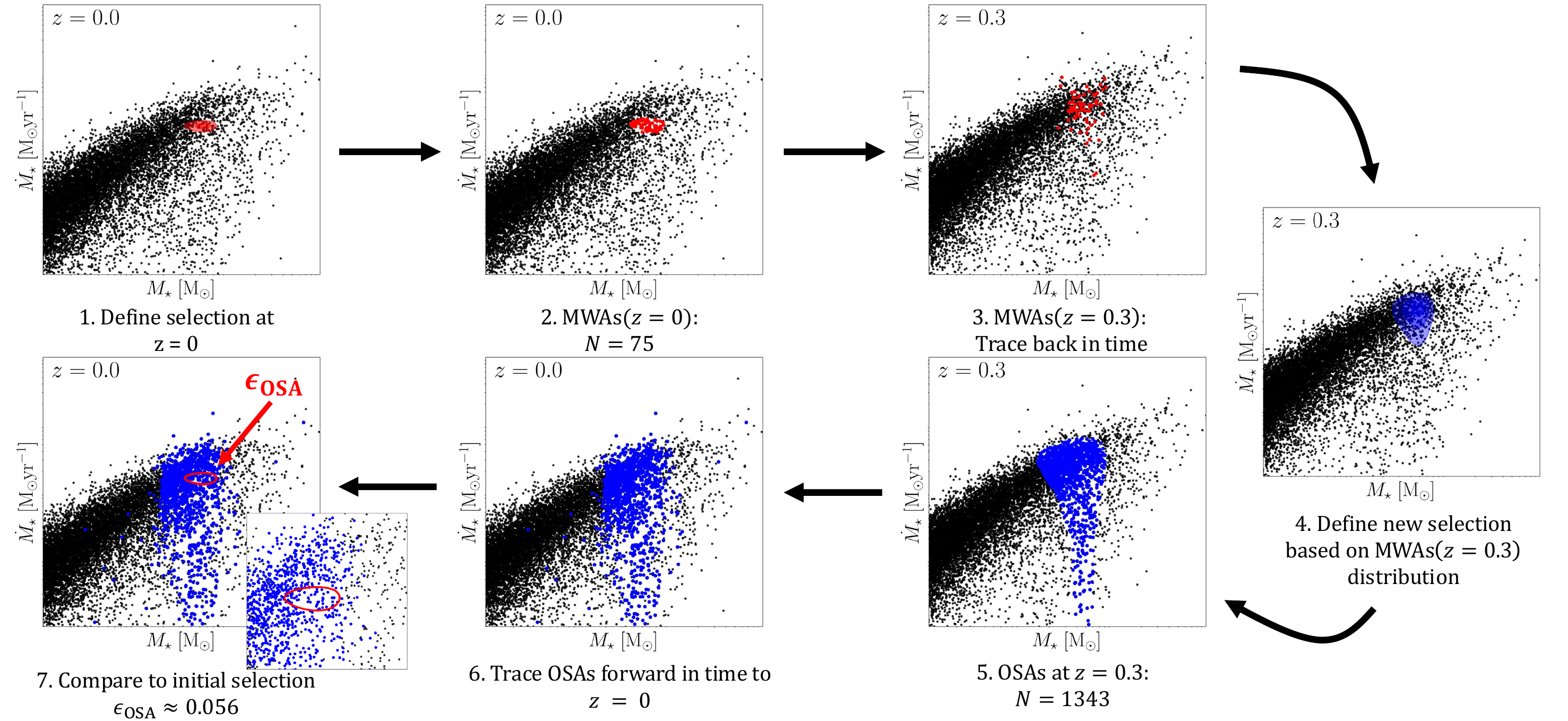}}
    \caption{Schematic diagram illustrating progenitor contamination and the process to calculate $\epsilon_\textrm{OSA}$, using EAGLE galaxies as a cartoon example.  \textbf{(1)}~We begin by defining our selection of MWAs(0) based on weights calculated from the MW's measured present-day SM and SFR as outlined in Section~\ref{sec:method2}.  Here, the red ``blob" is as described in Figure~\ref{fig:ellipses} as contours displaying the location of the MWAs(0).  \textbf{(2)}~Using the effective sample size (ess, $N$) calculated from our weights, giving us $N=75$ MWAs(0) in EAGLE, we highlight here in red the location of the top 75 MWAs(0) in SFR vs SM space at $z=0$. \textbf{(3)}~Once identified, the MWAs(0) are traced backwards in time along their main progenitor branches. The red points here show the top 75 MWAs(0.3), the MWAs from $z=0$ in step (2) traced back to $z=0.3$.  \textbf{(4)}~Based on the distribution of the MWAs(0.3), we use the average and standard deviation to calculate a new set of weights $\gamma(0.3)$.  These weights are used to define the \textit{observationally selected analogues} (OSAs) at $z=0.3$, which can be thought of as the analogues of the true Milky Way analogue progenitors at higher $z$.  Here, the ``pseudo selection box" defined by the MWAs(0.3) distribution is shown as blue contours.  \textbf{(5)}~The $\gamma(0.3)$ weights from step (4) give an ess of $N=1343$ OSAs at $z=0.3$, and here we highlight the locations of the 1343 galaxies with the highest $\gamma(0.3)$ factor in blue.  \textbf{(6)}~The 1343 OSAs from step (5) are traced forwards in time to $z=0$, shown again here by the blue points to have spread out in SFR vs SM space.  \textbf{(7)}~The OSAs from step (5) traced forward to $z=0$ are compared to the initial MWAs(0) selection, shown here by the red outline.  The ratio of the ess of these groups gives $\epsilon_\textrm{OSA}$, which in this example is calculated to be $\epsilon_\textrm{OSA}=75/1343\approx5.6\%$.}
    \label{fig:fPA_schematic}
\end{figure*}

To quantify contamination, we introduce the \textit{OSA selection efficiency}, $\epsilon_\textrm{OSA}$.  Mathematically, this efficiency is calculated from the ratio of MWAs(0) to OSAs at redshift $z$:
\begin{equation}
    \epsilon_\textrm{OSA}(z)=\frac{\textrm{No. of MWAs at }z=0}{\textrm{No. of OSAs at }z},
    \label{eq:fOSA}
\end{equation}
calculated using the effective sample sizes $N$ from Equation~\eqref{eq:ess}.  Conceptually, $\epsilon_\textrm{OSA}$ is found by defining the MWAs(0) group at $z=0$ with the $\gamma_2$ weights outlined in Section~\ref{sec:method2}, and tracing them back in time to redshift $z$ to identify the MWAs$(z)$ group.  We then select OSAs based on the average SM$(z)$ and SFR$(z)$ of the MWAs$(z)$, and use these values to determine the new $\gamma_2(z)$ weights.  The OSAs from redshift $z$ are then traced forward in time to $z=0$ and compared to the initial selection of MWAs(0).  The same method is repeated while only considering SM to calculate an efficiency for the progenitors of the SM-controlled sample, and only considering SFR for the SFR-controlled sample.  This process is more clearly outlined in the schematic of Figure~\ref{fig:fPA_schematic}.

\begin{figure}
    \centering
        \includegraphics[width=0.94\linewidth]{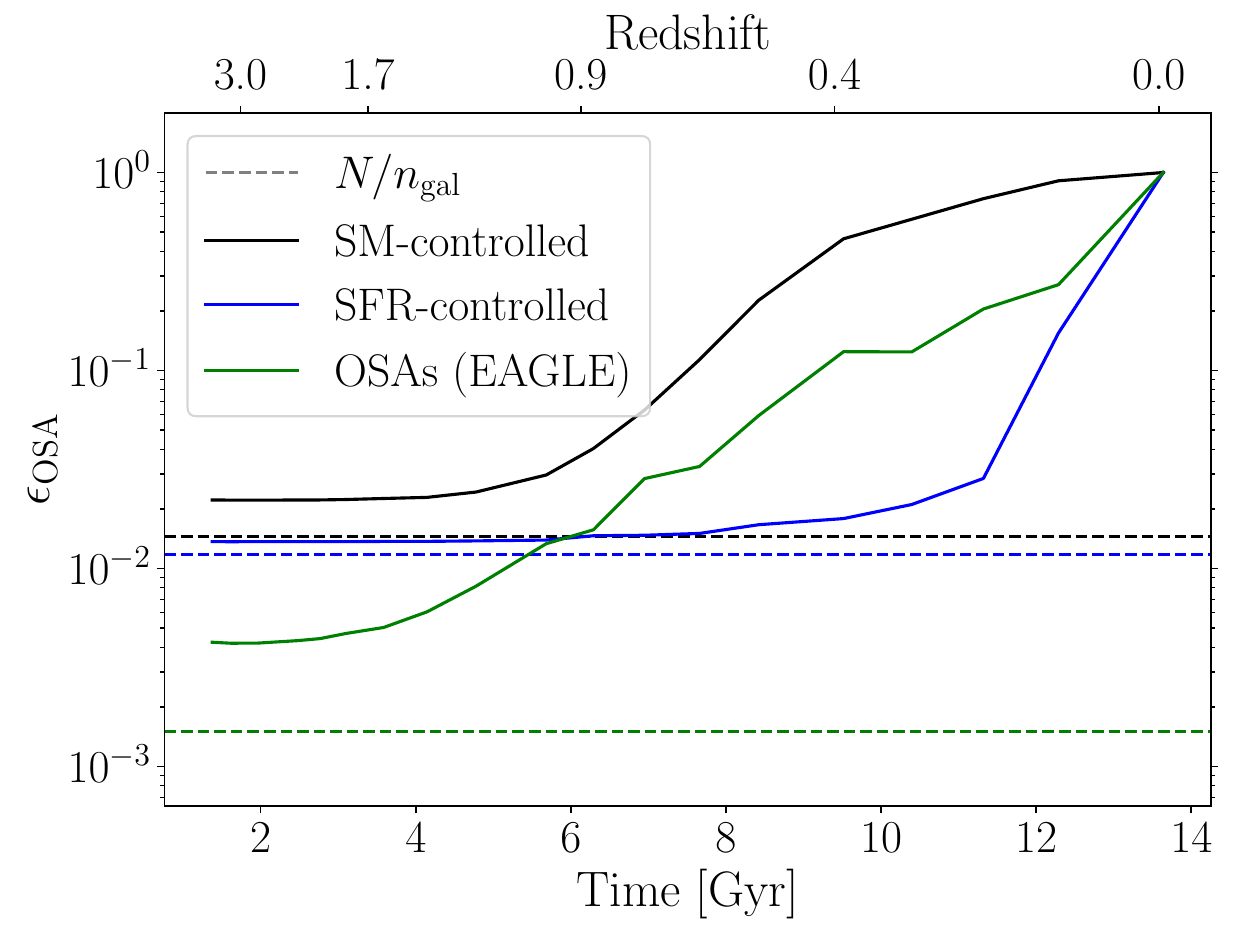}
        \includegraphics[width=0.94\linewidth]{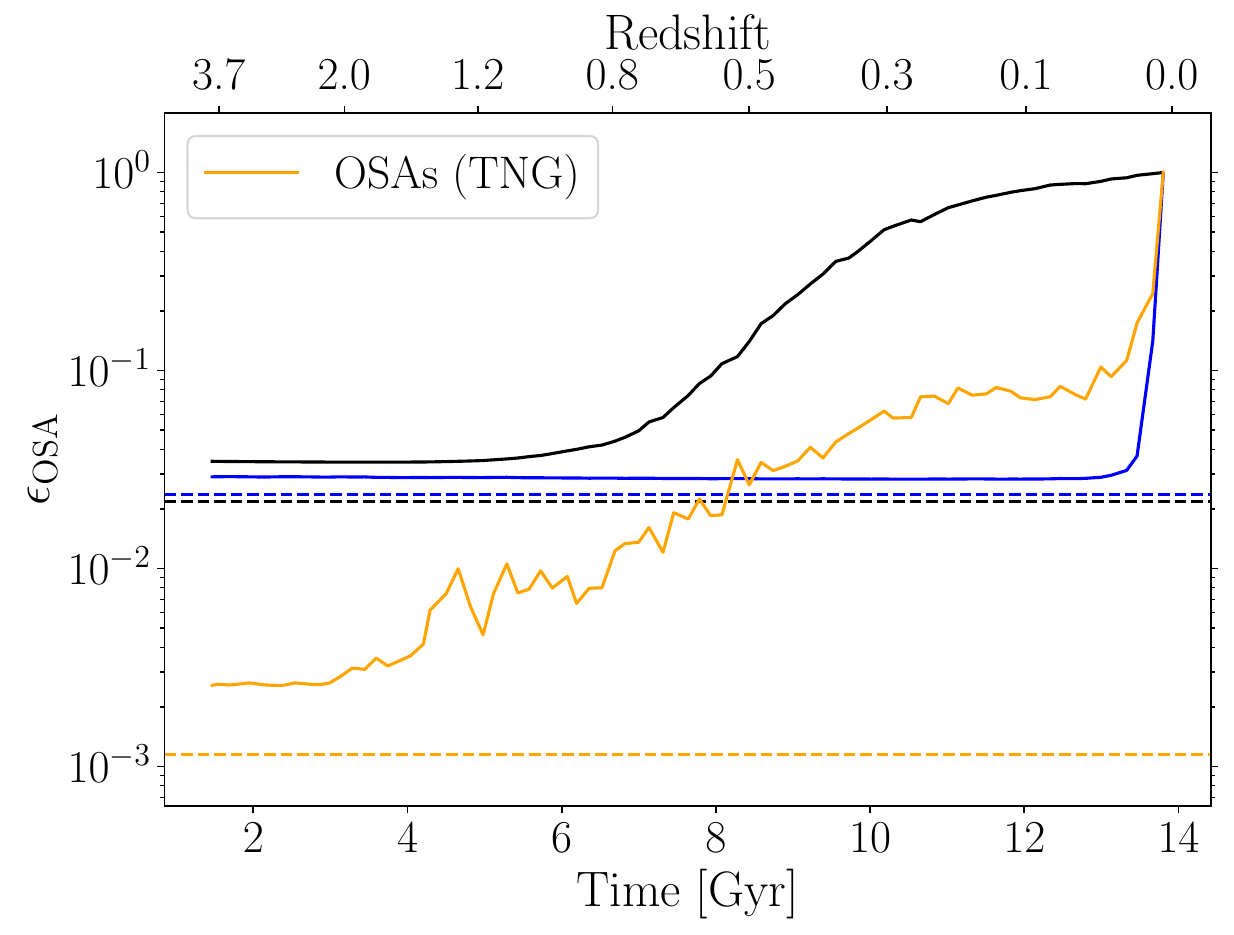}
    \caption{The selection efficiency of OSAs at each time step in EAGLE (\textbf{top}) and TNG (\textbf{bottom}) that become MWAs today, $\epsilon_\textrm{OSA}$.  The line colours are the same as in Figure~\ref{fig:hists}.  The solid lines show $\epsilon_\textrm{OSA}$ at each snapshot for each analogue group, and the dashed lines show the ratio of the effective sample sizes (ess), $N$, of each group to the total number of galaxies in the simulation, $n_\textrm{gal}$, representing the probability of randomly selecting a MWA(0) from the greater galaxy sample. In both simulations, the SM-controlled samples follow a smooth decline in $\epsilon_\textrm{OSA}$ going back in time, while the MWA group sees an immediate drop-off, with only $\sim10\%$ of OSAs one time step prior becoming MWAs today.   This dramatic decline in $\epsilon_\textrm{OSA}$ reveals that most galaxies that may have looked similar to the MW in the past do not ``grow up" to look like the MW today, and our OSA sample is contaminated with MWA-``imposters".  In fact, at earlier times, $\epsilon_\textrm{OSA}$ for all three analogue groups is seen to asymptote to $N/n_\textrm{gal}$, indicating that OSAs selected at higher redshift are no more MW-like today than any randomly selected galaxy.  The drastic discrepancy between $\epsilon_\textrm{OSA}$ for MWAs and for the SM-controlled group at higher $z$ must be caused by the inclusion of SFR as a selection criteria, as this is the only difference between the samples.  }
    \label{fig:fosa}
\end{figure}

The decrease in $\epsilon_\textrm{OSA}$ with increasing redshift compared to the selection efficiency of the SM- and SFR-controlled samples is shown in Figure~\ref{fig:fosa}.
Also included are dashed asymptotes representing the probability of randomly selecting a MWA(0) from the greater galaxy sample.  In both EAGLE and TNG, we see a substantial decline in $\epsilon_\textrm{OSA}$ from present day to the immediately preceding snapshot and continuing up until $\sim4$ Gyr ago, followed by a slower drop off until $\sim10$ Gyr ago, and finally a plateau stretching back to the beginning of the simulations.  This final plateau traces the $N/n_\textrm{gal}$ asymptote giving the ratio of the effective sample size $N$ of each analogue group to the total number of galaxies in the simulation $n_\textrm{gal}$, indicating that an OSA selected at these early times ends up no more MW-like today than any galaxy randomly drawn from the entire sample.  

Notice that $\epsilon_\textrm{OSA}$ decreases by a factor of 10 when we go back just one time step.  Comparing this jagged decline to the smooth transition in the case of the SM-controlled sample and to the even sharper drop in the SFR-controlled sample, it becomes obvious that the addition of SFR as a selection parameter is causing a steep increase in contamination at higher redshifts, and that selecting OSAs on SM alone is more efficient.  The exact reasons why SFR introduces this contamination will be discussed in Section~\ref{sec:disc}.

\section{Discussion\label{sec:disc}}
Now that we have explored all of our guiding questions, we will next contextualize their answers and discuss the resulting implications for studies of MWAs both in cosmological simulations and in observations. 

\subsection{Returning to the guiding questions\label{sec:As}}
At the beginning of this paper, we introduced three guiding questions to lead our analysis of the MWA evolutionary histories (see Section~\ref{sec:Qs}).  After conducting our investigation, we now answer these questions based on the results we have found:
\begin{enumerate}
    \item \textbf{What defines a Milky Way analogue?} Since this work focused on studying the SFHs of MWAs, we have chosen to select our MWAs on stellar mass (SM, $\SM$) and star formation rate (SFR, $\SFR$) based on the values found for the MW by \citetalias{LNB15}: $\SM=6.08\pm1.14\times10^{10}\ \MSun$ and $\SFR = 1.65 \pm 0.19\ \MSun\ \mathrm{yr}^{-1}$.  Our selection was determined using a $\chi^2$ distribution around the MW's SM and SFR (Equation~\eqref{eq:chi2}), which was then re-weighted against the background sample by normalization via a pdf to account for the overwhelming skew towards low-SM or low-SFR galaxies.  This method of MWA selection resulted in a set of weights, called $\gamma$, that provide a metric of similarity between each galaxy in the simulations to the MW itself.  In addition to our MWAs, we have identified two control groups, one selected on SM alone and one selected on SFR alone, in order to answer question (2).
    
    \item \textbf{How do the selection criteria used to pick out Milky Way analogues alter the sample of evolutionary tracks that are followed?} Comparing the evolutionary histories of the control groups to that of the MWAs led us to a few key results.  We found that all galaxies with a present-day SM similar to that of the MW (i.e., the SM-controlled group and the MWAs) evolve along the same SMH track regardless of present-day SFR, whereas galaxies which evolve only to the same present-day SFR as the MW (i.e., the SFR-controlled group) end with too-low SM to be considered MWAs(0).  Investigating the SFHs shows that MWAs have a burst in star formation at early times (i.e., evolve more closely to the SM-controlled group, which itself must burst early in order to have high final SM at present-day) and remain star forming at late times (i.e., evolve more closely to the SFR-controlled group, which itself exhibits relatively weak quenching); although, this turnover point when MWAs switch from following more closely with the SM-controlled sample to then following more closely with the SFR-controlled sample is different in both simulations, as are the discrepancies between the MWAs and each of the control groups.
    
    \item \textbf{How can the evolutionary histories of simulated Milky Way analogues better inform future observations?} It seems that the further back in time, the less ``unique" the MW becomes and thus the harder it is to pick true MWAs.  At earlier redshifts, many galaxies look similar to how the MW may have looked at such a point in its own evolution, contaminating the sample of observationally selected analogues (OSAs) with imposter galaxies that do not turn out like the MW today.  This makes it difficult for observational MWA analysts to trust their results, should they believe cosmological simulation evolution.  We find that selecting analogues on SM alone and eliminating SFR significantly reduces this contamination at higher redshift.
\end{enumerate}
So then, based on our analysis, how \textit{should} we be selecting MWAs?  

We have found from investigating the SMHs that using present-day SM as the sole selection criterion may be sufficient for identifying galaxies with a similar SM to what we expect for the MW in the past.  On the other hand, from the SFHs we see identifying galaxies with a similar SFR to what we expect for the MW in the past is not necessarily so straightforward.  

It seems that in terms of SFHs, selecting only on SM is more likely to pick out MWAs at early times, but selecting only on SFR is more likely to pick out MWAs at late times.  Further, this crossover from ``early" to ``late" occurs much earlier in TNG than in EAGLE, and the final difference between the MWAs and the SFR-controlled sample is also much larger in TNG than in EAGLE.  In fact, the SFHs of the EAGLE MWAs only diverge from the SM-controlled group at a point in their evolution at which they have already formed the vast majority of their stars, and so the elevated SFR at late times has a negligible effect overall.  In this case, present-day SM alone may be ``good enough" to identify MWAs at higher redshift.  

Additionally, we found that selecting OSAs on SM alone produced far less contamination at higher redshifts than selecting on both SM and SFR.  These results suggest that including SFR in the selection criteria may be redundant, or at a minimum that the inclusion of SFR may be too restrictive.  

In the following sections, we explore in more detail the contribution of SFR to picking out a MWA both today and in the past.  To accomplish this, we expand our SFR cutoff in the MWA ``selection box", including galaxies with higher and lower SFR than $\SFR=1.65\pm0.19\ \MSun\ \mathrm{yr}^{-1}$ as determined by \citetalias{LNB15}.

\subsection{Sliding the ``selection box" with shifted analogues\label{sec:SAs}}
To further investigate the restrictive effect of a small SFR window, we move our SFR ``pseudo selection box" at $z=0$ to higher and lower SFR values and study the evolutionary histories of the resulting new groups of galaxies.  In other words, rather than studying MWAs -- a group of galaxies that converge to the MW's SM and SFR today -- we study a group of galaxies that converge to the MW's SM today, but to a different value for present-day SFR.  We consider two groups that end with higher SFR than the MW and two groups that end with lower SFR in both simulations.  We keep the size of our selection window the same for these new groups (i.e., we keep the same uncertainty in SFR as we used to identify our original MWAs), and again use the population weights method (Section~\ref{sec:method3}).  

\begin{figure*}[hbtp!]
    \centering
        \includegraphics[width=\linewidth]{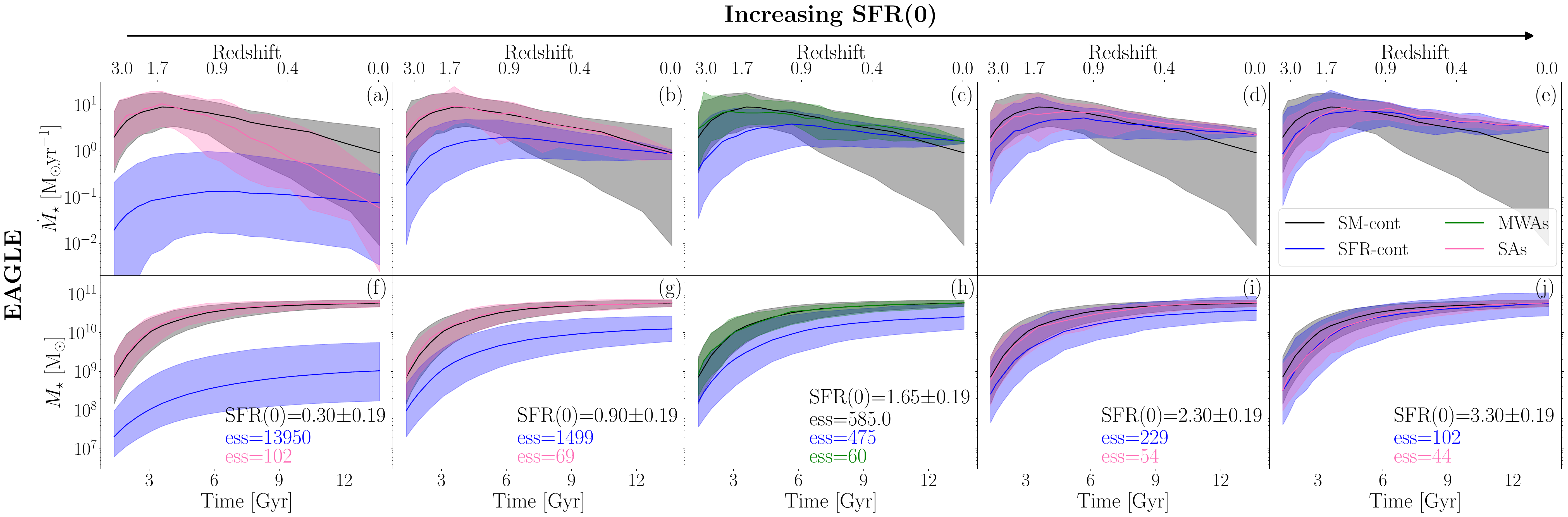}\\
        \vspace{10pt}
        \includegraphics[width=\linewidth]{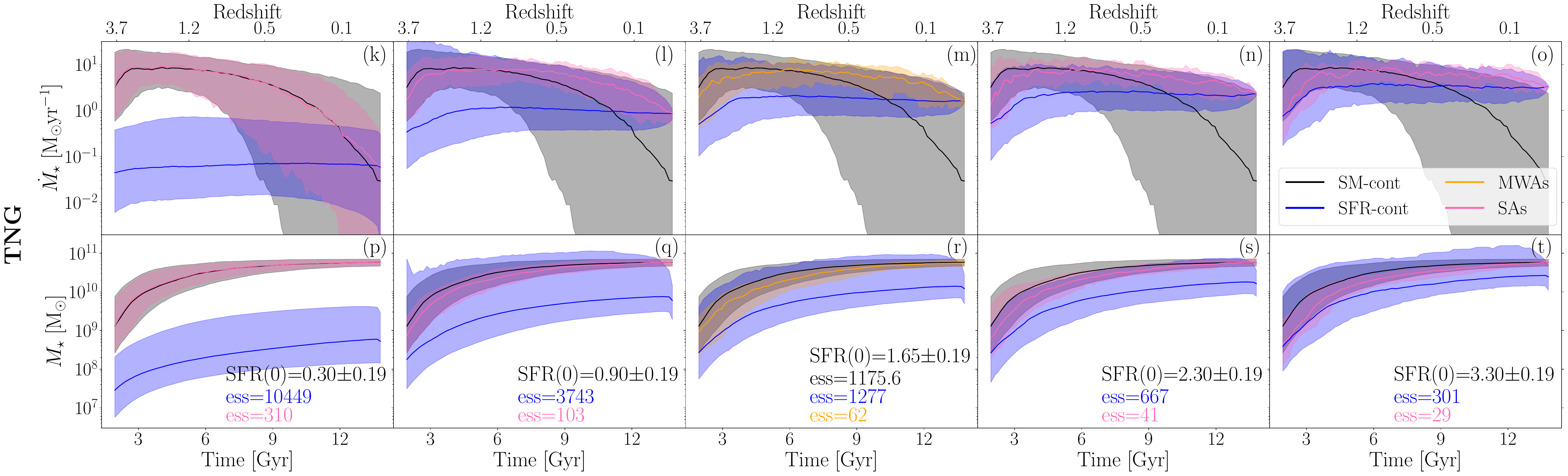}
    \caption{The star forming histories and stellar mass assembly histories of the shifted analogue (SA) groups in EAGLE (\textbf{top rows, panels (a)--(j)}) and TNG (\textbf{bottom rows, panels (k)--(t)}).  The lines, shaded regions, and colours are the same as in Figure~\ref{fig:hists}, with the new pink colour representing the shifted analogues -- the group of galaxies converging to the same present-day SM as the MW, but a different present-day SFR (SFR(0)), as indicated in each individual panel.  The shifted analogue selection SFR increases in each panel from left to right with the original MWAs in the centre panel.  The SFR-controlled samples have also been adjusted to converge to the indicated SFR(0).  We see that more extreme SFR(0) corresponds to more extreme changes in shifted analogue evolutionary history.  From this analysis (see Section~\ref{sec:SAs}), we see that the original MWAs in EAGLE (green) follow very close evolutionary tracks to the SM-controlled group compared to the shifted analogues selected with higher or lower SFR(0), indicating that present-day SM is sufficient as a stand-alone selection criterion.  In TNG, however, the MWAs (orange) do not closely follow the evolution of the SM-controlled sample -- rather, the lowest SFR(0) analogues (leftmost panels (a),(f),(k),(p)) trace the SM-controlled group perfectly -- indicating instead that present-day SFR cannot be left out as a selection parameter when picking MWAs.  Additionally, considering specifically the star forming histories in TNG, there is notable overlap between the original MWAs and the higher SFR(0) shifted analogues.  This overlap is the cause for the contamination driving down $\epsilon_\textrm{OSA}$ for TNG in Figure~\ref{fig:fosa}.}
    \label{fig:sas}
\end{figure*}

Figure~\ref{fig:sas}
shows the SFHs and SMHs for these new galaxy groups -- the \textit{shifted analogues} -- in pink, from lowest present-day SFR selection (which we call SFR(0)) on the left to highest on the right.  The original MWAs for both simulations are shown in the centre panels as green for EAGLE and orange for TNG.  The solid lines, shaded regions, and black and blue colouring are defined the same as in Figure~\ref{fig:hists}; however, the SFR-controlled samples have been adjusted to converge to the same SFR(0) as the shifted analogue groups, while the SM-controlled samples remain the same for all panels.  The exact values of SFR(0) used for each shifted analogue group as well as the effective sample sizes (calculated with Equation~(\ref{eq:ess})) for all groups are shown in the respective figure panels.  We see that with increasing SFR(0), the MWAs evolve less closely with the SM-controlled group and more closely with the SFR-controlled group.

As a general trend in the SMHs for both simulations (panels (f)--(j) for EAGLE, panels (p)--(t) for TNG), the higher the SFR(0), the higher the overall and final SM of the SFR-group, but the lower the overall and final SM of the shifted analogues.  In Figure~\ref{fig:sas}, this is seen as increasing SFR(0) corresponds to raising of the SFR-controlled group (blue) and lowering of the shifted analogue group (pink).  As SFR(0) increases (from left to right), the shifted analogue group (pink) pulls further away from the SM-controlled group (black) above and falls closer towards the SFR-controlled group (blue) below, indicating that SFR becomes a more important selection criteria for these shifted analogues.  In fact, on the highest SFR(0) end (the rightmost panels (j) and (t)), the shifted analogue SMHs line up more closely throughout time with the SFR-controlled group, rather than tracing the path of the SM-controlled group as they do on the lower SFR(0) end (lowest SFR(0) for TNG in panel (p), second lowest SFR(0) for EAGLE in panel (g)).  

The result of this trend tied to higher SFR(0) is that in both simulations, to produce galaxies with SM fixed to be similar to the MW at present day, as in the shifted analogue group (pink), galaxies with higher present-day SFR (rightmost panels (j) and (t)) must assemble their mass more slowly in order to counteract this late-time elevated SFR as opposed to galaxies with lower present-day SFR (leftmost panels (f) and (p)).  However, when present-day SM is not fixed, as in the SFR-controlled group (blue), galaxies with higher SFR(0) exhibit higher present-day SM and assemble their mass earlier.  This is to be expected, as galaxies that have been high star forming throughout their evolution should have more mass in stars.

Considering now the SFHs (panels (a)--(e) for EAGLE, panels (k)--(o) for TNG), we see that extreme values on either the high or low end for SFR(0) result in a more significant divergence from the SM-controlled group in the shifted analogue evolution.  The highest SFR(0) selection in both simulations (panels (e) and (o)) shows the shifted analogue groups (pink) diverging from the SM-controlled groups (black) very early and following more closely the path of the elevated SFR-controlled group (blue).  In the case of the lowest SFR(0) (panels(a) and (k)), the two simulations produce differing SFHs -- in both simulations, the SFR of the shifted analogues starts high, peaks early, and then rapidly quenches; the difference comes in the comparison between the shifted analogues and the SM-controlled sample.  In EAGLE (panel (a)), the depressed SFR-controlled group pulls the shifted analogues below the SM-controlled group; while in TNG (panel (k)), the shifted analogues align nearly perfectly with the SM-controlled group, further signifying  the intensity of quenching in this particular simulation.  

Taking a step back from MWAs, we consider for a moment how SM and SFR influence a galaxy's evolution more generally.  Collectively, the SMHs of the varying SFR-controlled samples (i.e., galaxies \textit{without} a fixed final SM; blue) demonstrate that galaxies with high present-day SFR most likely also have high SM (if we consider the MW to have high SM; panels (j) and (t)) and galaxies with low present-day SFR most likely also have low SM (panels (f) and (p)).  Now looking at the SFHs, in both cases of high (panels (d), (e), (n), (o)) or low (panels (a), (b), (k), (l)) present-day SFR, these galaxies are currently and have always been forming stars.  However, turning to the shifted analogues (i.e., now fixing present-day SM to be high; pink), only galaxies also fixed to have high SFR(0) will have elevated SFHs, and galaxies fixed to have lower SFR(0) will be forced to peak early -- as they must be high star forming at some point in order to end up with high mass in stars -- and then undergo rapid quenching.  Thus, it appears that a fixed present-day SM correlates to whether a galaxy exhibits quenching, as galaxies seem not to want to quench without this constraint, and a fixed present-day SFR correlates to a galaxy's final mass bracket.  

Finally, we note that when comparing simulations, varying SFR(0) results in similar evolutionary tracks in the shifted analogues, but different conclusions regarding the significance of SFR as a selection parameter compared to SM alone.  In EAGLE, the extreme choices in SFR(0) show drastic differences in evolutionary tracks of the shifted analogues -- on the low SFR(0) end (panels (a) and (f)), the SM assembly is too quick and star formation quenches too rapidly compared to the SM-controlled group (black), and on the high SFR(0) end (panels (e) and (j)) the SM assembles too slowly and the SFR does not quench enough compared to the SM-controlled sample.  However, the original MWAs (centre panels (c) and (h), green) trace closely the SM-controlled group for both the SMHs and SFHs as previously discussed, and lay in a sweet spot where present-day SFR and SM seem to correlate well, implying SFR may be a redundant selection criterion.  

In contrast, the extreme low end of SFR(0) (leftmost panels (k) and (p)) in TNG results in near perfect overlap between the shifted analogues (pink) and the SM-controlled group (black), while the original MWAs (centre panels (m) and (r), orange) assemble their SM too slowly and avoid strong quenching as compared to the SM-controlled group.  The MWAs do not lay in the same sweet spot as in EAGLE where present-day SFR and SM seem to correlate such that MWA evolution follows that of the SM-controlled group and the inclusion of present-day SFR as a selection criterion is redundant, and as such present-day SFR is necessary to include as a parameter in MWA selection.     

\subsection{Re-evaluating the selection parameters\label{sec:SFRbad}}
The results of our study of MWA SFHs in EAGLE and TNG produced conflicting views on the importance of present-day SFR as a MWA selection criterion.  The simulations imply different conclusions in the context of MWA evolutionary tracks -- in EAGLE, SM is sufficient as a stand-alone selection criterion for MWAs, but in TNG it is necessary to include SFR as a parameter.  Both simulations do agree, however, that including SFR as a parameter for OSAs drives down selection efficiency. 

If we consider the two results that (1) according to EAGLE's SFHs, SFR can safely be neglected as a selection parameter; and (2) according to both simulations, the SFR constraint in observational studies may not be useful, then the combination of these suggests that the inclusion of SFR as a selection parameter may be too restrictive.

To test the effect of loosening the SFR constraint, we have re-calculated $\epsilon_\mathrm{OSA}$, as in Equation~(\ref{eq:fOSA}), using a series of expanding SFR ``pseudo selection boxes", corresponding to a change in SFR(0) uncertainty ranging from 1 to 3 times the value of 0.19 provided by \citetalias{LNB15} in regular intervals of 0.2.  Here we remind the reader that the use of the term ``pseudo" is to signify that neither the MWAs nor the OSAs are identified using a harsh selection box, but are weighted with respect to this outlined uncertainty.  We also clarify that we are only expanding the range in SFR(0) used to find the true MWAs(0) (i.e., increasing the $\Delta_\mathrm{SFR}$ in Equation~\eqref{eq:chi2}) while leaving the range in SFR$(z)$ used to select the OSAs untouched (i.e., $\sigma_\mathrm{SFR}(z)$ in Equation~\eqref{eq:chi2-n} remains the same).  In other words, the ``mock observer" does not change the way in which they select OSAs at $z>0$; they continue to select OSAs using the $\mu_\mathrm{SM}(z)\pm\sigma_\mathrm{SFR}(z)$ we provide them as calculated from our MWAs$(z)$, and they are unaware of our loosening the SFR(0) constraint in selecting true MWAs(0).  

\begin{figure}
    \centering        \includegraphics[width= 0.94\linewidth]{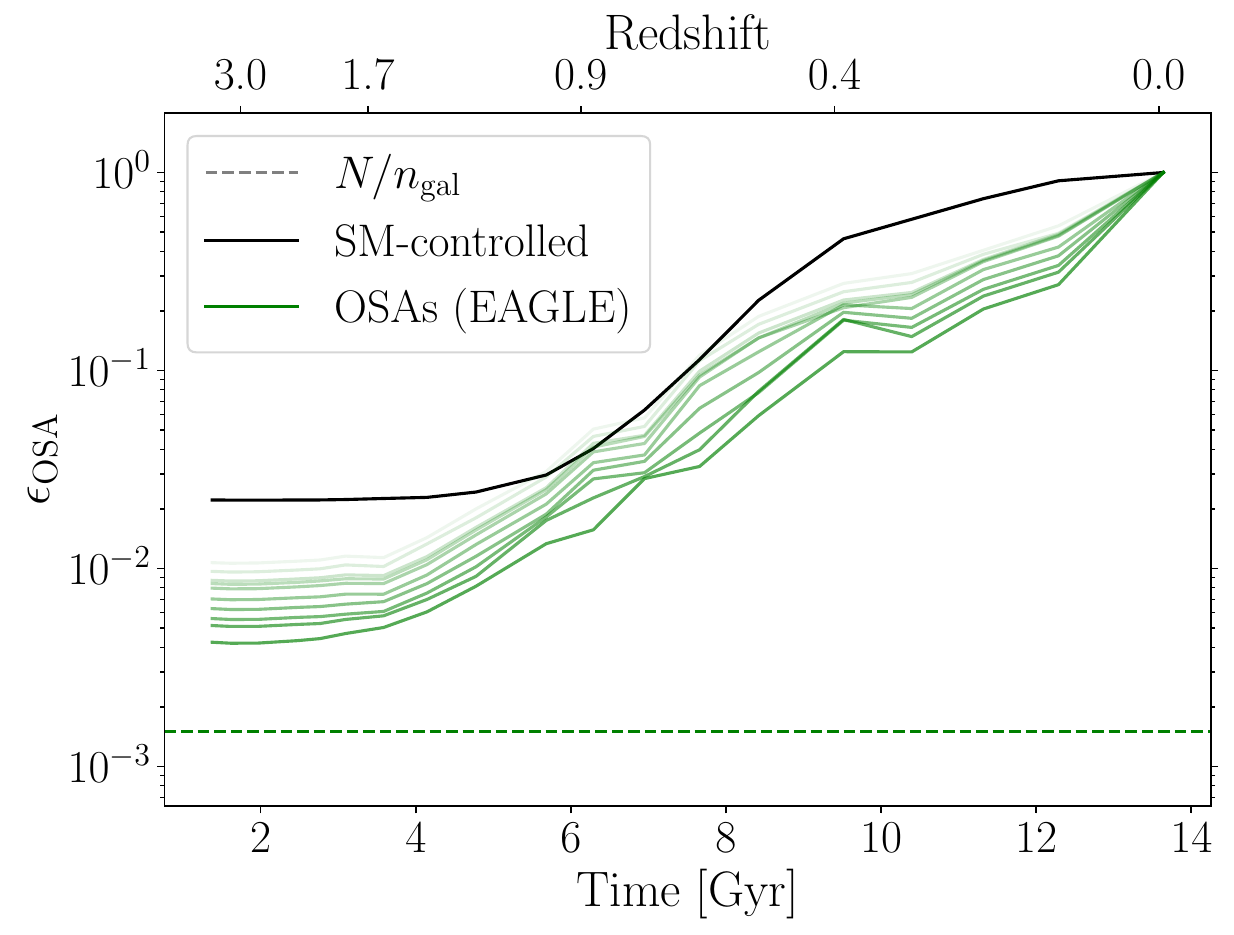}
    \includegraphics[width= 0.94\linewidth]{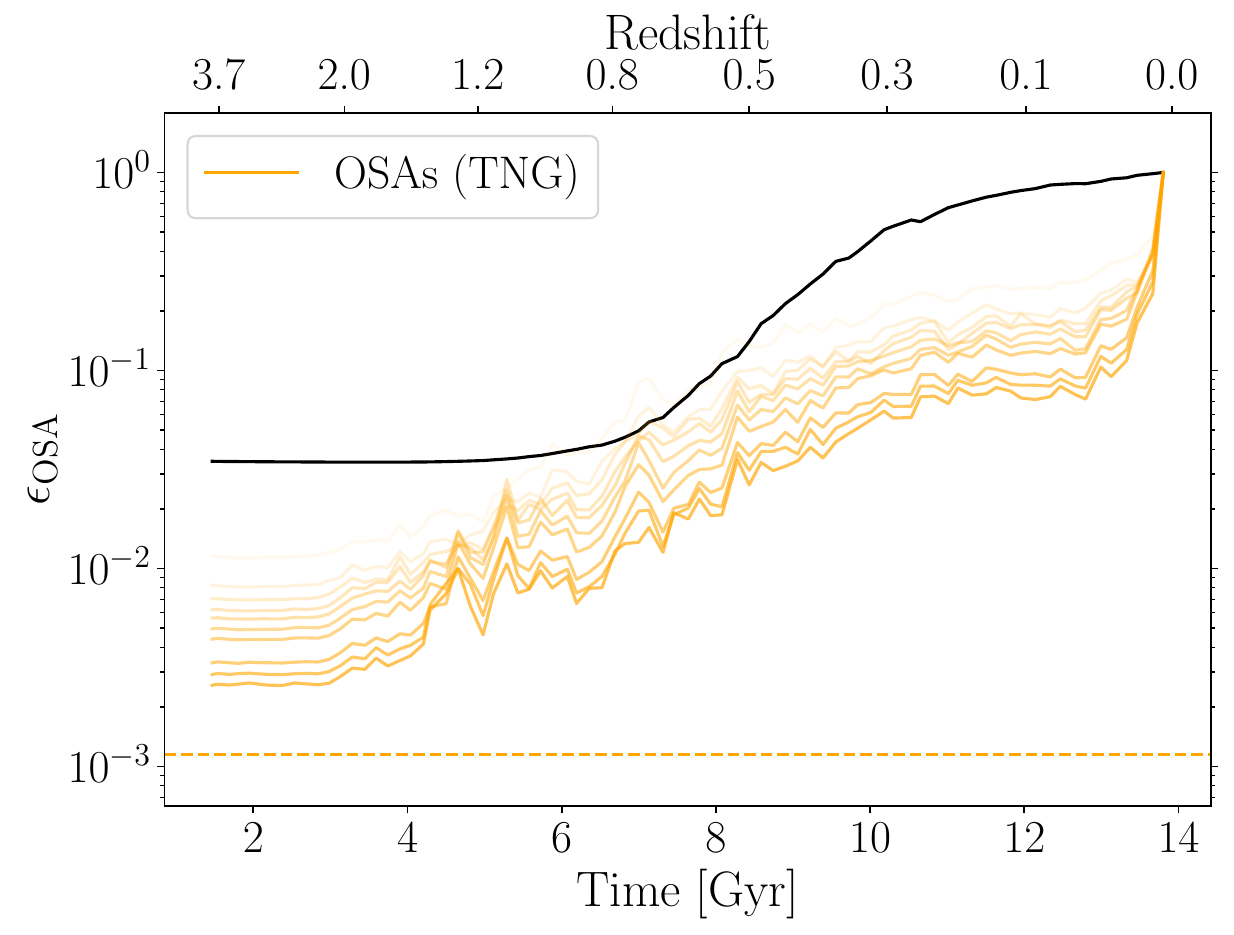}
    \caption{The reproduction of Figure~\ref{fig:fosa} with increasing the bounds on the SFR ``pseudo selection box"  scaling with line translucency for EAGLE (\textbf{top}) and TNG (\textbf{bottom}).  Here, increasing the bounds on the ``pseudo selection box" corresponds to a change in SFR uncertainty ranging from 1 (most opaque line) to 3 (most translucent line) times the value of 0.19 provided by \citetalias{LNB15} in regular intervals of 0.2.  The largest ``selection boxes" significantly reduce contamination at higher $z$, at some times resulting in a higher selection efficiency than the SM-controlled sample.}
    \label{fig:dfosa}
\end{figure}

The results are plotted in Figure~\ref{fig:dfosa},
where increasing transparency in the $\epsilon_\mathrm{OSA}$ line corresponds to increasing size in the SFR ``pseudo selection box".  We discover that loosening the SFR(0) constraint in this way results in a greater proportion of true MWAs$(z)$ and a smaller proportion of MWA-``imposters" -- in the sense that more OSAs satisfy the present-day SM and SFR criteria to be a MWA(0) -- significantly reducing contamination at higher redshifts.  Increasing the SFR(0) cutoff produces a smoother transition in $\epsilon_{OSA}$ going back in time, particularly for the largest selection box which produces selection nearly as efficient, and sometimes even better, than the SM-controlled group\footnote{Note that the OSA selection (translucent green/orange lines in Figure~\ref{fig:dfosa}) will not perfectly converge to the black SM-controlled line even with an infinitely large SFR(0) ``selection box" (i.e., making the SFR term in our initial MWA(0) selection vanish in Equation~\eqref{eq:chi2}). This is because the distribution of SFR still is present in the SFR$(z)$ term in the OSA selection (second term in Equation~\eqref{eq:chi2-n}). SFR-dependence only completely disappears if SFR is not considered either  selection \textit{at all} -- neither in MWA(0) selection nor OSA selection.}.  
Future studies of MWAs, especially in observations, may benefit from such an expanded selection window in SFR(0).  We also note that an expanded selection window should have minimal consequences on our study of MWA evolution, as the effects of incorporating slightly higher and lower SFRs(0) should essentially cancel, as evident from the panels directly to the right (panels (d) and (n)) and left (panels (b) and (l)) of the MWAs' SFHs in Figure~\ref{fig:sas}.

To understand why expanding the ``pseudo selection box" in SFR(0) increases selection efficiency, we first explain why $\epsilon_\mathrm{OSA}$ for the SFR-controlled group in Figure~\ref{fig:fosa} is so low, which ends up being simulation-dependent.  In EAGLE, instantaneous SFR tends to bounce around between snapshots due to high SFR variability in the simulation (see Figure~\ref{fig:dsfr}), pushing the limits of the MWA distribution further out and expanding the cutoff of the OSA ``pseudo selection box".  Consequently, many non-MWA(0) galaxies will get picked up as OSAs, driving $\epsilon_\mathrm{OSA}$ down.  

There is a different cause for the decline in $\epsilon_\mathrm{OSA}$ in TNG.  We previously noted that $\sigma_\textrm{SFR}$ for the MWAs in Figure~\ref{fig:dsfr} plateaus after going back $\sim4$ Gyr, indicating that SFR variability is not as strong as in EAGLE; rather, the MWAs as a group evolve cohesively.  This is illustrated in Figure~\ref{fig:sas}, where all groups but the one selected on the lowest SFR(0) exhibit little dispersion in their SFHs (see panels (l) -- (o)).  What is causing the $\epsilon_\mathrm{OSA}$ drop in TNG then is the overlap between these shifted analogue groups.  Looking at the SFHs of the shifted analogues to either side of the MWAs, we see that these groups all occupy much of the same space.  Thus, the OSA ``pseudo selection box" is picking up many of these galaxies with SFR(0) slightly above or below the MW's present-day value, hence why expanding the SFR cutoff significantly increases $\epsilon_\mathrm{OSA}$ efficiency.

\section{Summary \& conclusions\label{sec:end}}

The aim of this paper is to make use of the benefits of cosmological simulations to explore a group of MWAs through cosmic time.  We first select a subclass of galaxies that we have deemed to be similar to the MW on the basis of a choice set of selection parameters.  Once identified, we study how this class evolves in the simulations, providing us with an expectation for how the MW should have looked and behaved in its past.  Consequently, this information will have implications for future observational studies of MWAs at $z>0$ 
and can thus advise which galaxies should truly be considered MWAs. 

Our galaxy population comes from the cosmological, hydrodynamical simulations EAGLE and TNG.  We identify MWAs from this greater population using a $\chi^2$ distribution centred on the MW's SM and SFR given by Equation~(\ref{eq:chi2}), re-weighted against the background galaxies to offset a skew towards low SM or low SFR.  This process results in a set of MW-ness weights, $\gamma$, given by Equation~(\ref{eq:W}), providing a metric of similarity to the MW for each galaxy within the simulations.  In addition to our MWAs, we use the same method to identify two control groups, one selected on SM alone and the other on SFR alone, in order to determine the role of each parameter in altering the sample of MWA evolutionary tracks.  

With our analogues selected, we then trace each galaxy back through their evolution along their mpb in order to identify MWAs($z$).  The MWAs($z$) are the focus of the remainder of the paper through two different lenses: (1) using the time reversal offered by cosmological simulations to rewind, follow, and characterize MWA stellar mass and star forming evolution; and (2) analyzing mock observations of OSAs -- the analogues of the true progenitors of the MWAs as might be selected in an observational study -- to quantify the efficiency of selecting MWAs at higher redshift.  

The main results of our work are summarized as follows, noting again that our results refer only to MWAs as defined by their present-day SM and SFR, and our conclusions should not be extrapolated to MWAs of any other type:
\begin{itemize}
    \item The two simulations disagree on the MWAs' colour; \textbf{EAGLE MWAs successfully reproduce the MW's colour as a red spiral, while TNG's MWAs are too blue} (see Figure~\ref{fig:colour}).  In fact, TNG's galaxies are bluer overall, both in number and in magnitude (TNG's blue galaxies have a lower $g-r$).

    \item \textbf{MWAs' mass assembly histories are best predicted by present-day SM.}  The SMHs of the MWAs in Figure~\ref{fig:hists} (top panels) follow very closely to the SM-controlled group, with near-perfect overlap between the two groups in EAGLE, indicating all galaxies with a present-day SM comparable to the MW assemble their mass in stars similarly over time, regardless of final SFR.  

    \item \textbf{MWAs show a characteristic SFH profile predicted by present-day SM at early times and present-day SFR at late times, corresponding to an early peak in SFR that remains elevated through cosmic time}.  In Figure~\ref{fig:hists} (bottom panels), the SFHs of the MWAs align more closely with the SM-controlled group at early times, before crossing over to aligning more closely with the SFR-controlled group at late times.  This crossover point, where the tracer of star formation evolution changes from SM to SFR, occurs much earlier and results in a stronger divergence in TNG than in EAGLE.
    
    \item The divergence between the MWAs and SM-controlled sample mentioned above leads to another result: \textbf{in TNG, MWAs do not quench as strongly as most galaxies of the same present-day SM.}  The SFHs in Figure~\ref{fig:hists} (bottom right panel) show that galaxies with a MW-like SM (i.e., the SM-controlled group) in TNG rapidly quench early in their evolution, but the MWAs are the ones ``lucky" enough to bypass this shutoff of star formation.  The reason for this stark difference when including SFR in the selection criteria for TNG, allowing them to avoid quenching, will be the focus of a future publication.  We also note that this lack of quenching is why the TNG MWAs appear too blue in Figure~\ref{fig:colour}, as the galaxies have always been forming new stars and hence do not turn red.

    \item In contrast to the above point, all MW-SM galaxies in EAGLE experience slow quenching with the MWAs holding only a slightly elevated present-day SFR in comparison (see Figure~\ref{fig:hists}, bottom left panel).  Thus, it seems again that present-day SM is the best predictor for MWA evolution in EAGLE.  In fact, we conclude that\textbf{ present-day SM may be sufficient as a stand-alone selection parameter for MWAs in EAGLE, however in TNG it is necessary to include present-day SFR}.  

    \item \textbf{OSAs suffer from contamination by MWA-``imposters" -- galaxies that appear similar to MWAs(${z}$) in the past, but do not turn out like the MW today} (see Figure~\ref{fig:fosa}). In both simulations, this contamination is caused by a too-small cutoff in present-day SFR for the ``pseudo selection box", but for different reasons.   High SFR-variability in EAGLE causes OSA selection efficiency to decrease.  The MWAs($z$) in EAGLE produce a large population dispersion in their SFRs at early times (see Figure~\ref{fig:dsfr}), causing the OSA ``pseudo selection box" to expand and contaminating the sample with many galaxies that do not end up like the MW today.  This is seen from the broadening of the SFHs in EAGLE.  Meanwhile, the spread in TNG's SFHs remains comparatively constant, as they do not experience the same SFR-variability between time steps. In TNG, the contamination is then caused by galaxies that fit the SM-criterion to be considered MWAs but just miss the cutoff for final SFR (see Figure~\ref{fig:sas}).  These galaxies evolve along the same track as MWAs through cosmic time and enter the ``pseudo selection box" at earlier times, driving the efficiency down.

    \item OSA contamination poses two direct consequences: the MWAs shared a common evolution with a variety of galaxies that do not grow up to look or behave like the MW today, and thus observational studies of MWAs may be riddled with imposters.  Since the contamination is a direct result of the SFR constraints, \textbf{observational studies therefore may be more successful if they exclude SFR from their selection criteria, or at the very least loosen the constraint}, as we demonstrate in Figure~\ref{fig:dfosa}.
\end{itemize}

MWAs are valuable to astronomers; they are important for better understanding our home and its place in an extragalactic context.  The advent of large-scale, cosmological simulations offers capabilities for more detailed analyses and predictions, with use of time-reversal to study the formation and evolution of galaxies as well as a wealth of calculated properties to study.  

The ideas presented in this work can easily be extended to take full advantage of these benefits of cosmological simulations.  Although we have chosen to conduct our analysis based on MW-like galaxies selected on SM and SFR, there are many more properties of our Galaxy to include, such as dark matter or total mass, galaxy environment and neighbours, merger history, and morphology, to name a few.  Studies of MWAs selected using different parameters will help inform our understanding of the MW's evolutionary history and uniqueness, bearing in mind that each additional parameter included in selection will yield a smaller sample size.  There is also more to be learned from other ``flavours" of cosmological simulations, as our results are strongly linked to the underlying subgrid models, and so comparing MWAs across simulations can teach us both about the MW and the simulations' physical models (Kustec et al., in prep).  Another interesting extension of this work could be the analysis of the time-averaged SFHs, providing a more direct comparison with observations which usually consider time-averaged SFR.  The reason for not including this analysis in this work is that the timescales associated with different star formation processes are too short for the time resolution available to us, and thus would require simulations with more frequent snapshot output. 

We conclude by returning to the main question posed at the beginning of this paper: How \textit{should} MWAs be selected?  The answer is that it truly depends on what is being studied -- whether the galaxy population comes from observations or simulations, and if simulations, how the underlying subgrid physics are modelled; if the work is investigating MWA evolution or MWAs at a single snapshot in time, including when exactly the snapshot is taken; the properties of interest; and much more.  Our results suggest that, for a study interested in galaxies with similar SM and SFR to the MW, it seems using SM as the main selection criterion may do a ``good enough" job at selecting MWAs sufficiently, and using solely SM as the selection criterion provides, at the very least, an adequate and informed start.

\clearpage
\begin{acknowledgments}
The authors would like to thank Sina Babaei Zadeh, Ishika Bangari, Eshal Arshad, Matthew Kustec, Cissy Kuang, and Zehao Peng for helpful conversations and research that provided new insights and improved the quality of this work, as well as Vivian Yun Yan Tan for comments on the manuscript. AMS would like to thank Aryanna Schiebelbein-Zwack, Anika Slizewski, and Braden Gail for their support and valuable comments, as well as Bart Ripperda for guidance and advice in the preparation of this work. JSS would like to thank Rebecca Bleich for her support for the continued existence of the Milky Way.

AMS was supported by funding from the Dunlap Institute, a Queen Elizabeth II/Walter John Helm Graduate Scholarship in Science and Technology from the Government of Ontario, Walter C. Sumner Memorial Fellowship from the University of Toronto, and a Canada Graduate Scholarships -- Master's (CGS-M) Award and Postgraduate Scholarships - Doctoral (PGS-D) Award from the Natural Sciences and Engineering Research Council of Canada (NSERC). JSS was supported by funding from the Dunlap Institute, an NSERC Banting Postdoctoral Fellowship, NSERC Discovery Grant RGPIN-2023-04849, and a University of Toronto Connaught New Researcher Award. JTM was supported by funding from an NSERC Banting Postdoctoral Fellowship, the Dunlap Institute, and the Canadian Institute for Theoretical Astrophysics (CITA). NWM acknowledges the support of the Natural Sciences and Engineering Research Council of Canada (NSERC RGPIN-2023-04901). This work was performed in part at the Aspen Center for Physics, which is supported by National Science Foundation grant PHY-2210452. KGI was supported by funding from the Dunlap Institute and from NASA through a Hubble Fellowship.
\end{acknowledgments}

\vspace{5mm}

\software{numpy~\citep{numpy}, matplotlib~\citep{matplotlib},
            scipy~\citep{scipy}, eaglepy\footnote{https://github.com/jmackereth/eaglepy}, eagleSqlTools\footnote{https://github.com/kyleaoman/eagleSqlTools}, UMAP~\citep{umap}
          }

\appendix

\section{Key properties of Milky Way analogues at $z=0$ \label{app:z=0}}

In this paper, we made the choice to select MWAs on their SM and SFR; however, there are many other properties that may be interesting to astronomers depending on their research question, including morphology, galactic environment and neighbours, merger histories, and chemical enrichment.  The difficulty in selecting MWAs is that no selection is perfect -- no galaxy will be exactly the same as the MW, and so with each additional parameter the sample size will reduce.  Including too many properties in the selection criteria will result in no analogues at all, and so astronomers must pick and choose which parameters are most relevant to them.

In this appendix, we focus on two additional parameters: chemical enrichment and bulge-to-total ratio (B/T).  We investigate the distribution of these properties across our chosen MWA group to test their correlation with SM and SFR.  If any given parameter is shown to have values highly unique to the MWAs and this group correlates with SM and/or SFR, it would indicate that SM and/or SFR can act as a proxy for that property, and thus that a selection on SM and SFR produces a self-similar group of galaxies that share this property.  This property could then be compared to MW data to determine if the tight correlation in the simulations agrees with our observations for the MW, and thus if the simulations are accurately reproducing MW-like galaxies.  Conversely, if any given parameter is shown to have a large spread amongst the entire galaxy sample, this would serve as an indicator of a new potential parameter that could be included to more tightly constrain analogue selection.  

As such, this appendix serves two purposes: identifying key properties of our MWA group at  $z=0$ and consequently determining their uniqueness across a broader range of properties, as well as constraining our choice in selection parameters.

\subsection{Chemical enrichment\label{sec:chem}}
\begin{figure*}[htbp]
    \centering
    \includegraphics[width=\linewidth]{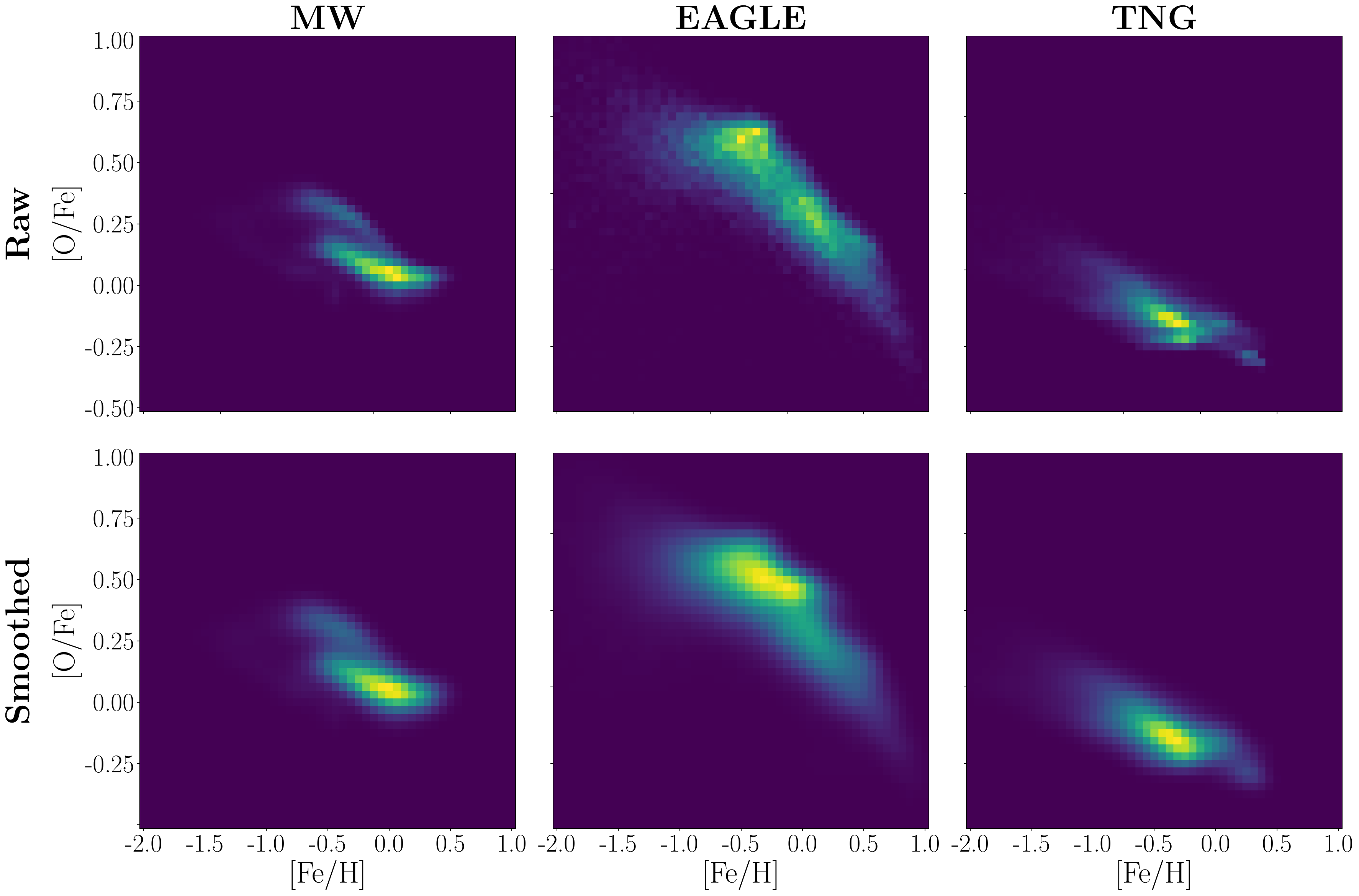}
    \caption{The $\alpha$-element abundance planes, specifically [O/Fe] vs [Fe/H] abundance, for the MW itself (\textbf{left column}) and for the most MW-like galaxies in EAGLE (\textbf{centre column}) and TNG (\textbf{right column}).  The data for the MW comes from apogee astroNN VAC DR17 \citep{Henry}.  The \textbf{top row} shows the raw 2D histograms for the abundance planes, while a Gaussian smoothing has been applied to same planes in the \textbf{bottom row} to reduce some of the sharp features to simplify the structures for UMAP.}
    \label{fig:tedsfav}
\end{figure*}

First, we consider the $\alpha$-element abundance planes (specifically, [O/Fe] vs [Fe/H]) of our total sample of 40,312 galaxies in EAGLE and 53,939 galaxies in TNG.  To study the impact of chemical enrichment on the MW-ness of a galaxy, we search for apparent consistencies between the $\alpha$-element abundance planes of our MWAs with respect to the greater galaxy sample.  We accomplish this by utilizing the dimensionality reduction package UMAP \citep{umap,numpy}, which reduces the abundance planes to a 2D projection, allowing simplified visualization of the similarities and differences in their physical structures.

The abundance planes are initially plotted as 2D density histograms of the [O/Fe] vs [Fe/H] abundances, with a sample plot of the $\alpha$-abundances for the most MW-like galaxy in both EAGLE and TNG, along with the the observed $\alpha$-abundance of the MW from the apogee astroNN VAC DR17 catalogue \citep{Henry} for comparison in Figure~\ref{fig:tedsfav}.  
Each 2D histogram is sorted into $50\times50=2500$ bins, corresponding to 2500 pixels per plot and thus 2500 dimensions, hence the need for a dimensionality reduction tool.  We also note that the 2D histograms are not normalized by bin count, so that plots indicate star/star particle density within the given galaxy rather than relative to all galaxies, and therefore UMAP will be comparing differences and similarities in structure alone and does not consider magnitude at all.  
Before running the abundance planes through UMAP, we first apply a Gaussian smoothing filter using \texttt{scipy.ndimage.gaussian\_filter} with $\sigma=1$ to reduce some of the sharpness in the density plots, which can also be seen on the example galaxies in the bottom 
row of Figure~\ref{fig:tedsfav}.  

\begin{figure*}[!hbt]
    \centering
    \includegraphics[width=\linewidth]{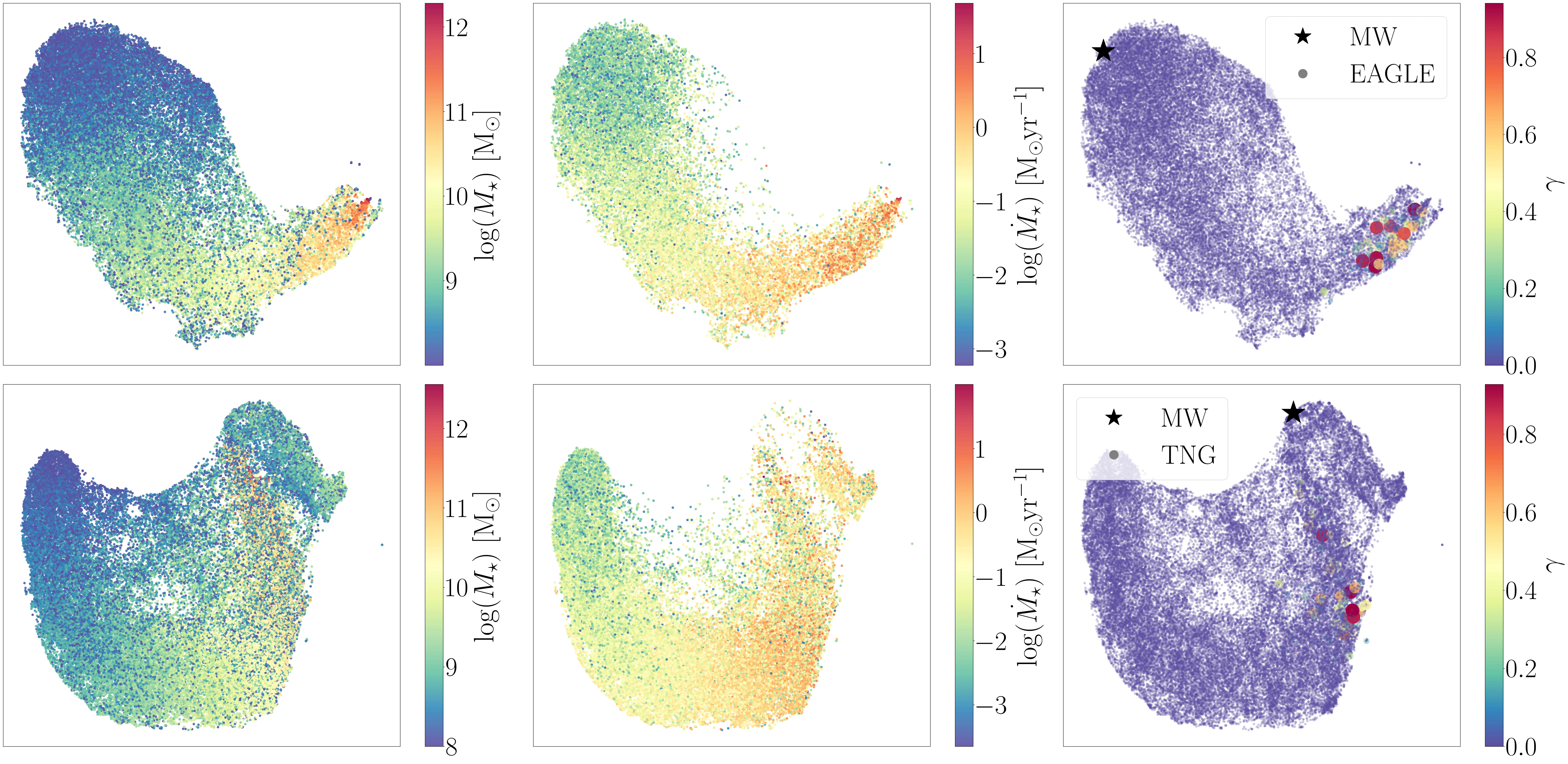}
    \caption{UMAP projections for the $\alpha$-abundance planes in EAGLE (\textbf{top row}) and TNG (\textbf{bottom row}) coloured by SM (\textbf{left column}), SFR (\textbf{centre column}), and $\gamma$ (\textbf{right column}).  The point size and opacity have also been scaled corresponding to increasing $\gamma$ in the right column.  The missing points in the centre panels coloured by SFR are a result of the logged colour bar excluding the many galaxies with a present-day SFR of 0.  The $x$- and $y$-axes here bear no physical meaning, and only the relative position of and distance between points tells us how similar a given set of abundance planes are.  The projections in both simulations take on a roughly \textit{U}-shape, with EAGLE's project resembling a ``tadpole" and TNG's resembling a ``duck".  There are notable trends in SM and SFR, both of which in general increase across the \textit{U} from left to right, excluding a population of galaxies in the top right of the TNG projection (the head of the ``duck") which appear to be low-SM but are forming stars at a high rate.  We thus find the MWAs along the right side of the \textit{U} -- where we would expect based on the trends in SM and SFR, indicating that selecting MWAs on only SM and SFR is sufficient for producing self-similar abundance planes.  The observed abundance plane for the MW from Figure~\ref{fig:tedsfav} has been added to the UMAP embedding and is depicted by the black star in the $\gamma$-coloured projections.  There is considerable discrepancy between the MW's abundance plane and those of the simulated MWAs in both simulations, with EAGLE's MWAs on the far opposite end of the projection space to the MW and the MW laying on the outskirts of TNG's MWA region.}
    \label{fig:umap}
\end{figure*}

We then plot the UMAP projection of the abundance planes with \texttt{n\_neighbors} $=10$, where \texttt{n\_neighbors} is the number of approximate nearest neighbours used to construct the initial high-dimensional graph, with a high value for \texttt{n\_neighbors} focusing on the global structure of the high-dimensional data while losing some finer detail, and a low value for \texttt{n\_neighbors} corresponding to a focus on the local structure.  This analysis was repeated on both simulations with a stronger focus on both the global structure (with \texttt{n\_neighbors} $=20$) and the local structure (with \texttt{n\_neighbors} $=3$) with no notable change in results.  The UMAP projections coloured by SM, SFR, and MW-ness can be seen in Figure~\ref{fig:umap} for both simulations.  

It is important to note that the $x$- and $y$-axes of the plots and the exact spatial positions of the points hold no physical meaning; rather, it is the relative spacing between points that illustrates the likeness of abundance planes.  In other words, two points that are spatially close on the UMAP projection represent two galaxies with similar structure in their $\alpha$-element abundance planes, and groupings of points represent groupings of abundance planes that all have similar structure, while points or groups spread out on the UMAP projection indicate galaxies or groupings of galaxies with very different structure in their abundance planes.  Thus, UMAP determines implicit axes in the projection space to organize points based on likeness to their neighbours, which will become more clear in our analysis of Figure~\ref{fig:umap}.  

The UMAP projections for both EAGLE (top row) and TNG (bottom row) show a vaguely $U$-shaped distribution of points representing the abundance planes.  To aid in our discussion of the structure of the projection, for EAGLE we refer to the shape of the projection as the ``tadpole", with the head in the top left and tail in the top right, and for TNG we call it the ``duck", with the head in the top right, breast in the bottom right, and tail on the left.  The colouring of the points in the left and middle columns of Figure~\ref{fig:umap} shows the trends in SM and SFR, respectively, across the projections, with a similar increase in SM and SFR from left to right (along the bodies of the tadpole and duck) in both simulations.  In EAGLE, this translates to a smooth transition between galaxies with low SM and SFR in the head of the tadpole to galaxies with high SM and SFR in the tail, and a similar transition in TNG from low-SM, low-SFR galaxies in the duck's tail to high-SM, high-SFR galaxies in the duck's breast and neck.  These trends indicate an implicit axis in UMAP projection space from high SM and SFR galaxies to low SM and SFR galaxies, appearing along the body of the tadpole from its head to its tail for EAGLE, and along the body of the duck from its tail to its neck for TNG.  An exception to this rule is the population of galaxies in the top right of the TNG projection (which are isolated in the head of the duck) that appear to be low SM but are actively forming stars at a higher rate.

These trends in SM and SFR lead us to expect our MWAs to be collected along the right side of the projection space, corresponding to the tadpole's tail for EAGLE and the duck's breast for TNG, based on the MW's own SM and SFR.  Indeed, the right column of Figure~\ref{fig:umap} colours the galaxies by MW-ness, with increasing point size and opacity also corresponding to increasing $\gamma$, and the MWAs fall generally where we would expect.  This clustering of the MWAs in line with the SM- and SFR-implicit axes suggests that SM and SFR may act as a proxy for $\alpha$ abundance.  We also remark that the MWAs appear fairly spread out along the region in which we expect to find them.  If the clustering of the MWAs was instead tighter within the expected region, it would signal that the $\alpha$ abundances of the MWAs are unique and may warrant a third selection criteria.  

Finally, we note that the observed $\alpha$ abundance plane for the MW as shown in Figure~\ref{fig:tedsfav} was afterwards fed into the existing UMAP embedding and is indicated in the MW-ness UMAP projections in the right column of Figure~\ref{fig:umap} as a black star.  The true $\alpha$ abundance plane does not fall within the MWA region for either simulation.  In EAGLE, the MW falls on the opposite end of the projection space as the MWAs, and is shown to be in the head of the tadpole.  TNG does a bit better, with the MW appearing at the very top of the head of the duck, towards the edge of the MWA region.  Interestingly, the position of the MW in the projection space corresponds to much too \textit{low} SM and SFR in the EAGLE simulation, but slightly too \textit{high} SM and SFR in the TNG simulation.  It seems that neither simulation is accurately reproducing the MW's $\alpha$ abundance plane, despite both being self-consistent in producing MWA $\alpha$-abundance planes.  The reasoning behind this discrepancy between simulation and observation is beyond the scope of this work, and should be addressed in future research.   

\subsection{B/T in EAGLE \label{app:B/T}}
The second additional property we consider is B/T, quantifying how ``bulge-y" (high B/T) or ``disk-y" (low B/T) a galaxy is.  We calculate B/T from $1-$D/T, where D/T is the disk-to-total ratio -- the mass fraction of stars that are rotationally-supported (as a proxy for the disk mass fraction).  D/T is given by \citet[their Equation (5)]{Thob2019} as
\begin{equation}
    \frac{D}{T}=1-\frac{B}{T}=1-2\frac{1}{\SM}\sum_{i,L_{z,i}<0} m_i,
    \label{eq:B/T}
\end{equation}
where $\SM$ is the mass of the galaxy, $m_i$ is the mass of each individual star particle, the sum is over all counter-rotating $(L_{z,i}<0)$ star particles within 30 pkpc, and $L_{z,i}$ is the component of the angular momentum of each individual star particle projected along the rotation axis.  Here, the mass fraction of the bulge, B/T, is given by twice the mass of stars that are counter-rotating with respect to the galaxy, as a popular means of estimating the disk fraction is to assume that the bulge component has no net angular momentum.  The MW, a disk galaxy, does indeed have a low B/T, measured by \citetalias{LNB15} to be B/T $= 0.15\pm0.02$.  As TNG does not directly output a field for B/T or D/T, the following analysis includes EAGLE galaxies only.  

\begin{figure}[htbp!]
    \centering
    \includegraphics[width=\linewidth]{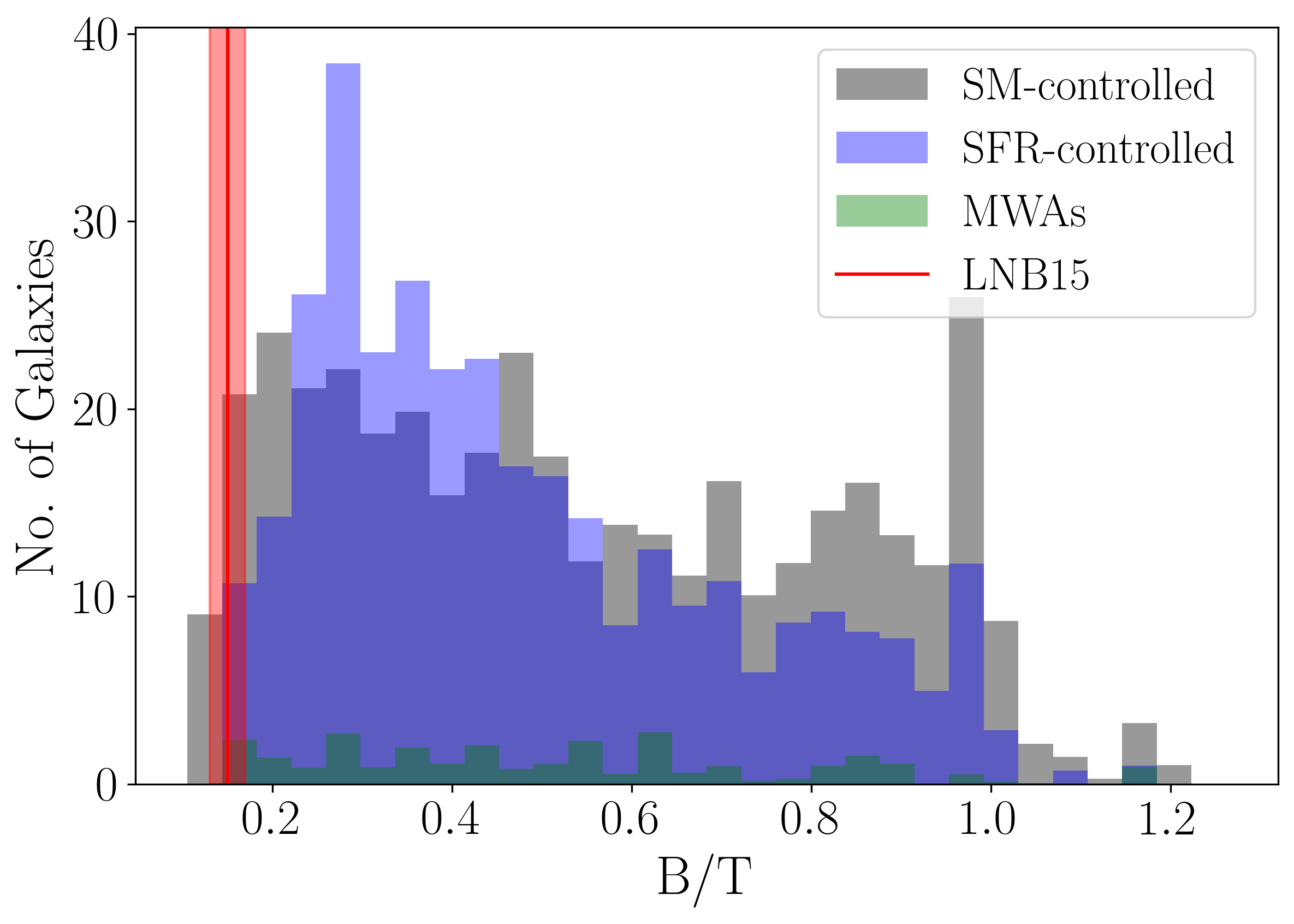}
    \caption{Weighted histograms showing the distribution of B/T in EAGLE for the SM-controlled group in black, the SFR-controlled group in blue, and the MWAs in green, with the weights applied being $\gamma_\textrm{SM}$, $\gamma_\textrm{SFR}$, and $\gamma$, respectively.  The observed B/T plus uncertainty for the MW from \citetalias{LNB15} is depicted by the red line and shaded region.  All three analogue groups saturate the entire B/T space, indicating SM and SFR are not sufficient selection parameters for reproducing the MW's B/T.  Thus, we consider a new selection of MWAs including B/T as a third parameter.}
    \label{fig:bt_dist}
\end{figure}

The distribution of B/T across the MWAs and both control samples is shown via histograms weighted by $\gamma$ (green), $\gamma_\textrm{SM}$ (black), and $\gamma_\textrm{SFR}$ (blue), compared to the MW's value from \citetalias{LNB15} (red line) in Figure~\ref{fig:bt_dist}.
Although there is a slight skew towards lower B/T values, where the MW lives, all three groups cover the full range of B/T (note that, while it should be true that B/T $\in[0,1]$, there are 601 galaxies in our sample with D/T $<0$, corresponding to B/T $>1$; from Equation~\ref{eq:B/T}, it is clear this happens when the second term is greater than 1).  While the MWAs are contained within the distribution of the SM- and SFR-controlled groups as was the case for chemical enrichment, these distributions span the entire B/T parameter space, contrary to chemical enrichment where the control samples occupied one small region of the UMAP projection space.  Thus, B/T must be added to the selection criteria.

\begin{figure}
    \centering
    \includegraphics[width=\linewidth]{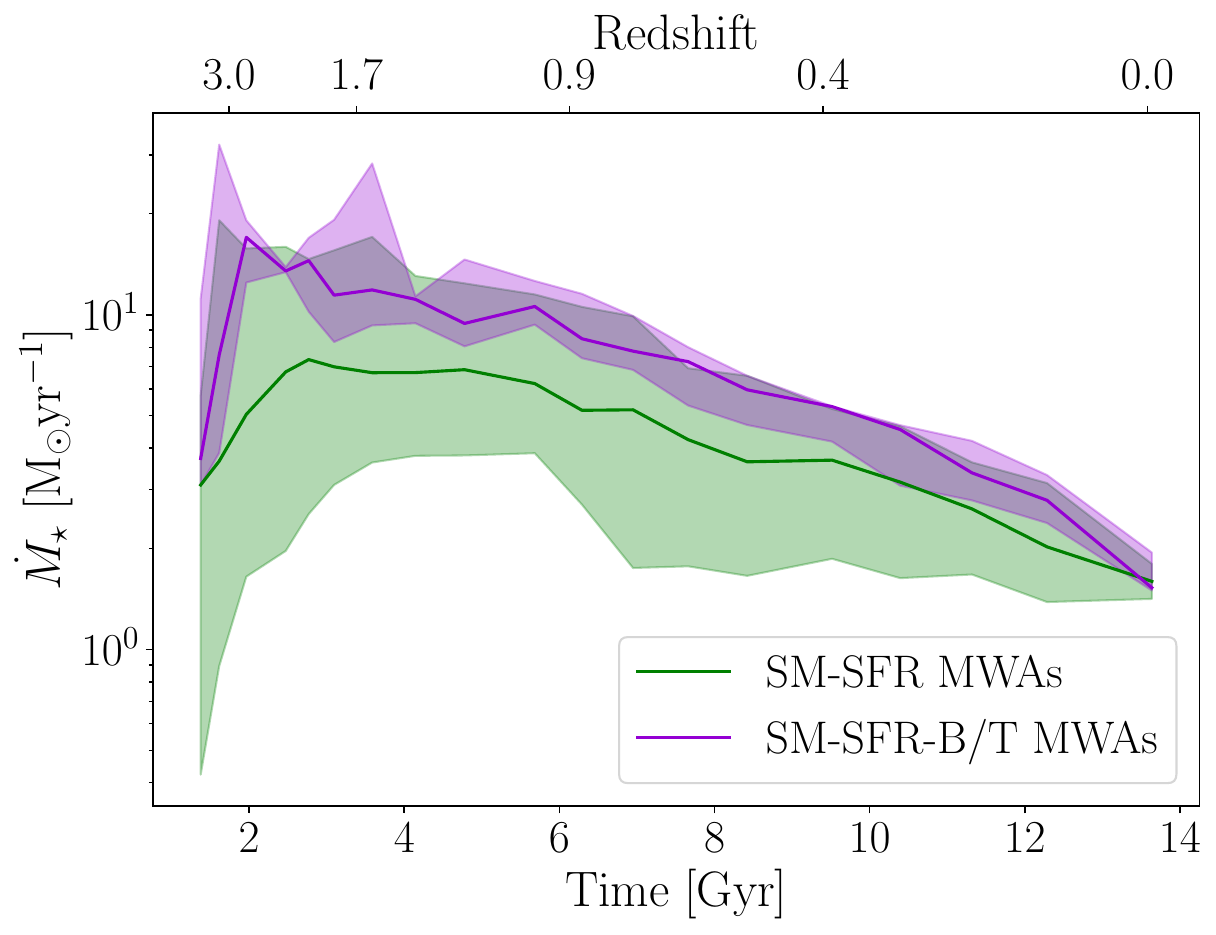}
    \includegraphics[width=\linewidth]{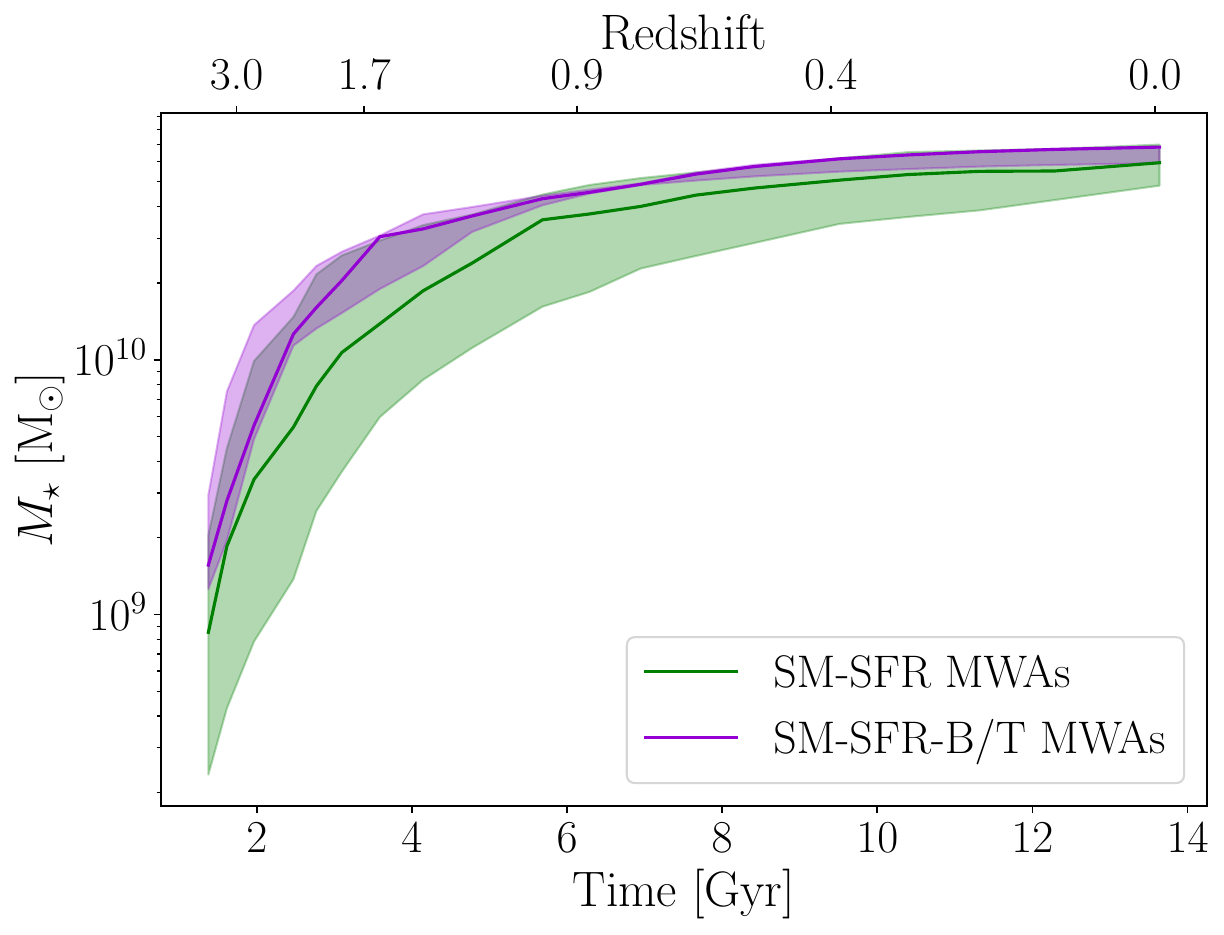}
    \caption{The star forming histories (\textbf{top}) and mass assembly histories (\textbf{bottom}) for the new MWAs selected on all of SM, SFR, and B/T in purple compared to the original MWAs selected on only SM and SFR in green.  The lines and shading are the same as in Figure~\ref{fig:hists}.  Generally, the MWAs that include B/T in their selection criteria follow similar evolutionary histories to the original MWAs, although they do peak higher in SFR and assemble their stellar mass more quickly.  Note that the effective sample size of the SM-SFR-B/T MWAs is only $\sim3$, and so our results do not hold much weight.  Rather, the limited insight we have from the small sample size in combination with the exclusion of most of the galaxy sample suggests that B/T should not be included in the selection criteria.}
    \label{fig:bt_hists}
\end{figure}

To include B/T in MWA selection, we update Equation~\ref{eq:chi2}:
\begin{multline}
    \chi_i^2=\left(\frac{M_{\star,i}-\rho_\textrm{SM}}{\Delta_\textrm{SM}}\right)^2+\left(\frac{\dot{M}_{\star,i}-\rho_\textrm{SFR}}{\Delta_\textrm{SFR}}\right)^2\\
    +\left(\frac{\mathrm{B/T}_i-\rho_\mathrm{B/T}}{\Delta_\mathrm{B/T}}\right)^2,
    \label{eq:chi2-B/T}
\end{multline}
where we have introduced into the metric the measured B/T and uncertainty for the MW from \citetalias{LNB15}, $\rho_\mathrm{B/T}\pm\Delta_\mathrm{B/T}=0.15\pm0.02$.  Using Equation~(\ref{eq:ess}), this third selection parameter results in an effective sample size of only $\sim3$ galaxies, illustrating the dangers of including too many selection criteria.  As such, our results should not be considered definitive.  

Despite the skepticism brought on by the small sample size, we present the SFH and SMH of the MWAs selected on SM, SFR, and B/T (purple) compared to those selected only on SM and SFR (green) in Figure~\ref{fig:bt_hists}.  Generally, the SM-SFR-B/T MWAs peak higher in SFR and assemble their SM quicker than the original MWAs, but the overall profiles of both the SFH and SMH are consistent with the SM-SFR MWAs, if only on the upper end.  The fact that the inclusion of B/T seems to preferentially select earlier assembly histories is in agreement with recent studies finding early disk formation of the MW \citep{Belokurov+2022,Rix+2022,Conroy+2022,Dillamore+2022,Semenov+2024}. We caution again that no strong conclusions can be drawn with such a small sample size, but note that the inclusion of B/T as a selection parameter did not drastically change our results.  Rather, it only served to introduce additional complications with no added benefit, and so, B/T may not be a defining characteristic of MWAs -- at least not in EAGLE.

\bibliography{MWA}{}

\begin{thebibliography}{}
\expandafter\ifx\csname natexlab\endcsname\relax\def\natexlab#1{#1}\fi
\providecommand{\url}[1]{\href{#1}{#1}}
\providecommand{\dodoi}[1]{doi:~\href{http://doi.org/#1}{\nolinkurl{#1}}}
\providecommand{\doeprint}[1]{\href{http://ascl.net/#1}{\nolinkurl{http://ascl.net/#1}}}
\providecommand{\doarXiv}[1]{\href{https://arxiv.org/abs/#1}{\nolinkurl{https://arxiv.org/abs/#1}}}

\bibitem[{Ade {et~al.}(2014)Ade, Aghanim, Alves, Armitage-Caplan, Arnaud, Ashdown, Atrio-Barandela, Aumont, Aussel, Baccigalupi, Banday, Barreiro, Barrena, Bartelmann, Bartlett, Bartolo, Basak, Battaner, Battye, Benabed, Benoît, Benoit-Lévy, Bernard, Bersanelli, Bertincourt, Bethermin, Bielewicz, Bikmaev, Blanchard, Bobin, Bock, Böhringer, Bonaldi, Bonavera, Bond, Borrill, Bouchet, Boulanger, Bourdin, Bowyer, Bridges, Brown, Bucher, Burenin, Burigana, Butler, Calabrese, Cappellini, Cardoso, Carr, Carvalho, Casale, Castex, Catalano, Challinor, Chamballu, Chary, Chen, Chiang, Chiang, Chon, Christensen, Churazov, Church, Clemens, Clements, Colombi, Colombo, Combet, Comis, Couchot, Coulais, Crill, Cruz, Curto, Cuttaia, Silva, Dahle, Danese, Davies, Davis, de~Bernardis, de~Rosa, de~Zotti, Déchelette, Delabrouille, Delouis, Démoclès, Désert, Dick, Dickinson, Diego, Dolag, Dole, Donzelli, Doré, Douspis, Ducout, Dunkley, Dupac, Efstathiou, Elsner, Enßlin, Eriksen, Fabre, Falgarone, Falvella, Fantaye,
  Fergusson, Filliard, Finelli, Flores-Cacho, Foley, Forni, Fosalba, Frailis, Fraisse, Franceschi, Freschi, Fromenteau, Frommert, Gaier, Galeotta, Gallegos, Galli, Gandolfo, Ganga, Gauthier, Génova-Santos, Ghosh, Giard, Giardino, Gilfanov, Girard, Giraud-Héraud, Gjerløw, González-Nuevo, Górski, Gratton, Gregorio, Gruppuso, Gudmundsson, Haissinski, Hamann, Hansen, Hansen, Hanson, Harrison, Heavens, Helou, Hempel, Henrot-Versillé, Hernández-Monteagudo, Herranz, Hildebrandt, Hivon, Ho, Hobson, Holmes, Hornstrup, Hou, Hovest, Huey, Huffenberger, Hurier, Ilić, Jaffe, Jaffe, Jasche, Jewell, Jones, Juvela, Kalberla, Kangaslahti, Keihänen, Kerp, Keskitalo, Khamitov, Kiiveri, Kim, Kisner, Kneissl, Knoche, Knox, Kunz, Kurki-Suonio, Lacasa, Lagache, Lähteenmäki, Lamarre, Langer, Lasenby, Lattanzi, Laureijs, Lavabre, Lawrence, Jeune, Leach, Leahy, Leonardi, León-Tavares, Leroy, Lesgourgues, Lewis, Li, Liddle, Liguori, Lilje, Linden-Vørnle, Lindholm, López-Caniego, Lowe, Lubin, Macías-Pérez, MacTavish,
  Maffei, Maggio, Maino, Mandolesi, Mangilli, Marcos-Caballero, Marinucci, Maris, Marleau, Marshall, Martin, Martínez-González, Masi, Massardi, Matarrese, Matsumura, Matthai, Maurin, Mazzotta, McDonald, McEwen, McGehee, Mei, Meinhold, Melchiorri, Melin, Mendes, Menegoni, Mennella, Migliaccio, Mikkelsen, Millea, Miniscalco, Mitra, Miville-Deschênes, Molinari, Moneti, Montier, Morgante, Morisset, Mortlock, Moss, Munshi, Murphy, Naselsky, Nati, Natoli, Negrello, Nesvadba, Netterfield, Nørgaard-Nielsen, North, Noviello, Novikov, Novikov, O’Dwyer, Orieux, Osborne, O’Sullivan, Oxborrow, Paci, Pagano, Pajot, Paladini, Pandolfi, Paoletti, Partridge, Pasian, Patanchon, Paykari, Pearson, Pearson, Peel, Peiris, Perdereau, Perotto, Perrotta, Pettorino, Piacentini, Piat, Pierpaoli, Pietrobon, Plaszczynski, Platania, Pogosyan, Pointecouteau, Polenta, Ponthieu, Popa, Poutanen, Pratt, Prézeau, Prunet, Puget, Pullen, Rachen, Racine, Rahlin, Räth, Reach, Rebolo, Reinecke, Remazeilles, Renault, Renzi, Riazuelo,
  Ricciardi, Riller, Ringeval, Ristorcelli, Robbers, Rocha, Roman, Rosset, Rossetti, Roudier, Rowan-Robinson, Rubiño-Martín, Ruiz-Granados, Rusholme, Salerno, Sandri, Sanselme, Santos, Savelainen, Savini, Schaefer, Schiavon, Scott, Seiffert, Serra, Shellard, Smith, Smoot, Souradeep, Spencer, Starck, Stolyarov, Stompor, Sudiwala, Sunyaev, Sureau, Sutter, Sutton, Suur-Uski, Sygnet, Tauber, Tavagnacco, Taylor, Terenzi, Texier, Toffolatti, Tomasi, Torre, Tristram, Tucci, Tuovinen, Türler, Tuttlebee, Umana, Valenziano, Valiviita, Tent, Varis, Vibert, Viel, Vielva, Villa, Vittorio, Wade, Wandelt, Watson, Watson, Wehus, Welikala, Weller, White, White, Wilkinson, Winkel, Xia, Yvon, Zacchei, Zibin, \& Zonca}]{Planck}
Ade, P. A.~R., Aghanim, N., Alves, M. I.~R., {et~al.} 2014, Astronomy \& Astrophysics, 571, A1, \dodoi{10.1051/0004-6361/201321529}

\bibitem[{Ade {et~al.}(2016)Ade, Aghanim, Arnaud, Ashdown, Aumont, Baccigalupi, Banday, Barreiro, Bartlett, Bartolo, Battaner, Battye, Benabed, Benoît, Benoit-Lévy, Bernard, Bersanelli, Bielewicz, Bock, Bonaldi, Bonavera, Bond, Borrill, Bouchet, Boulanger, Bucher, Burigana, Butler, Calabrese, Cardoso, Catalano, Challinor, Chamballu, Chary, Chiang, Chluba, Christensen, Church, Clements, Colombi, Colombo, Combet, Coulais, Crill, Curto, Cuttaia, Danese, Davies, Davis, de~Bernardis, de~Rosa, de~Zotti, Delabrouille, Désert, Valentino, Dickinson, Diego, Dolag, Dole, Donzelli, Doré, Douspis, Ducout, Dunkley, Dupac, Efstathiou, Elsner, Enßlin, Eriksen, Farhang, Fergusson, Finelli, Forni, Frailis, Fraisse, Franceschi, Frejsel, Galeotta, Galli, Ganga, Gauthier, Gerbino, Ghosh, Giard, Giraud-Héraud, Giusarma, Gjerløw, González-Nuevo, Górski, Gratton, Gregorio, Gruppuso, Gudmundsson, Hamann, Hansen, Hanson, Harrison, Helou, Henrot-Versillé, Hernández-Monteagudo, Herranz, Hildebrandt, Hivon, Hobson, Holmes,
  Hornstrup, Hovest, Huang, Huffenberger, Hurier, Jaffe, Jaffe, Jones, Juvela, Keihänen, Keskitalo, Kisner, Kneissl, Knoche, Knox, Kunz, Kurki-Suonio, Lagache, Lähteenmäki, Lamarre, Lasenby, Lattanzi, Lawrence, Leahy, Leonardi, Lesgourgues, Levrier, Lewis, Liguori, Lilje, Linden-Vørnle, López-Caniego, Lubin, Macías-Pérez, Maggio, Maino, Mandolesi, Mangilli, Marchini, Maris, Martin, Martinelli, Martínez-González, Masi, Matarrese, McGehee, Meinhold, Melchiorri, Melin, Mendes, Mennella, Migliaccio, Millea, Mitra, Miville-Deschênes, Moneti, Montier, Morgante, Mortlock, Moss, Munshi, Murphy, Naselsky, Nati, Natoli, Netterfield, Nørgaard-Nielsen, Noviello, Novikov, Novikov, Oxborrow, Paci, Pagano, Pajot, Paladini, Paoletti, Partridge, Pasian, Patanchon, Pearson, Perdereau, Perotto, Perrotta, Pettorino, Piacentini, Piat, Pierpaoli, Pietrobon, Plaszczynski, Pointecouteau, Polenta, Popa, Pratt, Prézeau, Prunet, Puget, Rachen, Reach, Rebolo, Reinecke, Remazeilles, Renault, Renzi, Ristorcelli, Rocha, Rosset,
  Rossetti, Roudier, d’Orfeuil, Rowan-Robinson, Rubiño-Martín, Rusholme, Said, Salvatelli, Salvati, Sandri, Santos, Savelainen, Savini, Scott, Seiffert, Serra, Shellard, Spencer, Spinelli, Stolyarov, Stompor, Sudiwala, Sunyaev, Sutton, Suur-Uski, Sygnet, Tauber, Terenzi, Toffolatti, Tomasi, Tristram, Trombetti, Tucci, Tuovinen, Türler, Umana, Valenziano, Valiviita, Tent, Vielva, Villa, Wade, Wandelt, Wehus, White, White, Wilkinson, Yvon, Zacchei, \& Zonca}]{Planck2015}
Ade, P. A.~R., Aghanim, N., Arnaud, M., {et~al.} 2016, Astronomy \& Astrophysics, 594, A13, \dodoi{10.1051/0004-6361/201525830}

\bibitem[{{Agertz} \& {Kravtsov}(2015)}]{Agertz+2015}
{Agertz}, O., \& {Kravtsov}, A.~V. 2015, \apj, 804, 18, \dodoi{10.1088/0004-637X/804/1/18}

\bibitem[{{Baldry} {et~al.}(2008){Baldry}, {Glazebrook}, \& {Driver}}]{Baldry+2008}
{Baldry}, I.~K., {Glazebrook}, K., \& {Driver}, S.~P. 2008, \mnras, 388, 945, \dodoi{10.1111/j.1365-2966.2008.13348.x}

\bibitem[{{Baldry} {et~al.}(2012){Baldry}, {Driver}, {Loveday}, {Taylor}, {Kelvin}, {Liske}, {Norberg}, {Robotham}, {Brough}, {Hopkins}, {Bamford}, {Peacock}, {Bland-Hawthorn}, {Conselice}, {Croom}, {Jones}, {Parkinson}, {Popescu}, {Prescott}, {Sharp}, \& {Tuffs}}]{Baldry+2012}
{Baldry}, I.~K., {Driver}, S.~P., {Loveday}, J., {et~al.} 2012, \mnras, 421, 621, \dodoi{10.1111/j.1365-2966.2012.20340.x}

\bibitem[{{Behroozi} {et~al.}(2019){Behroozi}, {Wechsler}, {Hearin}, \& {Conroy}}]{UM2}
{Behroozi}, P., {Wechsler}, R.~H., {Hearin}, A.~P., \& {Conroy}, C. 2019, \mnras, 488, 3143, \dodoi{10.1093/mnras/stz1182}

\bibitem[{{Behroozi} {et~al.}(2013){Behroozi}, {Wechsler}, \& {Conroy}}]{UM1}
{Behroozi}, P.~S., {Wechsler}, R.~H., \& {Conroy}, C. 2013, \apj, 770, 57, \dodoi{10.1088/0004-637X/770/1/57}

\bibitem[{{Belokurov} \& {Kravtsov}(2022)}]{Belokurov+2022}
{Belokurov}, V., \& {Kravtsov}, A. 2022, \mnras, 514, 689, \dodoi{10.1093/mnras/stac1267}

\bibitem[{{Bernardi} {et~al.}(2013){Bernardi}, {Meert}, {Sheth}, {Vikram}, {Huertas-Company}, {Mei}, \& {Shankar}}]{Bernardi+2013}
{Bernardi}, M., {Meert}, A., {Sheth}, R.~K., {et~al.} 2013, \mnras, 436, 697, \dodoi{10.1093/mnras/stt1607}

\bibitem[{{Bottinelli} {et~al.}(1985){Bottinelli}, {Gouguenheim}, {Paturel}, \& {de Vaucouleurs}}]{Bottinelli+1985}
{Bottinelli}, L., {Gouguenheim}, L., {Paturel}, G., \& {de Vaucouleurs}, G. 1985, \apjs, 59, 293, \dodoi{10.1086/191073}

\bibitem[{{Brandt} \& {Draine}(2012)}]{BrandtDraine2012}
{Brandt}, T.~D., \& {Draine}, B.~T. 2012, \apj, 744, 129, \dodoi{10.1088/0004-637X/744/2/129}

\bibitem[{{Brennan} {et~al.}(2017){Brennan}, {Pandya}, {Somerville}, {Barro}, {Bluck}, {Taylor}, {Wuyts}, {Bell}, {Dekel}, {Faber}, {Ferguson}, {Koekemoer}, {Kurczynski}, {McIntosh}, {Newman}, \& {Primack}}]{SCSAM4}
{Brennan}, R., {Pandya}, V., {Somerville}, R.~S., {et~al.} 2017, \mnras, 465, 619, \dodoi{10.1093/mnras/stw2690}

\bibitem[{Chabrier(2003)}]{Chabrier2003}
Chabrier, G. 2003, Publications of the Astronomical Society of the Pacific, 115, 763, \dodoi{10.1086/376392}

\bibitem[{{Conroy} {et~al.}(2022){Conroy}, {Weinberg}, {Naidu}, {Buck}, {Johnson}, {Cargile}, {Bonaca}, {Caldwell}, {Chandra}, {Han}, {Johnson}, {Speagle}, {Ting}, {Woody}, \& {Zaritsky}}]{Conroy+2022}
{Conroy}, C., {Weinberg}, D.~H., {Naidu}, R.~P., {et~al.} 2022, arXiv e-prints, arXiv:2204.02989, \dodoi{10.48550/arXiv.2204.02989}

\bibitem[{{Conselice}(2014)}]{Conselice2014}
{Conselice}, C.~J. 2014, \araa, 52, 291, \dodoi{10.1146/annurev-astro-081913-040037}

\bibitem[{Crain {et~al.}(2015)Crain, Schaye, Bower, Furlong, Schaller, Theuns, Vecchia, Frenk, McCarthy, Helly, Jenkins, Rosas-Guevara, White, \& Trayford}]{EAGLE2}
Crain, R.~A., Schaye, J., Bower, R.~G., {et~al.} 2015, Monthly Notices of the Royal Astronomical Society, 450, 1937, \dodoi{10.1093/mnras/stv725}

\bibitem[{{Dav{\'e}} {et~al.}(2019){Dav{\'e}}, {Angl{\'e}s-Alc{\'a}zar}, {Narayanan}, {Li}, {Rafieferantsoa}, \& {Appleby}}]{Simba}
{Dav{\'e}}, R., {Angl{\'e}s-Alc{\'a}zar}, D., {Narayanan}, D., {et~al.} 2019, \mnras, 486, 2827, \dodoi{10.1093/mnras/stz937}

\bibitem[{{Dav{\'e}} {et~al.}(2016){Dav{\'e}}, {Thompson}, \& {Hopkins}}]{Mufasa}
{Dav{\'e}}, R., {Thompson}, R., \& {Hopkins}, P.~F. 2016, \mnras, 462, 3265, \dodoi{10.1093/mnras/stw1862}

\bibitem[{{de Vaucouleurs} \& {Corwin}(1986)}]{deVacCorwin1986}
{de Vaucouleurs}, G., \& {Corwin}, H.~G., J. 1986, \apj, 308, 487, \dodoi{10.1086/164519}

\bibitem[{{Deason} \& {Belokurov}(2024)}]{DeasonBelokurov2024}
{Deason}, A.~J., \& {Belokurov}, V. 2024, \nar, 99, 101706, \dodoi{10.1016/j.newar.2024.101706}

\bibitem[{{Dillamore} {et~al.}(2022){Dillamore}, {Belokurov}, {Font}, \& {McCarthy}}]{Dillamore+2022}
{Dillamore}, A.~M., {Belokurov}, V., {Font}, A.~S., \& {McCarthy}, I.~G. 2022, \mnras, 513, 1867, \dodoi{10.1093/mnras/stac1038}

\bibitem[{{D'Souza} {et~al.}(2015){D'Souza}, {Vegetti}, \& {Kauffmann}}]{DSouza+2015}
{D'Souza}, R., {Vegetti}, S., \& {Kauffmann}, G. 2015, \mnras, 454, 4027, \dodoi{10.1093/mnras/stv2234}

\bibitem[{Durier \& Vecchia(2012)}]{DurierDallaVecchia2012}
Durier, F., \& Vecchia, C.~D. 2012, Monthly Notices of the Royal Astronomical Society, 419, 465, \dodoi{10.1111/j.1365-2966.2011.19712.x}

\bibitem[{{Fielder} {et~al.}(2021){Fielder}, {Newman}, {Andrews}, {Zasowski}, {Boardman}, {Licquia}, {Masters}, \& {Salim}}]{Fielder+2021}
{Fielder}, C.~E., {Newman}, J.~A., {Andrews}, B.~H., {et~al.} 2021, \mnras, 508, 4459, \dodoi{10.1093/mnras/stab2618}

\bibitem[{{Flynn} {et~al.}(2006){Flynn}, {Holmberg}, {Portinari}, {Fuchs}, \& {Jahrei{\ss}}}]{Flynn+2006}
{Flynn}, C., {Holmberg}, J., {Portinari}, L., {Fuchs}, B., \& {Jahrei{\ss}}, H. 2006, \mnras, 372, 1149, \dodoi{10.1111/j.1365-2966.2006.10911.x}

\bibitem[{{Font} {et~al.}(2020){Font}, {McCarthy}, {Poole-Mckenzie}, {Stafford}, {Brown}, {Schaye}, {Crain}, {Theuns}, \& {Schaller}}]{Font+2020}
{Font}, A.~S., {McCarthy}, I.~G., {Poole-Mckenzie}, R., {et~al.} 2020, \mnras, 498, 1765, \dodoi{10.1093/mnras/staa2463}

\bibitem[{{Gaia Collaboration} {et~al.}(2016){Gaia Collaboration}, {Prusti}, {de Bruijne}, {Brown}, {Vallenari}, {Babusiaux}, {Bailer-Jones}, {Bastian}, {Biermann}, {Evans}, {Eyer}, {Jansen}, {Jordi}, {Klioner}, {Lammers}, {Lindegren}, {Luri}, {Mignard}, {Milligan}, {Panem}, {Poinsignon}, {Pourbaix}, {Randich}, {Sarri}, {Sartoretti}, {Siddiqui}, {Soubiran}, {Valette}, {van Leeuwen}, {Walton}, {Aerts}, {Arenou}, {Cropper}, {Drimmel}, {H{\o}g}, {Katz}, {Lattanzi}, {O'Mullane}, {Grebel}, {Holland}, {Huc}, {Passot}, {Bramante}, {Cacciari}, {Casta{\~n}eda}, {Chaoul}, {Cheek}, {De Angeli}, {Fabricius}, {Guerra}, {Hern{\'a}ndez}, {Jean-Antoine-Piccolo}, {Masana}, {Messineo}, {Mowlavi}, {Nienartowicz}, {Ord{\'o}{\~n}ez-Blanco}, {Panuzzo}, {Portell}, {Richards}, {Riello}, {Seabroke}, {Tanga}, {Th{\'e}venin}, {Torra}, {Els}, {Gracia-Abril}, {Comoretto}, {Garcia-Reinaldos}, {Lock}, {Mercier}, {Altmann}, {Andrae}, {Astraatmadja}, {Bellas-Velidis}, {Benson}, {Berthier}, {Blomme}, {Busso}, {Carry}, {Cellino}, {Clementini},
  {Cowell}, {Creevey}, {Cuypers}, {Davidson}, {De Ridder}, {de Torres}, {Delchambre}, {Dell'Oro}, {Ducourant}, {Fr{\'e}mat}, {Garc{\'\i}a-Torres}, {Gosset}, {Halbwachs}, {Hambly}, {Harrison}, {Hauser}, {Hestroffer}, {Hodgkin}, {Huckle}, {Hutton}, {Jasniewicz}, {Jordan}, {Kontizas}, {Korn}, {Lanzafame}, {Manteiga}, {Moitinho}, {Muinonen}, {Osinde}, {Pancino}, {Pauwels}, {Petit}, {Recio-Blanco}, {Robin}, {Sarro}, {Siopis}, {Smith}, {Smith}, {Sozzetti}, {Thuillot}, {van Reeven}, {Viala}, {Abbas}, {Abreu Aramburu}, {Accart}, {Aguado}, {Allan}, {Allasia}, {Altavilla}, {{\'A}lvarez}, {Alves}, {Anderson}, {Andrei}, {Anglada Varela}, {Antiche}, {Antoja}, {Ant{\'o}n}, {Arcay}, {Atzei}, {Ayache}, {Bach}, {Baker}, {Balaguer-N{\'u}{\~n}ez}, {Barache}, {Barata}, {Barbier}, {Barblan}, {Baroni}, {Barrado y Navascu{\'e}s}, {Barros}, {Barstow}, {Becciani}, {Bellazzini}, {Bellei}, {Bello Garc{\'\i}a}, {Belokurov}, {Bendjoya}, {Berihuete}, {Bianchi}, {Bienaym{\'e}}, {Billebaud}, {Blagorodnova}, {Blanco-Cuaresma}, {Boch},
  {Bombrun}, {Borrachero}, {Bouquillon}, {Bourda}, {Bouy}, {Bragaglia}, {Breddels}, {Brouillet}, {Br{\"u}semeister}, {Bucciarelli}, {Budnik}, {Burgess}, {Burgon}, {Burlacu}, {Busonero}, {Buzzi}, {Caffau}, {Cambras}, {Campbell}, {Cancelliere}, {Cantat-Gaudin}, {Carlucci}, {Carrasco}, {Castellani}, {Charlot}, {Charnas}, {Charvet}, {Chassat}, {Chiavassa}, {Clotet}, {Cocozza}, {Collins}, {Collins}, {Costigan}, {Crifo}, {Cross}, {Crosta}, {Crowley}, {Dafonte}, {Damerdji}, {Dapergolas}, {David}, {David}, {De Cat}, {de Felice}, {de Laverny}, {De Luise}, {De March}, {de Martino}, {de Souza}, {Debosscher}, {del Pozo}, {Delbo}, {Delgado}, {Delgado}, {di Marco}, {Di Matteo}, {Diakite}, {Distefano}, {Dolding}, {Dos Anjos}, {Drazinos}, {Dur{\'a}n}, {Dzigan}, {Ecale}, {Edvardsson}, {Enke}, {Erdmann}, {Escolar}, {Espina}, {Evans}, {Eynard Bontemps}, {Fabre}, {Fabrizio}, {Faigler}, {Falc{\~a}o}, {Farr{\`a}s Casas}, {Faye}, {Federici}, {Fedorets}, {Fern{\'a}ndez-Hern{\'a}ndez}, {Fernique}, {Fienga}, {Figueras}, {Filippi},
  {Findeisen}, {Fonti}, {Fouesneau}, {Fraile}, {Fraser}, {Fuchs}, {Furnell}, {Gai}, {Galleti}, {Galluccio}, {Garabato}, {Garc{\'\i}a-Sedano}, {Gar{\'e}}, {Garofalo}, {Garralda}, {Gavras}, {Gerssen}, {Geyer}, {Gilmore}, {Girona}, {Giuffrida}, {Gomes}, {Gonz{\'a}lez-Marcos}, {Gonz{\'a}lez-N{\'u}{\~n}ez}, {Gonz{\'a}lez-Vidal}, {Granvik}, {Guerrier}, {Guillout}, {Guiraud}, {G{\'u}rpide}, {Guti{\'e}rrez-S{\'a}nchez}, {Guy}, {Haigron}, {Hatzidimitriou}, {Haywood}, {Heiter}, {Helmi}, {Hobbs}, {Hofmann}, {Holl}, {Holland}, {Hunt}, {Hypki}, {Icardi}, {Irwin}, {Jevardat de Fombelle}, {Jofr{\'e}}, {Jonker}, {Jorissen}, {Julbe}, {Karampelas}, {Kochoska}, {Kohley}, {Kolenberg}, {Kontizas}, {Koposov}, {Kordopatis}, {Koubsky}, {Kowalczyk}, {Krone-Martins}, {Kudryashova}, {Kull}, {Bachchan}, {Lacoste-Seris}, {Lanza}, {Lavigne}, {Le Poncin-Lafitte}, {Lebreton}, {Lebzelter}, {Leccia}, {Leclerc}, {Lecoeur-Taibi}, {Lemaitre}, {Lenhardt}, {Leroux}, {Liao}, {Licata}, {Lindstr{\o}m}, {Lister}, {Livanou}, {Lobel}, {L{\"o}ffler},
  {L{\'o}pez}, {Lopez-Lozano}, {Lorenz}, {Loureiro}, {MacDonald}, {Magalh{\~a}es Fernandes}, {Managau}, {Mann}, {Mantelet}, {Marchal}, {Marchant}, {Marconi}, {Marie}, {Marinoni}, {Marrese}, {Marschalk{\'o}}, {Marshall}, {Mart{\'\i}n-Fleitas}, {Martino}, {Mary}, {Matijevi{\v{c}}}, {Mazeh}, {McMillan}, {Messina}, {Mestre}, {Michalik}, {Millar}, {Miranda}, {Molina}, {Molinaro}, {Molinaro}, {Moln{\'a}r}, {Moniez}, {Montegriffo}, {Monteiro}, {Mor}, {Mora}, {Morbidelli}, {Morel}, {Morgenthaler}, {Morley}, {Morris}, {Mulone}, {Muraveva}, {Musella}, {Narbonne}, {Nelemans}, {Nicastro}, {Noval}, {Ord{\'e}novic}, {Ordieres-Mer{\'e}}, {Osborne}, {Pagani}, {Pagano}, {Pailler}, {Palacin}, {Palaversa}, {Parsons}, {Paulsen}, {Pecoraro}, {Pedrosa}, {Pentik{\"a}inen}, {Pereira}, {Pichon}, {Piersimoni}, {Pineau}, {Plachy}, {Plum}, {Poujoulet}, {Pr{\v{s}}a}, {Pulone}, {Ragaini}, {Rago}, {Rambaux}, {Ramos-Lerate}, {Ranalli}, {Rauw}, {Read}, {Regibo}, {Renk}, {Reyl{\'e}}, {Ribeiro}, {Rimoldini}, {Ripepi}, {Riva}, {Rixon},
  {Roelens}, {Romero-G{\'o}mez}, {Rowell}, {Royer}, {Rudolph}, {Ruiz-Dern}, {Sadowski}, {Sagrist{\`a} Sell{\'e}s}, {Sahlmann}, {Salgado}, {Salguero}, {Sarasso}, {Savietto}, {Schnorhk}, {Schultheis}, {Sciacca}, {Segol}, {Segovia}, {Segransan}, {Serpell}, {Shih}, {Smareglia}, {Smart}, {Smith}, {Solano}, {Solitro}, {Sordo}, {Soria Nieto}, {Souchay}, {Spagna}, {Spoto}, {Stampa}, {Steele}, {Steidelm{\"u}ller}, {Stephenson}, {Stoev}, {Suess}, {S{\"u}veges}, {Surdej}, {Szabados}, {Szegedi-Elek}, {Tapiador}, {Taris}, {Tauran}, {Taylor}, {Teixeira}, {Terrett}, {Tingley}, {Trager}, {Turon}, {Ulla}, {Utrilla}, {Valentini}, {van Elteren}, {Van Hemelryck}, {van Leeuwen}, {Varadi}, {Vecchiato}, {Veljanoski}, {Via}, {Vicente}, {Vogt}, {Voss}, {Votruba}, {Voutsinas}, {Walmsley}, {Weiler}, {Weingrill}, {Werner}, {Wevers}, {Whitehead}, {Wyrzykowski}, {Yoldas}, {{\v{Z}}erjal}, {Zucker}, {Zurbach}, {Zwitter}, {Alecu}, {Allen}, {Allende Prieto}, {Amorim}, {Anglada-Escud{\'e}}, {Arsenijevic}, {Azaz}, {Balm}, {Beck}, {Bernstein},
  {Bigot}, {Bijaoui}, {Blasco}, {Bonfigli}, {Bono}, {Boudreault}, {Bressan}, {Brown}, {Brunet}, {Bunclark}, {Buonanno}, {Butkevich}, {Carret}, {Carrion}, {Chemin}, {Ch{\'e}reau}, {Corcione}, {Darmigny}, {de Boer}, {de Teodoro}, {de Zeeuw}, {Delle Luche}, {Domingues}, {Dubath}, {Fodor}, {Fr{\'e}zouls}, {Fries}, {Fustes}, {Fyfe}, {Gallardo}, {Gallegos}, {Gardiol}, {Gebran}, {Gomboc}, {G{\'o}mez}, {Grux}, {Gueguen}, {Heyrovsky}, {Hoar}, {Iannicola}, {Isasi Parache}, {Janotto}, {Joliet}, {Jonckheere}, {Keil}, {Kim}, {Klagyivik}, {Klar}, {Knude}, {Kochukhov}, {Kolka}, {Kos}, {Kutka}, {Lainey}, {LeBouquin}, {Liu}, {Loreggia}, {Makarov}, {Marseille}, {Martayan}, {Martinez-Rubi}, {Massart}, {Meynadier}, {Mignot}, {Munari}, {Nguyen}, {Nordlander}, {Ocvirk}, {O'Flaherty}, {Olias Sanz}, {Ortiz}, {Osorio}, {Oszkiewicz}, {Ouzounis}, {Palmer}, {Park}, {Pasquato}, {Peltzer}, {Peralta}, {P{\'e}turaud}, {Pieniluoma}, {Pigozzi}, {Poels}, {Prat}, {Prod'homme}, {Raison}, {Rebordao}, {Risquez}, {Rocca-Volmerange}, {Rosen},
  {Ruiz-Fuertes}, {Russo}, {Sembay}, {Serraller Vizcaino}, {Short}, {Siebert}, {Silva}, {Sinachopoulos}, {Slezak}, {Soffel}, {Sosnowska}, {Strai{\v{z}}ys}, {ter Linden}, {Terrell}, {Theil}, {Tiede}, {Troisi}, {Tsalmantza}, {Tur}, {Vaccari}, {Vachier}, {Valles}, {Van Hamme}, {Veltz}, {Virtanen}, {Wallut}, {Wichmann}, {Wilkinson}, {Ziaeepour}, \& {Zschocke}}]{Gaia}
{Gaia Collaboration}, {Prusti}, T., {de Bruijne}, J.~H.~J., {et~al.} 2016, \aap, 595, A1, \dodoi{10.1051/0004-6361/201629272}

\bibitem[{{Garrison-Kimmel} {et~al.}(2014){Garrison-Kimmel}, {Boylan-Kolchin}, {Bullock}, \& {Lee}}]{Garrison-Kimmel+2014}
{Garrison-Kimmel}, S., {Boylan-Kolchin}, M., {Bullock}, J.~S., \& {Lee}, K. 2014, \mnras, 438, 2578, \dodoi{10.1093/mnras/stt2377}

\bibitem[{{Giodini} {et~al.}(2009){Giodini}, {Pierini}, {Finoguenov}, {Pratt}, {Boehringer}, {Leauthaud}, {Guzzo}, {Aussel}, {Bolzonella}, {Capak}, {Elvis}, {Hasinger}, {Ilbert}, {Kartaltepe}, {Koekemoer}, {Lilly}, {Massey}, {McCracken}, {Rhodes}, {Salvato}, {Sanders}, {Scoville}, {Sasaki}, {Smolcic}, {Taniguchi}, {Thompson}, \& {COSMOS Collaboration}}]{Giodini+2009}
{Giodini}, S., {Pierini}, D., {Finoguenov}, A., {et~al.} 2009, \apj, 703, 982, \dodoi{10.1088/0004-637X/703/1/982}

\bibitem[{{Grand} {et~al.}(2017){Grand}, {G{\'o}mez}, {Marinacci}, {Pakmor}, {Springel}, {Campbell}, {Frenk}, {Jenkins}, \& {White}}]{Grand+2017}
{Grand}, R. J.~J., {G{\'o}mez}, F.~A., {Marinacci}, F., {et~al.} 2017, \mnras, 467, 179, \dodoi{10.1093/mnras/stx071}

\bibitem[{{Guedes} {et~al.}(2011){Guedes}, {Callegari}, {Madau}, \& {Mayer}}]{Guedes+2011}
{Guedes}, J., {Callegari}, S., {Madau}, P., \& {Mayer}, L. 2011, \apj, 742, 76, \dodoi{10.1088/0004-637X/742/2/76}

\bibitem[{{Hammer} {et~al.}(2007){Hammer}, {Puech}, {Chemin}, {Flores}, \& {Lehnert}}]{Hammer+2007}
{Hammer}, F., {Puech}, M., {Chemin}, L., {Flores}, H., \& {Lehnert}, M.~D. 2007, \apj, 662, 322, \dodoi{10.1086/516727}

\bibitem[{Harris {et~al.}(2020)Harris, Millman, van~der Walt, Gommers, Virtanen, Cournapeau, Wieser, Taylor, Berg, Smith, Kern, Picus, Hoyer, van Kerkwijk, Brett, Haldane, del Río, Wiebe, Peterson, Gérard-Marchant, Sheppard, Reddy, Weckesser, Abbasi, Gohlke, \& Oliphant}]{numpy}
Harris, C.~R., Millman, K.~J., van~der Walt, S.~J., {et~al.} 2020, Nature, 585, 357, \dodoi{10.1038/s41586-020-2649-2}

\bibitem[{Hopkins(2013)}]{Hopkins2013}
Hopkins, P.~F. 2013, Monthly Notices of the Royal Astronomical Society, 428, 2840, \dodoi{10.1093/mnras/sts210}

\bibitem[{{Hopkins} {et~al.}(2018){Hopkins}, {Wetzel}, {Kere{\v{s}}}, {Faucher-Gigu{\`e}re}, {Quataert}, {Boylan-Kolchin}, {Murray}, {Hayward}, {Garrison-Kimmel}, {Hummels}, {Feldmann}, {Torrey}, {Ma}, {Angl{\'e}s-Alc{\'a}zar}, {Su}, {Orr}, {Schmitz}, {Escala}, {Sanderson}, {Grudi{\'c}}, {Hafen}, {Kim}, {Fitts}, {Bullock}, {Wheeler}, {Chan}, {Elbert}, \& {Narayanan}}]{FIRE}
{Hopkins}, P.~F., {Wetzel}, A., {Kere{\v{s}}}, D., {et~al.} 2018, \mnras, 480, 800, \dodoi{10.1093/mnras/sty1690}

\bibitem[{{Horta} \& {Schiavon}(2024)}]{HortaSchiavon2024}
{Horta}, D., \& {Schiavon}, R.~P. 2024, arXiv e-prints, arXiv:2404.16939, \dodoi{10.48550/arXiv.2404.16939}

\bibitem[{{Hubble}(1926)}]{Hubble1926}
{Hubble}, E.~P. 1926, \apj, 64, 321, \dodoi{10.1086/143018}

\bibitem[{Hunter(2007)}]{matplotlib}
Hunter, J.~D. 2007, Computing in Science \& Engineering, 9, 90, \dodoi{10.1109/MCSE.2007.55}

\bibitem[{{Iyer} {et~al.}(2020){Iyer}, {Tacchella}, {Genel}, {Hayward}, {Hernquist}, {Brooks}, {Caplar}, {Dav{\'e}}, {Diemer}, {Forbes}, {Gawiser}, {Somerville}, \& {Starkenburg}}]{Iyer2020}
{Iyer}, K.~G., {Tacchella}, S., {Genel}, S., {et~al.} 2020, \mnras, 498, 430, \dodoi{10.1093/mnras/staa2150}

\bibitem[{{Kormendy} \& {Ho}(2013)}]{Kormendy+2013}
{Kormendy}, J., \& {Ho}, L.~C. 2013, \araa, 51, 511, \dodoi{10.1146/annurev-astro-082708-101811}

\bibitem[{{Kruijssen} {et~al.}(2019){Kruijssen}, {Pfeffer}, {Reina-Campos}, {Crain}, \& {Bastian}}]{Kruijssen+2019}
{Kruijssen}, J.~M.~D., {Pfeffer}, J.~L., {Reina-Campos}, M., {Crain}, R.~A., \& {Bastian}, N. 2019, \mnras, 486, 3180, \dodoi{10.1093/mnras/sty1609}

\bibitem[{{Leung} \& {Bovy}(2019)}]{Henry}
{Leung}, H.~W., \& {Bovy}, J. 2019, \mnras, 483, 3255, \dodoi{10.1093/mnras/sty3217}

\bibitem[{{Li} \& {White}(2009)}]{LiWhite2009}
{Li}, C., \& {White}, S. D.~M. 2009, \mnras, 398, 2177, \dodoi{10.1111/j.1365-2966.2009.15268.x}

\bibitem[{{Libeskind} {et~al.}(2020){Libeskind}, {Carlesi}, {Grand}, {Khalatyan}, {Knebe}, {Pakmor}, {Pilipenko}, {Pawlowski}, {Sparre}, {Tempel}, {Wang}, {Courtois}, {Gottl{\"o}ber}, {Hoffman}, {Minchev}, {Pfrommer}, {Sorce}, {Springel}, {Steinmetz}, {Tully}, {Vogelsberger}, \& {Yepes}}]{Libeskind+2020}
{Libeskind}, N.~I., {Carlesi}, E., {Grand}, R. J.~J., {et~al.} 2020, \mnras, 498, 2968, \dodoi{10.1093/mnras/staa2541}

\bibitem[{{Licquia} {et~al.}(2015){Licquia}, {Newman}, \& {Brinchmann}}]{LNB15}
{Licquia}, T.~C., {Newman}, J.~A., \& {Brinchmann}, J. 2015, \apj, 809, 96, \dodoi{10.1088/0004-637X/809/1/96}

\bibitem[{{Lovisari} {et~al.}(2015){Lovisari}, {Reiprich}, \& {Schellenberger}}]{Lovisari+2015}
{Lovisari}, L., {Reiprich}, T.~H., \& {Schellenberger}, G. 2015, \aap, 573, A118, \dodoi{10.1051/0004-6361/201423954}

\bibitem[{{Marinacci} {et~al.}(2018){Marinacci}, {Vogelsberger}, {Pakmor}, {Torrey}, {Springel}, {Hernquist}, {Nelson}, {Weinberger}, {Pillepich}, {Naiman}, \& {Genel}}]{TNG100-5}
{Marinacci}, F., {Vogelsberger}, M., {Pakmor}, R., {et~al.} 2018, \mnras, 480, 5113, \dodoi{10.1093/mnras/sty2206}

\bibitem[{{Marshall} {et~al.}(2008){Marshall}, {Fux}, {Robin}, \& {Reyl{\'e}}}]{Marshall+2008}
{Marshall}, D.~J., {Fux}, R., {Robin}, A.~C., \& {Reyl{\'e}}, C. 2008, \aap, 477, L21, \dodoi{10.1051/0004-6361:20078967}

\bibitem[{{McConnell} \& {Ma}(2013)}]{McConnellMa2013}
{McConnell}, N.~J., \& {Ma}, C.-P. 2013, \apj, 764, 184, \dodoi{10.1088/0004-637X/764/2/184}

\bibitem[{McInnes {et~al.}(2018)McInnes, Healy, Saul, \& Gro{\ss}berger}]{umap}
McInnes, L., Healy, J., Saul, N., \& Gro{\ss}berger, L. 2018, Journal of Open Source Software, 3, 861, \dodoi{10.21105/joss.00861}

\bibitem[{{Moster} {et~al.}(2013){Moster}, {Naab}, \& {White}}]{Moster+2013}
{Moster}, B.~P., {Naab}, T., \& {White}, S. D.~M. 2013, \mnras, 428, 3121, \dodoi{10.1093/mnras/sts261}

\bibitem[{{Naiman} {et~al.}(2018){Naiman}, {Pillepich}, {Springel}, {Ramirez-Ruiz}, {Torrey}, {Vogelsberger}, {Pakmor}, {Nelson}, {Marinacci}, {Hernquist}, {Weinberger}, \& {Genel}}]{TNG100-4}
{Naiman}, J.~P., {Pillepich}, A., {Springel}, V., {et~al.} 2018, \mnras, 477, 1206, \dodoi{10.1093/mnras/sty618}

\bibitem[{{Nelson} {et~al.}(2015){Nelson}, {Pillepich}, {Genel}, {Vogelsberger}, {Springel}, {Torrey}, {Rodriguez-Gomez}, {Sijacki}, {Snyder}, {Griffen}, {Marinacci}, {Blecha}, {Sales}, {Xu}, \& {Hernquist}}]{Illustris1}
{Nelson}, D., {Pillepich}, A., {Genel}, S., {et~al.} 2015, Astronomy and Computing, 13, 12, \dodoi{10.1016/j.ascom.2015.09.003}

\bibitem[{{Nelson} {et~al.}(2018){Nelson}, {Pillepich}, {Springel}, {Weinberger}, {Hernquist}, {Pakmor}, {Genel}, {Torrey}, {Vogelsberger}, {Kauffmann}, {Marinacci}, \& {Naiman}}]{TNG100-3}
{Nelson}, D., {Pillepich}, A., {Springel}, V., {et~al.} 2018, \mnras, 475, 624, \dodoi{10.1093/mnras/stx3040}

\bibitem[{{Nelson} {et~al.}(2019){Nelson}, {Springel}, {Pillepich}, {Rodriguez-Gomez}, {Torrey}, {Genel}, {Vogelsberger}, {Pakmor}, {Marinacci}, {Weinberger}, {Kelley}, {Lovell}, {Diemer}, \& {Hernquist}}]{TNG100-0}
{Nelson}, D., {Springel}, V., {Pillepich}, A., {et~al.} 2019, Computational Astrophysics and Cosmology, 6, 2, \dodoi{10.1186/s40668-019-0028-x}

\bibitem[{{Oesch} {et~al.}(2015){Oesch}, {Bouwens}, {Illingworth}, {Franx}, {Ammons}, {van Dokkum}, {Trenti}, \& {Labb{\'e}}}]{Oesch+2015}
{Oesch}, P.~A., {Bouwens}, R.~J., {Illingworth}, G.~D., {et~al.} 2015, \apj, 808, 104, \dodoi{10.1088/0004-637X/808/1/104}

\bibitem[{{Papovich} {et~al.}(2015){Papovich}, {Labb{\'e}}, {Quadri}, {Tilvi}, {Behroozi}, {Bell}, {Glazebrook}, {Spitler}, {Straatman}, {Tran}, {Cowley}, {Dav{\'e}}, {Dekel}, {Dickinson}, {Ferguson}, {Finkelstein}, {Gawiser}, {Inami}, {Faber}, {Kacprzak}, {Kawinwanichakij}, {Kocevski}, {Koekemoer}, {Koo}, {Kurczynski}, {Lotz}, {Lu}, {Lucas}, {McIntosh}, {Mehrtens}, {Mobasher}, {Monson}, {Morrison}, {Nanayakkara}, {Persson}, {Salmon}, {Simons}, {Tomczak}, {van Dokkum}, {Weiner}, \& {Willner}}]{Papovich+2015}
{Papovich}, C., {Labb{\'e}}, I., {Quadri}, R., {et~al.} 2015, \apj, 803, 26, \dodoi{10.1088/0004-637X/803/1/26}

\bibitem[{{Patel} {et~al.}(2013){Patel}, {Fumagalli}, {Franx}, {van Dokkum}, {van der Wel}, {Leja}, {Labb{\'e}}, {Brammer}, {Skelton}, {Momcheva}, {Whitaker}, {Lundgren}, {Muzzin}, {Quadri}, {Nelson}, {Wake}, \& {Rix}}]{Patel+2013}
{Patel}, S.~G., {Fumagalli}, M., {Franx}, M., {et~al.} 2013, \apj, 778, 115, \dodoi{10.1088/0004-637X/778/2/115}

\bibitem[{{Pillepich} {et~al.}(2018{\natexlab{a}}){Pillepich}, {Springel}, {Nelson}, {Genel}, {Naiman}, {Pakmor}, {Hernquist}, {Torrey}, {Vogelsberger}, {Weinberger}, \& {Marinacci}}]{TNG2}
{Pillepich}, A., {Springel}, V., {Nelson}, D., {et~al.} 2018{\natexlab{a}}, \mnras, 473, 4077, \dodoi{10.1093/mnras/stx2656}

\bibitem[{{Pillepich} {et~al.}(2018{\natexlab{b}}){Pillepich}, {Nelson}, {Hernquist}, {Springel}, {Pakmor}, {Torrey}, {Weinberger}, {Genel}, {Naiman}, {Marinacci}, \& {Vogelsberger}}]{TNG100-2}
{Pillepich}, A., {Nelson}, D., {Hernquist}, L., {et~al.} 2018{\natexlab{b}}, \mnras, 475, 648, \dodoi{10.1093/mnras/stx3112}

\bibitem[{{Porter} {et~al.}(2014){Porter}, {Somerville}, {Primack}, \& {Johansson}}]{SCSAM3}
{Porter}, L.~A., {Somerville}, R.~S., {Primack}, J.~R., \& {Johansson}, P.~H. 2014, \mnras, 444, 942, \dodoi{10.1093/mnras/stu1434}

\bibitem[{{Rix} {et~al.}(2022){Rix}, {Chandra}, {Andrae}, {Price-Whelan}, {Weinberg}, {Conroy}, {Fouesneau}, {Hogg}, {De Angeli}, {Naidu}, {Xiang}, \& {Ruz-Mieres}}]{Rix+2022}
{Rix}, H.-W., {Chandra}, V., {Andrae}, R., {et~al.} 2022, \apj, 941, 45, \dodoi{10.3847/1538-4357/ac9e01}

\bibitem[{Robert~C.(1989)}]{Kennicutt1989}
Robert~C., J.~K. 1989, The Astrophysical Journal, 344, 685, \dodoi{10.1086/167834}

\bibitem[{{Sawala} {et~al.}(2016){Sawala}, {Frenk}, {Fattahi}, {Navarro}, {Bower}, {Crain}, {Dalla Vecchia}, {Furlong}, {Helly}, {Jenkins}, {Oman}, {Schaller}, {Schaye}, {Theuns}, {Trayford}, \& {White}}]{Sawala+2016}
{Sawala}, T., {Frenk}, C.~S., {Fattahi}, A., {et~al.} 2016, \mnras, 457, 1931, \dodoi{10.1093/mnras/stw145}

\bibitem[{Schaye(2004)}]{Schaye2004}
Schaye, J. 2004, The Astrophysical Journal, 609, 667, \dodoi{10.1086/421232}

\bibitem[{Schaye \& Vecchia(2007)}]{SchayeDallaVecchia2008}
Schaye, J., \& Vecchia, C.~D. 2007, Monthly Notices of the Royal Astronomical Society, 383, 1210, \dodoi{10.1111/j.1365-2966.2007.12639.x}

\bibitem[{Schaye {et~al.}(2015)Schaye, Crain, Bower, Furlong, Schaller, Theuns, Vecchia, Frenk, McCarthy, Helly, Jenkins, Rosas-Guevara, White, Baes, Booth, Camps, Navarro, Qu, Rahmati, Sawala, Thomas, \& Trayford}]{EAGLE1}
Schaye, J., Crain, R.~A., Bower, R.~G., {et~al.} 2015, Monthly Notices of the Royal Astronomical Society, 446, 521, \dodoi{10.1093/mnras/stu2058}

\bibitem[{Schmidt(1959)}]{Schmidt1959}
Schmidt, M. 1959, The Astrophysical Journal, 129, 243, \dodoi{10.1086/146614}

\bibitem[{{Semenov} {et~al.}(2024{\natexlab{a}}){Semenov}, {Conroy}, {Chandra}, {Hernquist}, \& {Nelson}}]{Semenov+2024}
{Semenov}, V.~A., {Conroy}, C., {Chandra}, V., {Hernquist}, L., \& {Nelson}, D. 2024{\natexlab{a}}, \apj, 962, 84, \dodoi{10.3847/1538-4357/ad150a}

\bibitem[{{Semenov} {et~al.}(2024{\natexlab{b}}){Semenov}, {Conroy}, {Smith}, {Puchwein}, \& {Hernquist}}]{Semenov+2014}
{Semenov}, V.~A., {Conroy}, C., {Smith}, A., {Puchwein}, E., \& {Hernquist}, L. 2024{\natexlab{b}}, arXiv e-prints, arXiv:2409.18173, \dodoi{10.48550/arXiv.2409.18173}

\bibitem[{{Shen} {et~al.}(2003){Shen}, {Mo}, {White}, {Blanton}, {Kauffmann}, {Voges}, {Brinkmann}, \& {Csabai}}]{Shen+2003}
{Shen}, S., {Mo}, H.~J., {White}, S. D.~M., {et~al.} 2003, \mnras, 343, 978, \dodoi{10.1046/j.1365-8711.2003.06740.x}

\bibitem[{{Somerville} \& {Dav{\'e}}(2015)}]{SommervilleDave2015}
{Somerville}, R.~S., \& {Dav{\'e}}, R. 2015, \araa, 53, 51, \dodoi{10.1146/annurev-astro-082812-140951}

\bibitem[{{Somerville} {et~al.}(2008){Somerville}, {Hopkins}, {Cox}, {Robertson}, \& {Hernquist}}]{SCSAM1}
{Somerville}, R.~S., {Hopkins}, P.~F., {Cox}, T.~J., {Robertson}, B.~E., \& {Hernquist}, L. 2008, \mnras, 391, 481, \dodoi{10.1111/j.1365-2966.2008.13805.x}

\bibitem[{{Somerville} {et~al.}(2015){Somerville}, {Popping}, \& {Trager}}]{SCSAM2}
{Somerville}, R.~S., {Popping}, G., \& {Trager}, S.~C. 2015, \mnras, 453, 4337, \dodoi{10.1093/mnras/stv1877}

\bibitem[{Springel(2005)}]{GADGET}
Springel, V. 2005, Monthly Notices of the Royal Astronomical Society, 364, 1105, \dodoi{10.1111/j.1365-2966.2005.09655.x}

\bibitem[{{Springel}(2010)}]{Arepo}
{Springel}, V. 2010, \mnras, 401, 791, \dodoi{10.1111/j.1365-2966.2009.15715.x}

\bibitem[{Springel {et~al.}(2005)Springel, White, Jenkins, Frenk, Yoshida, Gao, Navarro, Thacker, Croton, Helly, Peacock, Cole, Thomas, Couchman, Evrard, Colberg, \& Pearce}]{Springel+2005}
Springel, V., White, S. D.~M., Jenkins, A., {et~al.} 2005, Nature, 435, 629, \dodoi{10.1038/nature03597}

\bibitem[{{Springel} {et~al.}(2018){Springel}, {Pakmor}, {Pillepich}, {Weinberger}, {Nelson}, {Hernquist}, {Vogelsberger}, {Genel}, {Torrey}, {Marinacci}, \& {Naiman}}]{TNG100-1}
{Springel}, V., {Pakmor}, R., {Pillepich}, A., {et~al.} 2018, \mnras, 475, 676, \dodoi{10.1093/mnras/stx3304}

\bibitem[{{Tan} {et~al.}(2024){Tan}, {Muzzin}, {Marchesini}, {Sok}, {Sarrouh}, \& {Marsan}}]{Tan+2024}
{Tan}, V. Y.~Y., {Muzzin}, A., {Marchesini}, D., {et~al.} 2024, \apj, 964, 177, \dodoi{10.3847/1538-4357/ad2c90}

\bibitem[{{Tang} {et~al.}(2021){Tang}, {Lin}, {Wang}, \& {Napolitano}}]{Tang+2021}
{Tang}, L., {Lin}, W., {Wang}, Y., \& {Napolitano}, N.~R. 2021, \mnras, 508, 3321, \dodoi{10.1093/mnras/stab2722}

\bibitem[{Thob {et~al.}(2019)Thob, Crain, McCarthy, Schaller, Lagos, Schaye, Talens, James, Theuns, \& Bower}]{Thob2019}
Thob, A. C.~R., Crain, R.~A., McCarthy, I.~G., {et~al.} 2019, Monthly Notices of the Royal Astronomical Society, 485, 972, \dodoi{10.1093/mnras/stz448}

\bibitem[{{van Dokkum} {et~al.}(2013){van Dokkum}, {Leja}, {Nelson}, {Patel}, {Skelton}, {Momcheva}, {Brammer}, {Whitaker}, {Lundgren}, {Fumagalli}, {Conroy}, {F{\"o}rster Schreiber}, {Franx}, {Kriek}, {Labb{\'e}}, {Marchesini}, {Rix}, {van der Wel}, \& {Wuyts}}]{vanDokkum+2013}
{van Dokkum}, P.~G., {Leja}, J., {Nelson}, E.~J., {et~al.} 2013, \apjl, 771, L35, \dodoi{10.1088/2041-8205/771/2/L35}

\bibitem[{Virtanen {et~al.}(2020)Virtanen, Gommers, Oliphant, Haberland, Reddy, Cournapeau, Burovski, Peterson, Weckesser, Bright, van~der Walt, Brett, Wilson, Millman, Mayorov, Nelson, Jones, Kern, Larson, Carey, İlhan Polat, Feng, Moore, VanderPlas, Laxalde, Perktold, Cimrman, Henriksen, Quintero, Harris, Archibald, Ribeiro, Pedregosa, van Mulbregt, Vijaykumar, Bardelli, Rothberg, Hilboll, Kloeckner, Scopatz, Lee, Rokem, Woods, Fulton, Masson, Häggström, Fitzgerald, Nicholson, Hagen, Pasechnik, Olivetti, Martin, Wieser, Silva, Lenders, Wilhelm, Young, Price, Ingold, Allen, Lee, Audren, Probst, Dietrich, Silterra, Webber, Slavič, Nothman, Buchner, Kulick, Schönberger, de~Miranda~Cardoso, Reimer, Harrington, Rodríguez, Nunez-Iglesias, Kuczynski, Tritz, Thoma, Newville, Kümmerer, Bolingbroke, Tartre, Pak, Smith, Nowaczyk, Shebanov, Pavlyk, Brodtkorb, Lee, McGibbon, Feldbauer, Lewis, Tygier, Sievert, Vigna, Peterson, More, Pudlik, Oshima, Pingel, Robitaille, Spura, Jones, Cera, Leslie, Zito, Krauss,
  Upadhyay, Halchenko, \& Vázquez-Baeza}]{scipy}
Virtanen, P., Gommers, R., Oliphant, T.~E., {et~al.} 2020, Nature Methods, 17, 261, \dodoi{10.1038/s41592-019-0686-2}

\bibitem[{{Vogelsberger} {et~al.}(2020){Vogelsberger}, {Marinacci}, {Torrey}, \& {Puchwein}}]{Vogelsberger+2020}
{Vogelsberger}, M., {Marinacci}, F., {Torrey}, P., \& {Puchwein}, E. 2020, Nature Reviews Physics, 2, 42, \dodoi{10.1038/s42254-019-0127-2}

\bibitem[{{Vogelsberger} {et~al.}(2014){Vogelsberger}, {Genel}, {Springel}, {Torrey}, {Sijacki}, {Xu}, {Snyder}, {Bird}, {Nelson}, \& {Hernquist}}]{Illustris2}
{Vogelsberger}, M., {Genel}, S., {Springel}, V., {et~al.} 2014, \nat, 509, 177, \dodoi{10.1038/nature13316}

\bibitem[{{Wang} {et~al.}(2015){Wang}, {Dutton}, {Stinson}, {Macci{\`o}}, {Penzo}, {Kang}, {Keller}, \& {Wadsley}}]{Wang+2015}
{Wang}, L., {Dutton}, A.~A., {Stinson}, G.~S., {et~al.} 2015, \mnras, 454, 83, \dodoi{10.1093/mnras/stv1937}

\bibitem[{{Weinberger} {et~al.}(2017){Weinberger}, {Springel}, {Hernquist}, {Pillepich}, {Marinacci}, {Pakmor}, {Nelson}, {Genel}, {Vogelsberger}, {Naiman}, \& {Torrey}}]{TNG1}
{Weinberger}, R., {Springel}, V., {Hernquist}, L., {et~al.} 2017, \mnras, 465, 3291, \dodoi{10.1093/mnras/stw2944}

\bibitem[{Wiersma {et~al.}(2009{\natexlab{a}})Wiersma, Schaye, \& Smith}]{Wiersma2009a}
Wiersma, R. P.~C., Schaye, J., \& Smith, B.~D. 2009{\natexlab{a}}, Monthly Notices of the Royal Astronomical Society, 393, 99, \dodoi{10.1111/j.1365-2966.2008.14191.x}

\bibitem[{Wiersma {et~al.}(2009{\natexlab{b}})Wiersma, Schaye, Theuns, Vecchia, \& Tornatore}]{Wiersma2009b}
Wiersma, R. P.~C., Schaye, J., Theuns, T., Vecchia, C.~D., \& Tornatore, L. 2009{\natexlab{b}}, Monthly Notices of the Royal Astronomical Society, 399, 574, \dodoi{10.1111/j.1365-2966.2009.15331.x}

\bibitem[{{Yin} {et~al.}(2009){Yin}, {Hou}, {Prantzos}, {Boissier}, {Chang}, {Shen}, \& {Zhang}}]{Yin+2009}
{Yin}, J., {Hou}, J.~L., {Prantzos}, N., {et~al.} 2009, \aap, 505, 497, \dodoi{10.1051/0004-6361/200912316}

\bibitem[{{York} {et~al.}(2000){York}, {Adelman}, {Anderson}, {Anderson}, {Annis}, {Bahcall}, {Bakken}, {Barkhouser}, {Bastian}, {Berman}, {Boroski}, {Bracker}, {Briegel}, {Briggs}, {Brinkmann}, {Brunner}, {Burles}, {Carey}, {Carr}, {Castander}, {Chen}, {Colestock}, {Connolly}, {Crocker}, {Csabai}, {Czarapata}, {Davis}, {Doi}, {Dombeck}, {Eisenstein}, {Ellman}, {Elms}, {Evans}, {Fan}, {Federwitz}, {Fiscelli}, {Friedman}, {Frieman}, {Fukugita}, {Gillespie}, {Gunn}, {Gurbani}, {de Haas}, {Haldeman}, {Harris}, {Hayes}, {Heckman}, {Hennessy}, {Hindsley}, {Holm}, {Holmgren}, {Huang}, {Hull}, {Husby}, {Ichikawa}, {Ichikawa}, {Ivezi{\'c}}, {Kent}, {Kim}, {Kinney}, {Klaene}, {Kleinman}, {Kleinman}, {Knapp}, {Korienek}, {Kron}, {Kunszt}, {Lamb}, {Lee}, {Leger}, {Limmongkol}, {Lindenmeyer}, {Long}, {Loomis}, {Loveday}, {Lucinio}, {Lupton}, {MacKinnon}, {Mannery}, {Mantsch}, {Margon}, {McGehee}, {McKay}, {Meiksin}, {Merelli}, {Monet}, {Munn}, {Narayanan}, {Nash}, {Neilsen}, {Neswold}, {Newberg}, {Nichol}, {Nicinski},
  {Nonino}, {Okada}, {Okamura}, {Ostriker}, {Owen}, {Pauls}, {Peoples}, {Peterson}, {Petravick}, {Pier}, {Pope}, {Pordes}, {Prosapio}, {Rechenmacher}, {Quinn}, {Richards}, {Richmond}, {Rivetta}, {Rockosi}, {Ruthmansdorfer}, {Sandford}, {Schlegel}, {Schneider}, {Sekiguchi}, {Sergey}, {Shimasaku}, {Siegmund}, {Smee}, {Smith}, {Snedden}, {Stone}, {Stoughton}, {Strauss}, {Stubbs}, {SubbaRao}, {Szalay}, {Szapudi}, {Szokoly}, {Thakar}, {Tremonti}, {Tucker}, {Uomoto}, {Vanden Berk}, {Vogeley}, {Waddell}, {Wang}, {Watanabe}, {Weinberg}, {Yanny}, {Yasuda}, \& {SDSS Collaboration}}]{SDSS}
{York}, D.~G., {Adelman}, J., {Anderson}, John~E., J., {et~al.} 2000, \aj, 120, 1579, \dodoi{10.1086/301513}

\end{thebibliography}
\bibliographystyle{aasjournal}

\end{document}